\documentclass[
 preprint,
 amsmath,amssymb,
 aps,
floatfix,
]{revtex4-2}
\usepackage[utf8]{inputenc}
\usepackage{amsmath}
\usepackage{amssymb}
\usepackage[hidelinks]{hyperref}
\usepackage{graphicx}
\usepackage{float}
\usepackage{bm}
\usepackage{titlesec}
\usepackage{physics}
\usepackage{mathtools}
\usepackage{appendix}

\usepackage{siunitx}
\usepackage[capitalise]{cleveref}
\usepackage{longtable}
\usepackage{booktabs}
\usepackage{makecell}
\usepackage{threeparttable}
\usepackage{setspace}
\usepackage{tabularx}

\usepackage{color}
\definecolor{blue}{rgb}{0.,0,1.}
\definecolor{red}{rgb}{1,0,0}
\definecolor{green}{rgb}{0,0.5,0}
\definecolor{pink}{rgb}{0.6,0,0.6}

\newcommand{\kevap}{k_{Q\text{-}\mathrm{He}}}
\newcommand{\betatb}{\beta_{Q\text{-}Q}}
\newcommand{\kB}{k_\mathrm{B}}
\newcommand{\dcpolarizability}{\alpha_\mathrm{s}}
\newcommand{\dcpolarizabilityani}{\Delta\alpha_\mathrm{s}}
\newcommand{\ionienergy}{I_0}
\newcommand{\ionirate}{W_\mathrm{i}}
\newcommand{\ionirateTenK}{W_{\mathrm{i},\SI{10}{K}}}
\newcommand{\Rayleighrate}{W_\mathrm{R}}
\newcommand{\rotRamanrate}{W_\mathrm{RR}}

\newcommand{\trapdepth}{T_{\mathrm{trap}}}
\newcommand{\extratrapdepth}{\Delta T_{\mathrm{trap}}}
\newcommand{\electronmass}{m_\mathrm{e}}
\newcommand{\boilingpoint}{T_\mathrm{B}}
\newcommand{\Ramanlossrate}{\Gamma_\mathrm{RR}}
\newcommand{\DeltaERRS}{\Delta E_\mathrm{RR}}
\newcommand{\QQinelasticrate}{\nu_{Q\text{-}Q,i}}
\newcommand{\QHeinelasticrate}{\nu_{Q\text{-}\mathrm{He},i}}
\newcommand{\nQcell}{n_{Q,\mathrm{LR}}}
\newcommand{\Edoth}{\dot E_{\mathrm{h}}}
\newcommand{\trappotJM}{U_{\tilde J, m}}

\DeclareSIUnit\angstrom{\protect \text {Å}}

\begin{document}
\title{Dynamics of a buffer-gas-loaded, deep optical trap for molecules}

\author{Ashwin Singh}
\email{ashwin\_singh@berkeley.edu}
\author{Lothar Maisenbacher}
\author{Ziguang Lin}
\author{\\Jeremy J. Axelrod}
\altaffiliation[Also at ]{Lawrence Berkeley National Laboratory, Berkeley, California 94720, USA}
\author{Cristian D. Panda}
\author{Holger M\"{u}ller}
\affiliation{Department of Physics, University of California, Berkeley, Berkeley, CA 94720, USA}

\date{\today}

\begin{abstract}
We describe an approach to optically trapping small, chemically stable molecules at cryogenic temperatures by buffer-gas loading a deep optical dipole trap. The \SI{\sim 10}{K} trap depth will be produced by a tightly focused, 1064-nm cavity capable of reaching intensities of hundreds of \si{GW/cm^2}. Molecules will be directly buffer-gas loaded into the trap using a helium buffer gas at \SI{1.5}{K}. The very far-off-resonant, quasielectrostatic trapping mechanism is insensitive to a molecule's internal state, energy level structure, and its electric and magnetic dipole moment. Here, we theoretically investigate the trapping and loading dynamics, as well as the heating and loss rates, and conclude that $10^4$--$10^6$ molecules are likely to be trapped. Our trap would open new possibilities in molecular spectroscopy, studies of cold chemical reactions, and precision measurement, amongst other fields of physics.
\end{abstract}

\maketitle

\section{Introduction}

\subsection{Background}

Cooling and trapping of atoms and ions has enabled unparalleled quantum control of both internal and external degrees of freedom \cite{2001_Metcalf_Book_Laser_cooling_and_Trapping}. It has led to advances in quantum information processing \cite{2021_Postler_Universal_Quantum_Gates}, quantum simulation \cite{2012_Bloch_Quantum_Simulations_with_ultracold_Quantum_Gases}, studies of cold phases of matter \cite{2012_Lewenstein_Ultracold_Atoms_in_Optical_Lattices}, spectroscopy and atomic clocks \cite{2015_Ludlow_Optical_Atomic_Clocks}, and tests of the Standard Model \cite{2018_Muller_alpha}, amongst other areas of physics. Molecules possess a rich internal energy level structure not seen in atoms, consisting of rotational and vibrational transitions in addition to electronic transitions. This complexity has generated great interest in cooling and trapping molecules \cite{2009_Krems_Cold_Molecules_Theory_Experiment_Applications, 2009_Friedrich_Why_Are_Cold_Molecules_So_Hot}. Trapped polar molecules have been proposed as potential qubits for quantum information processing \cite{2002_DeMille_QC_Polar_Molecules_Theory}, with the possibility of using their rotational states for quantum error correction \cite{2020_Albert_Robust_Encoding_of_a_Qubit_in_a_Molecule}. Cold, trapped molecules can exhibit unique phases of matter \cite{2017_Bohn_Cold_Molecules_Chemistry_Materials_Review, 2000_Lewenstein_BEC_Polar_Molecules_Theory, 2020_Ye_KRb_Degenerate_Gas}, and large polar molecules have applications in tests of Standard Model physics \cite{2018_Safronova_New_Physics_Atoms_Molecules,2018_ACME_EDM,2019_Ye_EDM_Review}. Furthermore, there is much interest in studying the cold chemistry of trapped molecules \cite{2017_Bohn_Cold_Molecules_Chemistry_Materials_Review, 2016_Balakrishnan_Controlled_Cold_Chemistry_Review, 2021_Heazlewood_Cold_Chemistry_Review, 2020_Puzzarini_Grand_Challenges_in_Astrochemistry}. For this reason, it is a goal of atomic, molecular and optical physics to develop a trap for arbitrary chemical species. 

In recent decades, significant progress has been made towards this ambitious goal. Crossed or merged molecular beams have become established techniques used to interrogate cold collisions over short interaction times \cite{1987_Herschbach_1987_Nobel,1987_Lee_1987_Nobel,1999_Phaneuf_Merged_beams}. Buffer-gas cooling has enabled the production of a broad spectrum of molecular samples at temperatures of order \SI{1}{K} \cite{2008_Doyle_buffer_gas_loading_review}, which has allowed for cold spectroscopic experiments \cite{2016_Spaun_Buffer_Gas_Molecule_Spectroscopy, 2016_Changala_Buffer_Gas_Molecule_Spectroscopy_Details_Fast_Shutter}, as well as trapping low-field-seeking paramagnetic molecules in magnetic traps \cite{1998_Doyle_CaH_Magnetic_Trap, 2007_Doyle_Magnetic_Trap_NH, 2019_Segev_Magnetic_Trapping_O2} and molecular ions in ion traps \cite{2021_Wester_Cl-H2_Complex_Studies, 2021_Wester_HeH+}. Polar molecules like ND$_3$ \cite{2000_Bethlem_Meijer_Electrostatic_Trapping_Ammonia}, OH \cite{2005_van_de_Meerakker_Meijer_Electrostatic_Trapping_OH}, CH$_3$F \cite{2005_van_de_Meerakker_Meijer_Electrostatic_Trapping_OH}, and CH$_2$O \cite{2016_Prehn_Rempe_Electrostatic_Trapping_Opeoelectrical_Cooling_CH2O} have been loaded from low-field-seeking states in buffer-gas beams into low-density electrostatic traps. It has been proposed that polar molecules in their true, high-field seeking ground state could be deeply trapped in an intense microwave trap addressing rotational transitions in the molecules \cite{2004_DeMille_Microwave_Trap_For_Cold_Polar_Molecules}, and progress has been made on trapping ultracold atoms in this kind of trap \cite{2019_Wright_Tarbutt_Microwave_Trap_for_Atoms_and_Molecules}. Other molecules, particularly bialkalis, have been trapped through photoassociation of ultracold atoms in optical traps \cite{2016_Moses_Bialkali_Review, 2020_Liu_Kang-kuen-ni_light_assisted_pathways, 2021_Cornish_RbCs_Qubit}. For a limited number of molecules, with a bivalent metal connected to a ligand to produce an “alkali-like” energy structure \cite{2020_Mitra_Laser_Cooling_CaCOH3, 2010_Shuman_DeMille_Laser_Cooling_SrF, 2018_Anderegg_Laser_Cooling_Trapping_CaF, 2020_Doyle_CaOH_Laser_Cooling, 2020_Ye_Sub-Doppler_Laser_Cooling_YO, 2018_Lim_Laser_Cool_YbF, 2017_Doyle_Laser_cooling_SrOH, 2020_Augenbraun_Laser_Cooling_YbOH, 2020_Zelevinsky_Laser_Cooling_BaH}, laser cooling has been achieved, albeit requiring multiple repump lasers addressing loss channels into other rovibrational states \cite{2021_Fitch_Tarbutt_Laser_Cooled_Molecules_Review}.

Unfortunately, none of these techniques is universally applicable, and access to cold molecules remains limited. Magnetic and electrostatic traps are limited to molecules with a large magnetic or electric dipole moment, respectively, and only trap molecules in excited, low-field-seeking states, which leads to loss through state-changing collisions.
Optical dipole traps (ODTs), based on the attractive force of an infrared laser beam, are now widely used. Still, even with hundreds of watts of laser power, they are limited to \si{mK} trap depths, and thus to the few species of molecules that have been laser-cooled.

\subsection{Buffer-gas-loaded dipole trap}

\begin{figure}
     \centering
         \includegraphics{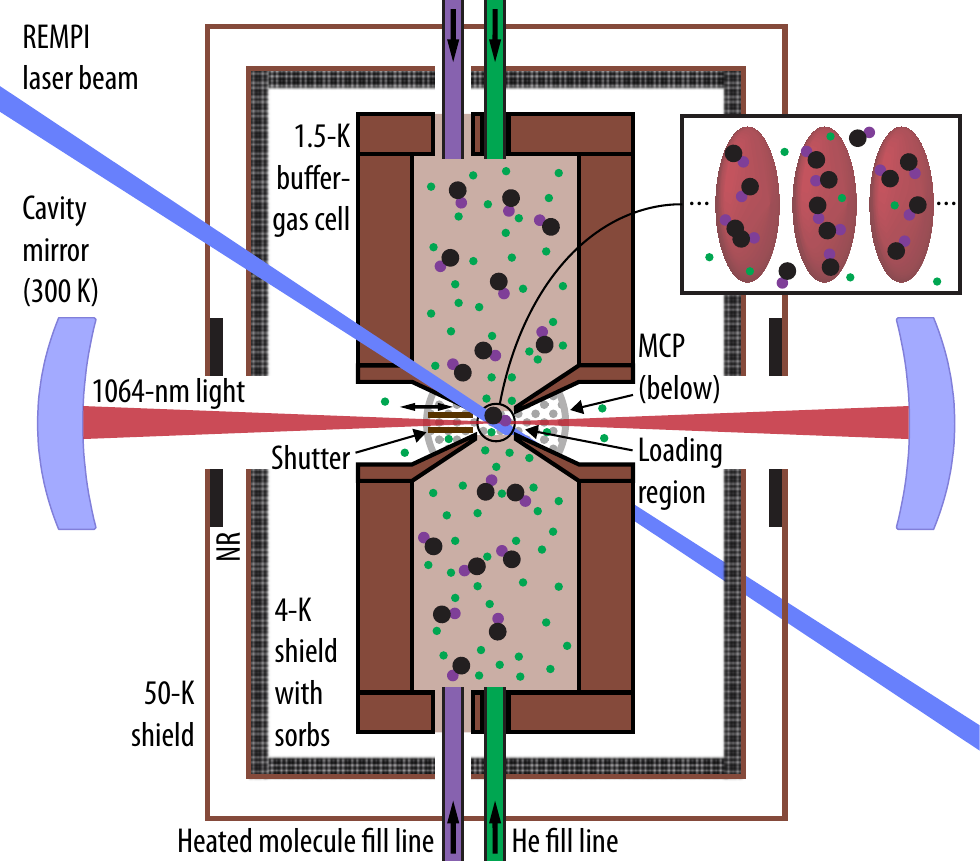}
        \caption{\footnotesize
        Schematic design of the buffer-gas-loaded, deep optical trap for molecules.
        The optical trap is formed by 1064-nm light (red beam) inside a 100-mm-long, near-concentric build-up cavity, whose mirrors are held at room temperature.
        Helium buffer gas (He; shown in green) at \SI{1.5}{K} and molecules (black and purple) at a higher temperature flow through fill lines into two opposing cryogenic cells, held at \SI{1.5}{K}.
        The molecules thermalize with the buffer gas and flow, along with the buffer gas, through conical apertures into the loading region.
        Here, the molecules are loaded through buffer-gas collisions into, and subsequently trapped in, the distinct lattice sites formed by the antinodes of the standing wave of the cavity (inset on top right).
        Shutters in front of the cells can quickly interrupt the flow out of the cells to create an isolated sample of trapped molecules.
        Resonance-enhanced multiphoton ionization (REMPI) with a second laser beam (blue beam) ionizes the molecules, which are then detected with a microchannel plate (MCP).
        Also shown are radiation shields at \SI{50}{K} and \SI{4}{K} (with attached charcoal sorbs used for cryopumping of He), and non-reflective (NR) material to absorb scattered light.
        The drawing is to scale.
        }
        \label{fig: cell design}
\end{figure}

Here, we consider trapping neutral molecules in their ground state with a deep optical dipole trap, taking an additional step towards the ultimate goal of a universal trap. Our proposed design is shown in \cref{fig: cell design}. This trap, originally envisioned decades ago \cite{1995_Friedrich_Alignment_Trapping_Spheroidal_Wave_Eqn_Theory}, is made possible by the development of high-intensity cavities able to generate continuous-wave (CW) laser intensities over \SI{400}{GW/cm^2} with 1064-nm light \cite{2020_Turnbaugh_microscope_paper, 2019_Schwartz_LPP, 2017_Schwartz_LPP}. At such high intensities, a very far-off-resonant, quasielectrostatic dipole trap has a trap depth of order \SI{10}{K} for most molecules. Buffer-gas cooling with $^4$He in the trap volume is therefore sufficient to load molecules into the trap. After equilibration, the buffer gas is pumped out of the chamber, leaving a trapped sample in the laser beam. These methods rely only on a molecule's DC polarizability, which is nonzero for any species, and do not require a particular energy level structure or magnetic or polar molecules. This not only allows a large number of species that cannot be trapped by existing methods to be loaded into our dipole trap, but also for two or more different species to be trapped at the same time.

Although our methods should be applicable to molecules of any symmetry, we will limit ourselves to discussing linear molecules for simplicity. \cref{tab: molecule list} summarizes our results. Molecules with typical mean DC polarizabilities ($\dcpolarizability \gtrsim\SI{2}{\angstrom^3}$, averaged over all orientations), high ionization energies ($\ionienergy\sim\SI{12}{eV}$), few atoms ($\leq 3$), and low DC polarizability anisotropies ($\dcpolarizabilityani\lesssim\SI{1}{\angstrom^3}$) are good candidates for our trap, but other molecules could be trapped with small modifications to the trap design. With so many molecules to consider, for clarity we will frequently refer to a hypothetical molecule $Q$ as a representative of molecules we wish to trap. $Q$ is a small, chemically stable (SCS) molecule with characteristics similar to many of the molecules in \cref{tab: molecule list}. We choose the polarizability and ionization energy of $Q$ to be $\dcpolarizability = \SI{2}{\angstrom^3}$ and $\ionienergy = \SI{12}{eV}$, respectively, based on the species listed in \cref{tab: molecule list}. Furthermore, we assume $Q$'s polarizability to be isotropic (i.e., zero $\dcpolarizabilityani$), and take $Q$ to have a boiling point of $\boilingpoint = \SI{200}{K}$ and a molecular mass of $m_Q = \SI{30}{u}$. These are typical values for SCS molecules, and we note that the trapping is not sensitive to the exact values.

\subsection{Outline}

The experimental design, including the cavity and the buffer-gas cells, is discussed in \cref{sec: Cavity Design}.
\cref{sec: Dipole Trapping of Molecules} treats the very far-off-resonant, quasielectrostatic dipole trapping of molecules.
\cref{sec: Effects of High Intensity Light on Trapped Molecules} considers the effects of the high-intensity light on the molecules in the trap.
The dynamics of buffer-gas cooling and loading into the trap are the subject of \cref{sec: Buffer Gas Dynamics}.
Finally, the ionization-based detection of molecules is reviewed in \cref{sec: Detection}, and a summary and outlook are given in \cref{sec: Outlook}.

\newcommand{\myskip}{\hskip 9 pt}

\begin{table*}
\begin{spacing}{1}
\centering
\begin{threeparttable}
\caption{\footnotesize{Candidate small, chemically stable (SCS) molecules, including our hypothetical molecule $Q$. All molecules listed are linear symmetric tops. Entries are calculated assuming a laser intensity of $I=\SI{300}{GW/cm^2}$ at a wavelength of $\lambda = \SI{1064}{nm}$.
The mean value $\dcpolarizability$ and the anisotropy $\dcpolarizabilityani$ of the DC polarizability determine the trap depth $\trapdepth$, with $\extratrapdepth$ being the contribution to $\trapdepth$ from molecular alignment by the intense trap light (see \cref{Appendix: Rotational State Hybridization and Extra Trap Depth}).
The Rayleigh and rotational Raman scattering rates, $\Rayleighrate$ and $\rotRamanrate$, are proportional to the square of $\dcpolarizability$ and $\dcpolarizabilityani$, respectively (see \cref{subsec: Rayleigh Scattering} and \cref{subsec: Raman Scattering}}).
The ionization energy $\ionienergy$ is used to estimate the ionization rate $\ionirate$ (see \cref{subsec: Ionization of Isolated Molecules}).
Additionally, the mass $m$ and boiling point $\boilingpoint$ are given.
Resonance-enhanced multiphoton ionization (REMPI) schemes for detection of each molecule are shown (see \cref{sec: Detection}), where $n+l$ refers to a scheme absorbing a total of $n+l$ photons at the given wavelength(s).
Entries and relevant spectra are taken from a variety of public databases and individual publications \cite{CCCBDB,NIST_Webbook,HITRAN,MPI_Spectral_Atlas_Database,Bridge1966,1978_Bogaard_Polarizability_Anisotropies}.}
\footnotesize
\begin{tabular*}{\textwidth}{l @{\myskip} r @{\myskip} r @{\myskip} r @{\myskip} r @{\myskip} r @{\myskip} r @{\myskip} r @{\myskip} r @{\myskip}r @{\myskip} r @{\extracolsep{\fill}} l}
\toprule
Species & \makecell[r]{$\dcpolarizability$\\(\si{\angstrom^3})} & \makecell[r]{$\dcpolarizabilityani$\\(\si{\angstrom^3})} & \makecell[r]{$\trapdepth$\\(\si{K})} & \makecell[r]{$\extratrapdepth$\\(\si{K})}& \makecell[r]{$\ionienergy$\\(\si{eV})} & \makecell[r]{$\Rayleighrate$\\(\si{s^{-1}})} & \makecell[r]{$\rotRamanrate$\\(\si{s^{-1}})} & \makecell[r]{$\ionirate$\\(s$^{-1}$)} & \makecell[r]{$m$\\(\si{u})} & \makecell[r]{$\boilingpoint$\\(\si{K})} & REMPI scheme \\
\midrule
\multicolumn{12}{l}{Suitable for cooling, trapping, and REMPI detection with the experimental parameters proposed here}\\
\midrule
N$_2$ & 1.7 & 0.7 & 7.8 & 0.1 & 15.6 & 479 & 17 & \num{1.6e-18} & 28 & 77 & 2+2, \SI{283}{nm} \cite{2014_McGuire_REMPI_N2_2_2}\tnote{a} \\
CO & 2.0 & 0.5 & 8.9 & 0.0 & 14.0 & 624 & 10 & \num{3.7e-15} & 28 & 82 & 2+1, \SI{230}{nm} \cite{1997_Peng_REMPI_CO} \\
O$_2$ & 1.6 & 1.1 & 7.3 & 0.2 & 12.1 & 399 & 43 & \num{2.5e-09} & 32 & 90 & 2+1, 215--\SI{240}{nm} \cite{1992_Yokelson_REMPI_O2} \\
HCl & 2.5 & 0.3 & 11.5 & 0.0 & 12.7 & 1035 & 4 & \num{3.3e-10} & 36 & 188 & 2+1, 208--\SI{260}{nm} \cite{1991_Green_REMPI_HCl} \\
Xe$^\dagger$ & 4.0 & 0.0 & 18.2 & 0.0 & 12.1 & 2626 & 0 & \num{2.2e-09} & 131 & 165 & 2+1, \SI{224}{nm} \cite{2020_Galea_REMPI_Xe} \\
\midrule
\multicolumn{12}{l}{Lower mean DC polarizability $\dcpolarizability$, requiring higher intensity $I$}\\
\midrule
H & 0.7 & 0.0 & 3.0 & 0.0 & 13.6 & 73 & 0 & \num{1.1e-12} & 1 & 21 & 2+1, \SI{243}{nm} \cite{1964_Zernik_REMPI_H} \\
H$_2$ & 0.8 & 0.3 & 3.6 & 0.0 & 15.4 & 101 & 4 & \num{2.0e-18} & 2 & 20 & 2+1, \SI{202}{nm} \cite{1982_Marinero_REMPI_H2} \\
$^4$He$^\ddagger$ & 0.2 & 0.0 & 0.9 & 0.0 & 24.6 & 7 & 0 & \num{7.8e-44} & 4 & 4 & --- \\
\midrule
\multicolumn{12}{l}{Might require longer wavelength because of larger rotational Raman scattering rate $\rotRamanrate$}\\
\multicolumn{12}{l}{and/or ionization rate $\ionirate$}\\
\midrule
CO$_2$ & 2.5 & 2.1 & 13.8 & 2.4 & 13.8 & 1029 & 160 & \num{6.8e-13} & 44 & 195 & 3+1, 280--\SI{330}{nm} \cite{1993_Taylor_REMPI_CO2} \\
N$_2$O & 3.0 & 3.0 & 17.6 & 3.9 & 12.9 & 1471 & 319 & \num{4.1e-12} & 44 & 185 & 1+2+1, \SI{204}{nm} \cite{1993_Hanisco_REMPI_N2O}\tnote{b} \\
Cl$_2$ & 4.6 & 2.6 & 25.2 & 4.2 & 11.5 & 3479 & 246 & \num{6.4e-07} & 70 & 239 & 2+1, 220--\SI{260}{nm} \cite{1996_Al-Kahali_REMPI_Cl2} \\
CS$_2$ & 8.7 & 9.5 & 63.5 & 23.7 & 10.1 & 12529 & 3255 & \num{1.2e-03} & 76 & 319 & 1+1, 208--\SI{217}{nm} \cite{2008_Hu_REMPI_CS2} \\
NO & 1.7 & 0.8 & 7.8 & 0.1 & 9.3 & 472 & 26 & \num{2.0e-01} & 30 & 121 & 1+1, \SI{226}{nm} \cite{2006_Fulton_REMPI_NO} \\
\midrule
\multicolumn{12}{l}{Hypothetical molecule $Q$ (see text)}\\
\midrule
$Q$ & 2.0 & 0.0 & 9.1 & 0.0 & 12.0 & 655 & 0 & \num{2.8e-09} & 30 & 200 & --- \\
\bottomrule
\end{tabular*}
\begin{tablenotes}
\item[$^\dagger$]Atomic Xe is included as its properties make it a good test species for experimental designs.
\item[$^\ddagger$]Included for reference.
\item[a]Also: 2+1, \SI{202}{nm} \cite{2009_Salumbides_REMPI_N2}.
\item[b]Photodissociation of N$_2$O into N$_2$ and O, followed by 2+1 REMPI of N$_2$, using the same laser beam at \SI{204}{nm}.
\end{tablenotes}
\label{tab: molecule list}
\end{threeparttable}
\end{spacing}
\end{table*}

\section{Experimental design}\label{sec: Cavity Design}

\subsection{Near-concentric build-up cavity for 1064-nm light}

The high-intensity, 1064-nm trap light is produced by a near-concentric build-up cavity, modeled on a demonstrated cavity designed by our group \cite{2020_Turnbaugh_microscope_paper}. The demonstrated cavity has a finesse of $\mathcal{F} \approx 37000$ and a power enhancement factor $P_{\mathrm{circ}}/P_{\mathrm{in}} \approx 9000$ (accounting for coupling inefficiencies and technical losses). By coupling input powers $P_{\mathrm{in}} \approx \SI{15}{W}$ into the fundamental TEM$_{00}$ mode of the cavity, circulating powers of $P_{\mathrm{circ}} \geq \SI{125}{kW}$ have been achieved. With the 20-mm-long, symmetric cavity operating \SI{10}{\micro\meter} from concentricity, the mode has been focused to a waist of $w_0\approx \SI{8.7}{\micro\meter}$ ($1/e^2$ intensity radius), resulting in an intensity $\geq\SI{400}{GW/cm^2}$ at the antinodes of the cavity's standing intensity wave \cite{2020_Turnbaugh_microscope_paper}.

Compared to \cite{2020_Turnbaugh_microscope_paper}, we require a five-fold increase in the cavity length to \SI{100}{mm}, which creates space for the cryogenic system surrounding the cavity focus.
To achieve the same mode waist, the longer cavity needs to be aligned closer to concentricity, which makes it more sensitive to misalignments. In particular, thermal mirror deformation from laser-induced, local heating has a tendency to increase the mode waist and hence decrease the intensity in the cavity. While a detailed analysis is beyond the scope of this work, using mirrors made from ultra-low expansion glass (ULE), coated with an ultra-low-absorption reflective coating (with an absorption as low as 0.4 ppm at \SI{1064}{nm} (unpublished measurement by our group)), will allow us to achieve an intensity of $I = \SI{300}{GW/cm^2}$ with a mode waist of $w_0=\SI{8}{\micro\meter}$, based on extrapolating the data of \cite{2020_Turnbaugh_microscope_paper, 2020_Axelrod_reversal_of_ponderomotive_potential}.

Our cavity mirrors will be kept outside the cryogenic system to prevent buildup of frozen molecules on their surfaces.
The high intensity and high finesse of the cavity limits the use of optical elements like windows in the cavity's optical path.
Therefore, as shown in \cref{fig: cell design}, our cryogenic system will have apertures that let in the cavity light. To prevent diffraction losses, the diameter of these apertures is here chosen to be four times the $1/e^2$ intensity beam diameter.

\subsection{Buffer-gas cells}

A 1.5-K buffer gas is used to both cool and load the molecules into the trap.
Many aspects of the design of buffer-gas cells benefit from the extensive work done in other experiments \cite{1998_Doyle_CaH_Magnetic_Trap,2007_Doyle_Magnetic_Trap_NH,2018_Truppe_A_buffer_gas_beam_source_for_short_intense_and_slow_molecular_pulses,2019_Anderegg_thesis_ultracold_molecules,2020_Baum_CaOH_MOT_cryogenic_techniques}.
However, as opposed to creating a buffer-gas beam \cite{2019_Anderegg_thesis_ultracold_molecules,2020_Baum_CaOH_MOT_cryogenic_techniques} or loading our trap within a buffer-gas cell \cite{1998_Doyle_CaH_Magnetic_Trap,2007_Doyle_Magnetic_Trap_NH}, we here opt to populate a millimeter-scale loading region between two cells, centered on the dipole trap, with buffer gas and cold molecules.
This minimizes the amount of gas pumpout required and allows for optical access.
Furthermore, since our experiment is critically dependent on highly reflective cavity mirrors whose sensitivity to ablation byproducts is unknown, we will not source molecules in the cells using laser ablation, and instead will flow molecules into the buffer-gas cells through heated fill lines.
The resulting dual-buffer-gas-cell geometry is depicted in \cref{fig: cell design}, and uses two closely spaced, cylindrically symmetric, opposing 1.5-K cells of $\sim$30-mm dimensions and conical apertures with a diameter of \SI{5}{mm}.
This design was validated to achieve the required densities and sufficient thermalization using the direct simulation Monte Carlo (DSMC) method \cite{2021_Takahashi_Simulation_of_Cryogenic_Buffer_Gas_Beams}.
Vacuum is maintained outside the loading region by differential pumping with activated charcoal sorbs on the 4-K shield at \SI{6}{L/s/cm^2} \cite{2007_Day_Basics_and_Applications_of_Cryopumps}. Pumping out the loading region is done by shutting off flow from the buffer-gas cells using rapidly actuating, cryogenic shutters. Based on the demonstrated shutter of \cite{2016_Changala_Buffer_Gas_Molecule_Spectroscopy_Details_Fast_Shutter}, we assume a shutter can actuate in \SI{1}{ms}, which sets the pumpout timescale from the loading region in our experiment. 

\subsection{Heat load on cryogenic system}

The cells will be maintained at \SI{1.5}{K} by thermal contact to a $^4$He-filled 1-K pot, which is pre-cooled and radiation-shielded by 4-K and 50-K stages of a pulse-tube cryocooler. Commercially available cryocoolers are able to provide \SI{100}{mW} of cooling power at \SI{1.5}{K}, which sets the heat-load budget for the experiment. 

The conductive heat load from the heated molecule fill lines has been shown to be manageable by previous experiments, including one using a fill line for water held at \SI{280}{K} connected to a similar buffer-gas cell \cite{2022_Doyle_CaOH_MOT}.
Convective heat loads will be managed by differential pumping with activated charcoal. Radiative heat loads on the buffer-gas cells through the apertures in the radiation shields are estimated to be less than \SI{10}{mW}.

Another major heat load is scattering of the high-intensity cavity light off the mirrors through the apertures onto the buffer-gas cells. Polishing the cells to enhance their reflectivity, and carefully designing the radiation shields, including adding non-reflective (NR) material to the 50-K shields (see \cref{fig: cell design}), manages this heat load (see \cref{Appendix: Scattered Heat Load}).

\section{Dipole trapping of molecules}\label{sec: Dipole Trapping of Molecules}

\subsection{Trap depth}\label{subsec: trap depth}
Most small, chemically stable (SCS) molecules have limited activity in the optical and near-infrared (NIR) spectrum. Therefore, a dipole trap using light at a wavelength of $\lambda = \SI{1064}{nm}$ is red-detuned by several harmonics from the first electronic transition of a typical SCS molecule. In this regime, the commonly used rotating wave approximation does not apply, and the light produces a quasielectrostatic trap by creating a dipole potential \cite{1995_Takekoshi_QUEST_theory, 1996_Takekoshi_QUEST, 2000_Hans_Engler_QUEST_thesis, 2000_Grimm_optical_trapping_review}
\begin{equation}\label{eq: trap depth}
    U \approx -\frac{\dcpolarizability}{2}\langle|\mathbf{E}|^2\rangle = -\frac{\dcpolarizability}{4}E_0^2 = -\frac{\dcpolarizability I}{2\epsilon_0 c},
\end{equation}
where $\dcpolarizability$ is the mean DC polarizability \cite{2013_LeKien_dynaic_polarizability_tutorial_derivation} and $\langle \cdot \rangle$ denotes a time average over the optical period $2\pi/\omega = \lambda/c$ ($\omega$: optical angular frequency, $c$: speed of light). The oscillating electric field is given by $\mathbf{E} = E_0\cos(\omega t)\mathbf{\hat{E}}$, with $\mathbf{\hat{E}}$ the normalized polarization vector and $E_0$ the electric field amplitude. The intensity is $I=\epsilon_0 c \langle |\mathbf{E}|^2\rangle = \epsilon_0 c E_0^2/2$  ($\epsilon_0$: permittivity of free space). From \cref{eq: trap depth} the intensity required for a \SI{10}{K} trap depth depends simply on $\dcpolarizability$ as $I|_{T_\mathrm{trap} = \SI{10}{K}} \approx (\SI{659}{GW/cm^2})/(\dcpolarizability\,\text{[\si{\angstrom^3}]})$.

The trap depth of several \si{K} is comparable to the rotational energy level spacing of many SCS molecules. The trap may therefore significantly hybridize the rotational levels and align the molecules' most polarizable axis with the optical polarization, leading to an increase in the trap depth that is not accounted for in \cref{eq: trap depth} \cite{1995_Friedrich_Alignment_Trapping_Spheroidal_Wave_Eqn_Theory, 1995_Friedrich_Spheroidal_Wave_Eqn_Theory_Details,2022_Friedrich_ElectroOptic_Trap_for_Molecules}. The trap depth $\trapdepth=\max{(|U|/k_B)}$ ($\kB$: Boltzmann's constant) as listed in \cref{tab: molecule list} therefore includes a correction to \cref{eq: trap depth} of $\extratrapdepth$, which depends on the polarizability anisotropy $\dcpolarizabilityani$ and is discussed in \cref{Appendix: Rotational State Hybridization and Extra Trap Depth}. This correction is usually only a few percent of $\trapdepth$, but for highly anisotropic molecules like CS$_2$ it can be substantial.

As seen in \cref{tab: molecule list}, most molecules have $\trapdepth\gtrsim\SI{10}{K}$ at our intensity $I=\SI{300}{GW/cm^2}$, more than six times the buffer gas temperature. Species with particularly low polarizabilities, like H and H$_2$, may still be trapped deeply if much higher intensities can be generated. Helium buffer gas atoms are also attracted to the trap center, but with a smaller trap depth of $\trapdepth = \SI{0.95}{K}$ due to their low polarizability. We ignore this small effect in the rest of this publication.

Molecules typically have a rich spectrum of pure rovibrational transitions. Although the trap light is blue-detuned from all pure rovibrational transitions, these transitions do not contribute significantly to the dynamic polarizability at \SI{1064}{nm} \cite{2007_Kongsted_dynamic_polarizability_methane, 2013_Tomza_Anisotropic_Polarizability_Trap_Depth_Heating_Rb2, 2020_He_Scattering_and_Absorption_Cross_Sections_Atmospheric_Gases_UV_Vis}, and hence do not lead to antitrapping. Typical SCS molecules are in their electronic ground state and predominantly in their rovibrational ground state at our chosen buffer gas temperature of $T = \SI{1.5}{K}$.

\subsection{Optical potential}\label{subsec: optical potential}

The trap itself is characterised firstly by the trap depth $\trapdepth = \max{|U|}/\kB$. Since our trap light is the Gaussian TEM$_{00}$ mode of an optical cavity, the spatial dependence of the trapping potential, in cylindrical coordinates measured relative to the focus of the cavity (placed at the origin), is 
\begin{align}
    U(r, z) &= -\kB \trapdepth\frac{e^{-2r^2/w^2(z)}}{1+z^2/z_\mathrm{R}^2}\cos^2\left(kz+\tan^{-1}\left(\frac{z}{z_\mathrm{R}}\right)+\frac{kr^2}{2R(z)}\right)
    \nonumber\\
    &\approx -\kB \trapdepth\frac{w_0^2}{w^2(z)}e^{-2r^2/w^2(z)}\cos^2(kz),
    \label{eq: U trap}
\end{align}
where the approximation is valid near the focus of the trap. Here, $k = 2 \pi / \lambda$, $w(z) = w_0\sqrt{1 + z^2/z_\mathrm{R}^2}$, $R(z) = z + z_\mathrm{R}^2/z$, and $z_\mathrm{R} = kw_0^2/2$ is the Rayleigh range. 

Appreciable molecule loading will only occur in the region where the trap potential is large compared to the buffer gas temperature. The trap volume is therefore characterized approximately by the volume within which $U(r,z)\geq 3\kB T$, which is sensitive to the dimensionless trap depth parameter $\eta_0 = \trapdepth/T$:

\begin{equation}\label{eq: V trap order of magnitude}
    V \approx \frac{2\pi w_0^2 z_\mathrm{R}}{9}\left[\sqrt{\frac{\eta_0}{3}-1}\left(\frac{\eta_0}{3}+5\right)-6\tan^{-1}\left(\sqrt{\frac{\eta_0}{3}-1}\right)\right].
\end{equation}
We note that $V$ thus scales as $w_0^4$. This volume is split non-uniformly into $4\sqrt{\frac{\eta_0}{3}-1}z_\mathrm{R}/\lambda$ distinct lattice sites, spaced axially by $\lambda/2$.

We propose $Q$ be loaded from a buffer gas at \SI{1.5}{K} into our 1064-nm dipole trap with a peak intensity of \SI{300}{GW/cm^2} and an $\SI{8}{\micro\meter}$ waist. The resulting trap depth is \SI{9.4}{K}, and $\eta_0 = 6.3$. The corresponding approximate trap volume is $V = 1.8 w_0^2z_\mathrm{R} = \SI{2.2e-8}{cm^{-3}}$, and there are 743 distinct lattice sites with a site depth greater than $3\kB T$. 

\section{Effects of high-intensity light on trapped molecules}\label{sec: Effects of High Intensity Light on Trapped Molecules}

In this section, we review Rayleigh and Raman scattering, photoionization, and photodissociation of molecules, and determine the class of molecules which is unaffected by these processes.

\subsection{Rayleigh scattering}\label{subsec: Rayleigh Scattering}

In a quasielectrostatic trap, photon scattering is dominated by Rayleigh scattering \cite{1995_Takekoshi_QUEST_theory}, with a cross section of $\sigma_\mathrm{R} = 8\pi^3\dcpolarizability^2/(3\epsilon_0^2\lambda^4)$ \cite{2002_Andrews_an_introduction_to_laser_spectroscopy}. Even at the high intensity of $I=\SI{300}{GW/cm^2}$, the Rayleigh scattering rate $\Rayleighrate=I\sigma_R/\hbar\omega$ is only $\mathrel{\sim\!10^3}$\,\si{s^{-1}} for the molecules in \cref{tab: molecule list}. 

The maximum amount by which the Rayleigh scattering can be enhanced by the cavity is given by the Purcell factor $F_P = 6\lambda^2\mathcal{F}/(\pi^3w_0^2)$ \cite{2010_Motsch_CavityRayleigh}, which is 119 for our cavity. Taking this into account, the heating rate from photon scattering is \SI{<0.1}{K/s} for all molecules in \cref{tab: molecule list} \cite{2000_Grimm_optical_trapping_review}, which is negligible in our experiment. Similarly, the low scattering rate means the maximum density limit from reabsorption of scattered light is $\mathrel{>\!10^{35}}$\,\si{cm^{-3}}, which can be ignored \cite{1991_Sesko_Density_Limit_from_Repulsive_Scattering}. In reality, $F_P=119$ overestimates the cavity enhancement of the Rayleigh scattering, as effects such as the recoil shift and collisional broadening will shift the scattered photons off resonance with the cavity.

\subsection{Raman scattering}\label{subsec: Raman Scattering}

A small fraction of photon scattering events are inelastic. These spontaneous Raman scattering events lead to rotational or vibrational excitation of trapped molecules.
Rotational Raman scattering (RRS) occurs in molecules with a nonzero polarizability anisotropy $\dcpolarizabilityani$. The molecule absorbs a cavity photon and emits a photon of a different energy, the energy difference accounting for a change in rotational state of the molecule. In linear molecules with zero electronic angular momentum about the symmetry axis, the Raman selection rule is $\Delta J = 0,\pm 2$, where $J$ is the rotational angular momentum of the molecule \cite{1950_Herzberg}. RRS from the $J=0$ ground state of molecules occurs at a rate \cite{Bridge1966, 1974_Penney_Rotational_Raman_Scattering}
\begin{equation}\label{eq: RRS rate}
    \rotRamanrate = \sigma_{RR}\frac{I}{\hbar\omega} = \frac{512\pi^5}{135}(1+2\rho)\frac{1}{(\lambda')^4}\left(\frac{\dcpolarizabilityani}{4\pi\epsilon_0}\right)^2 \frac{I}{\hbar\omega},
\end{equation}
where $\lambda'\approx\lambda$ is the wavelength of the scattered photon, $\omega = 2\pi c/\lambda$ is the incident angular frequency, and $\rho=3/4$ is the depolarization ratio of the Raman transition \cite{1970_Dallemand_Depolarization_Ratio} ($\hbar=h/2\pi$: reduced Planck constant). In rare cases, e.g., for NO, the molecular ground state has nonzero electronic angular momentum about the symmetry axis, in which case \cref{eq: RRS rate} is modified by different selection rules and Placzek-Teller coefficients \cite{1970_Shotton_RRS_NO}. 

Molecules with $\rotRamanrate\lesssim\SI{70}{s^{-1}}$, such as N$_2$, CO, O$_2$ and HCl, are ideal first candidates for our proposed experiment, as they can be isolated from the buffer gas, and potentially evaporatively cooled, faster than they are rotationally heated (see \cref{subsubsec: evaporative effects and other considerations}). The RRS cross section scales with $\dcpolarizabilityani^2$ and $\lambda^{-4}$, so molecules with higher polarizability anisotropies can still be trapped and isolated as described in this work using a trap with $\lambda>\SI{1064}{nm}$ (see \cref{tab: molecule list}).

Vibrational Raman scattering rates in molecules are typically less than \SI{\sim 1}{s^{-1}} at \SI{300}{GW/cm^2} of 1064-nm light, and can be ignored \cite{1966_Porto_Angular_Distribution_Raman_Scattering, 1973_Fenner_Vibrational_Raman_Cross_Sections_of_Simple_Gases, 2022_Carvalho_Vibrational_Raman_Scattering_Cross_Sections}.

\subsection{Ionization}\label{subsec: Ionization of Isolated Molecules}

The molecules suitable for trapping in \cref{tab: molecule list} have ionization energies $\ionienergy>\SI{12}{eV}$  and no transitions at any low harmonics of the 1064-nm trap light. Under these conditions, we can approximately estimate the ionization rate $\ionirate$ of isolated molecules in the trap. A $Q$ molecule with $\ionienergy=\SI{12}{eV}$ trapped with $I=\SI{300}{GW/cm^2}$ has a Keldysh parameter of 14, meaning non-resonance-enhanced ionization occurs via multiphoton ionization (MPI) \cite{1965_keldysh_ionization}. The ionization rate $\ionirate$ scales as $\ionirate = \sigma_nI^n$, where $n=11$ is the number of photons needed to reach the continuum and $\sigma_n$ is the MPI cross section.

It is difficult to estimate $\sigma_n$ \textit{ab initio}. As a starting point, we use Popruzhenko's formula \cite{2008_Popruzhenko_coulomb_correction_to_Keldysh_high_freq} for hydrogen-like atoms, which is a Perelomov-Popov-Terent'ev (PPT) ionization formula \cite{1966_Perelomov_update_Keldysh_preexp}. The formula matches (within two orders of magnitude) experimental results for ionization rates of noble gases \cite{1983_LHuillier_MPI_noble_gases_1064, 1988_Perry_multiphoton_ionisation_coefficients_noble_gases} and air \cite{2020_Woodbury_PRL_AbsoluteMeasuremntLaserIonization}. PPT formulas also match \textit{ab initio} estimates of ionization rates for polyatomic molecules \cite{2015_Zhao_molecular_ionization_rates_formula}. Popruzhenko's formula predicts $\ionirate=\SI{2.8e-9}{s^{-1}}$ for $Q$. $\ionirate$ scales down superexponentially with increasing $\ionienergy$, meaning molecules with slightly higher ionization energies have substantially smaller ionization rates, as seen in \cref{tab: molecule list}.

The ionization rate can be minimized by either using a molecule with a high ionization energy, or a molecule with a high polarizability which requires less laser intensity for trapping. To compare these approaches, we compute the ionization rate at an intensity for which the trap depth is \SI{10}{K}, $\ionirateTenK = \ionirate(I|_{T_\mathrm{trap} = \SI{10}{K}})$. This is plotted for different values of the mean DC polarizability $\dcpolarizability$ and ionization energy $\ionienergy$ in \cref{fig: pPAIR}, ignoring the correction to the trap depth $\extratrapdepth$ discussed in \cref{Appendix: Rotational State Hybridization and Extra Trap Depth}. The near-vertical contours roughly represent lines where the ionization energy corresponds to an integer multiple of the photon energy, so the absorption of an extra photon becomes necessary to ionize the molecule, thereby greatly decreasing the ionization rate.
By comparison, the higher intensity necessary for lower polarizabilities has a much smaller influence on the ionization rate. Thus, ionization energy is a far more important quantity than polarizability in the selection of suitable molecules, granted that a sufficiently high laser intensity can be achieved.
\begin{figure}
    \centering
    \includegraphics[width = 0.5\linewidth]{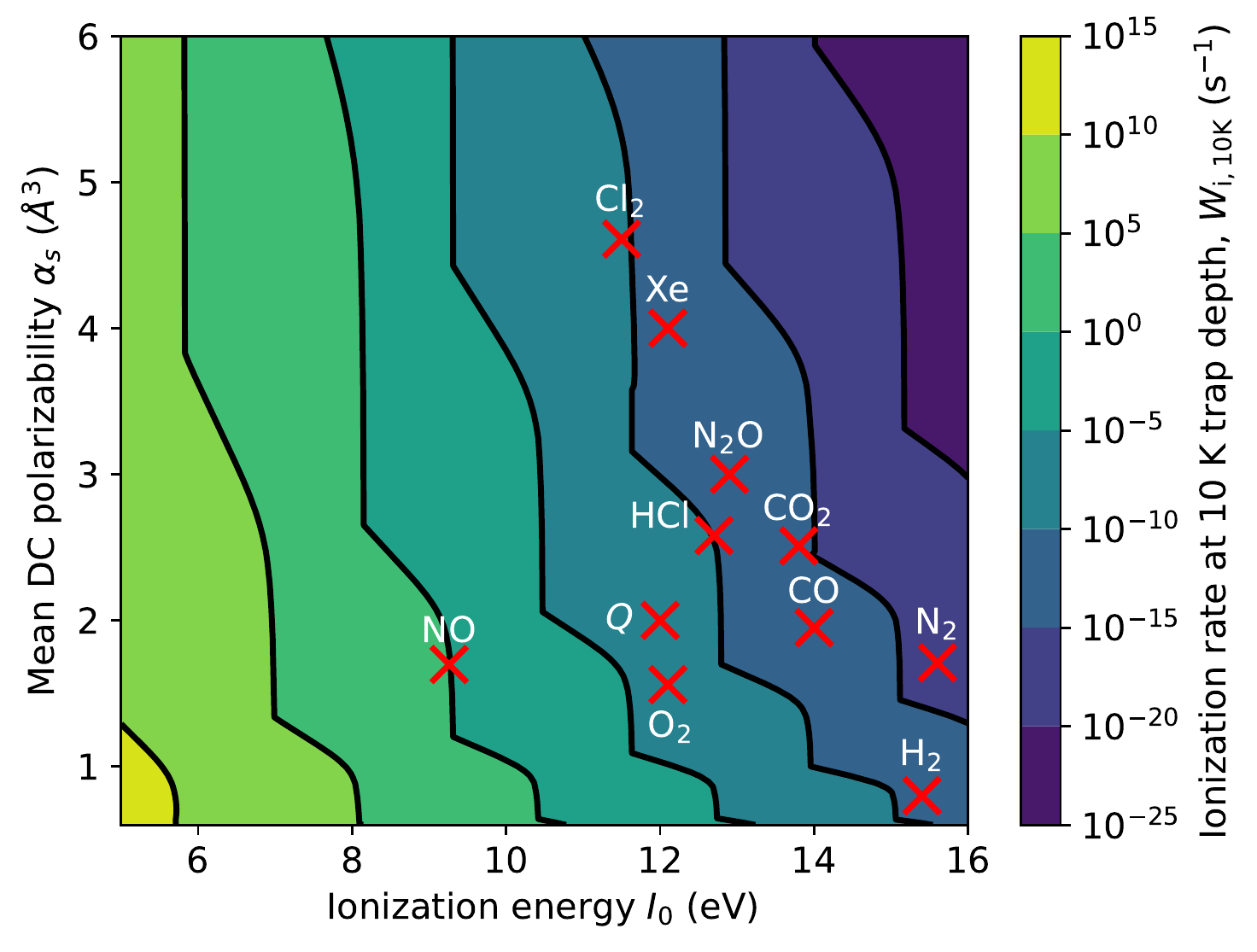}
    \caption{Ionization rate at \SI{10}{K} trap depth, $\ionirateTenK$, against ionization energy $\ionienergy$ and mean DC polarizability $\dcpolarizability$. The molecules of \cref{tab: molecule list} (except CS$_2$, which is outside the shown region) are marked with red crosses on the plot. Evidently, $\ionirateTenK$ is much more sensitive to the ionization energy than the polarizability.}
    \label{fig: pPAIR}
\end{figure}

An alternate way to reduce ionization rates is to trap molecules at a longer wavelength. The ionization rate depends primarily on the number of photons needed to reach the ionization continuum, so molecules with low ionization energies may be suitable for trapping with light at $\lambda>\SI{1064}{nm}$.

Popruzhenko's formula provides a first estimate of the ionization rate, but can be inaccurate in general. There are many examples where multiphoton ionization occurs through an intermediate multiphoton resonance, which can enhance the ionization rate by several orders of magnitude when compared to Popruzhenko's formula \cite{1985_Lompre_multiphoton_ionisation_He_532, 1971_Chin_multiphoton_ionisation_of_molecules, 1983_Lhuillier_multiphoton_dissociation_ionisation_molecules}. Resonance-enhanced multiphoton ionization (REMPI) is now a routine experimental technique for ionization of different molecules, as will be discussed in \cref{sec: Detection}. It is challenging to estimate REMPI cross sections a priori, particularly due to the lack of detailed spectra of general molecules that are free of spectral broadening. For this reason, we seek molecules with an \textit{ab initio} ionization rate estimate of $\mathrel{\lesssim\!10^{-9}}$\,\si{s^{-1}} ($\ionienergy\gtrsim\SI{12}{eV})$ in Popruzhenko's formula as first candidates for our trap, so that even if we ignore the contribution of electronic resonances to the ionization rate, we are still unlikely to see ionization in the trap. We note that limits on the ionization rate can be placed with a room-temperature experiment using our trap light, which is especially helpful for determining the suitability of molecules with higher \textit{ab initio} ionization rates such as NO or CS$_2$ (see \cref{tab: molecule list}).

In a high-density gas, laser-induced breakdown becomes the dominant mechanism for ionization in the beam. We have studied breakdown in high-intensity, CW laser beams in detail, and the results are shown in \cref{appendix: Laser Induced Breakdown}. We have determined that the buffer gas is too low-density for breakdown to be a concern during trap loading.

\subsection{Dissociation}\label{subsec: Dissociation}

Photodissociation of molecules can occur by two distinctively different mechanisms depending on the character of the light that drives it. The first mechanism is direct electronic excitation to a dissociative state. In principle, any molecular state located above any bond's dissociation threshold is dissociative. In practice, though, excited states above the dissociation continuum can be long-lived, as the probability of direct tunneling out of a quasibound excited state into the dissociation continuum is extremely small due to the Franck-Condon principle \cite{1950_Herzberg}. Only certain electronic excitations lead to dissociation. In SCS molecules, the dissociative states are either highly excited quasibound states, which have been probed through absorption of single vacuum-ultraviolet photons \cite{2002_Hanf_UV_VUV_dissociation_CCl4, 1996_Heck_UV_photofragmentation_methane}, or excited states above the ionization continuum, as has been seen in multiphoton dissociation experiments using NIR and visible pulsed lasers \cite{1983_Lhuillier_multiphoton_dissociation_ionisation_molecules, 2006_Xu_neutral_dissociation_methane_spectroscopy, 2008_Song_neutral_dissociation_methane, 1981_Carney_mechanism_of_multiphoton_ionisation_H2S}.
In particular, \cite{1983_Lhuillier_multiphoton_dissociation_ionisation_molecules} finds that dissociation using NIR lasers scales highly nonlinearly with the peak intensity, as expected for multiphoton absorption. Thus, following the arguments on ionization in \cref{subsec: Ionization of Isolated Molecules} showing that simultaneous absorption of enough photons to reach these highly excited states is unlikely, and provided $Q$'s electronic transitions are far from resonant with the low harmonics of the 1064-nm light, we expect not to be limited by this form of dissociation.

The second mechanism, infrared multiphoton dissociation (IRMPD), is usually observed in polyatomic molecules or ions using mid- to long-wavelength infrared light, such as from a CO$_2$ laser at \SI{10}{\micro\meter} \cite{2020_Maitre_IRMPD_review}, resonant with pure rovibrational transitions.
IRMPD is a heating mechanism, so the dissociation probability depends on the total energy absorbed rather than peak intensity, provided the light is near a resonance and above a relatively low threshold intensity \cite{1978_Woodin_IRMPD_original_fluence_NOT_intensity}. As the latter is easily exceeded with our trap light, we must ensure that the IRMPD will not occur at \SI{1064}{nm} for our molecules of choice.

The three-step mechanism of IRMPD is explained in detail in \cite{1977_Black_IRMPD_SF6_Mechanism, 1980_Harrison_IRMPD_Mechanism_Review_polyatomic_molecules, 1976_Book_TunableLasersAndApplications_w_chapter_on_IRMPD}. Firstly, light tuned to a pure rovibrational transition of a molecule repeatedly excites a vibrational mode, until the anharmonicity of the internuclear potential shifts the transition frequency off resonance with the light. This step is impossible in homonuclear diatomic molecules, which do not admit dipole-allowed, pure rovibrational transitions. Secondly, once excited to a high vibrational energy, in a polyatomic molecule, the different molecular vibrational modes are split into a number of closely spaced energy levels, and a ``quasicontinuum'' of energy levels forms. In the quasicontinuum, heating and ergodic mixing of vibrational quanta occur. The quasicontinuum cannot form in any diatomic molecules, and is unlikely to form in triatomic molecules, because there are not enough different vibrational modes to form a closely spaced structure \cite{1980_Harrison_IRMPD_Mechanism_Review_polyatomic_molecules, 1984_Bloembergen_IRMPE_IRMPD_small_molecules}. Finally, once the molecule's total vibrational energy is higher than the dissociation energy, the molecule soon dissociates through accumulation of excitations in a particular dissociative bond. The molecules in \cref{tab: molecule list} all have $\leq 3$ atoms, and therefore are not susceptible to IRMPD.

For molecules with many atoms, IRMPD may still be suppressed because 1064-nm light is several times more energetic than a single vibrational quantum in any molecule. In the dipole and harmonic approximations, rovibrational selection rules only allow molecules to absorb one vibrational quantum of energy at a time \cite{2016_Gupta_Interaction_of_Radiation_and_Matter_and_Electronic_Spectra}, so absorption in high harmonics of rovibrational transitions is unlikely, and the IRMPD mechanism cannot begin. In polyatomic molecules that do not contain H atoms, absorption at optical frequencies appears to be much smaller than Rayleigh scattering \cite{2020_He_Scattering_and_Absorption_Cross_Sections_Atmospheric_Gases_UV_Vis}, so it is likely that the trap light is far enough off-resonant from the fundamental vibrational modes to suppress IRMPD. For molecules containing H atoms, NIR frequencies correspond roughly to the third harmonic of a fundamental vibrational mode, and absorption has been seen in room-temperature experiments \cite{1978_Giver_CH4_Rovib_Overtone_Spectroscopy, 2005_Rueda_Overtone_Spectroscopy_Methanol}. However, we are not aware of any corresponding data at cold temperatures.
Thus, for molecules with more than three atoms the overall likelihood of IRMPD in our trap is unclear, and we therefore do not include them as candidate molecules in \cref{tab: molecule list}. 

\section{Buffer gas dynamics}\label{sec: Buffer Gas Dynamics}

\subsection{Buffer-gas cooling}\label{subsec: Buffer Gas Cooling}

The first step toward loading the optical trap is cooling the molecules to cryogenic temperatures through buffer-gas cooling \cite{2008_Doyle_buffer_gas_loading_review}. Molecules of all masses and initial temperatures appear to be amenable to buffer-gas cooling \cite{2008_Doyle_buffer_gas_loading_review, 2012_Hutzler_Buffer_Gas_Beam_Review}. As seen in \cref{tab: molecule list}, we focus on molecules with a boiling point $\boilingpoint$ around \SI{300}{K} or lower, so they can be loaded into the cold buffer gas in gas phase through a heated fill line without significant thermal load on the cryogenic system. In the design shown in \cref{fig: cell design}, buffer-gas cooling occurs in two opposing 1.5-K cells to thermalize the molecules to $T=\SI{1.5}{K}$ before buffer-gas loading the trap. This section focuses only on the cooling in the cells, and the loading dynamics are left to \cref{subsec: Buffer Gas Loading}.

A cold gas of helium at temperature $T=\SI{1.5}{K}$ is pumped into cells, also at $T$, at a moderate to low density which we set to be $n_{\mathrm{He}} = 10^{15}$\,\si{cm^{-3}}. $Q$ is pumped into the cells through heated fill lines to make up a small fraction of the number density, which we set to be 1/100. Through collisions with He atoms, $Q$ comes from a warm temperature $T_i$ to $T$, however it does not solidify as long as it does not collide with the cell walls. To produce a cold gas of $Q$, we therefore need the mean free path $l$ of $Q$ in the buffer gas to be short compared to the cell dimensions. $l$ is given by \cite{2020_Gantner_low_density_buffer_gas_sim_vs_expt}
\begin{equation}\label{eq: mean free path}
    l = \frac{v_Q}{\nu} = \frac{1}{n_{\mathrm{He}}\sigma}\frac{v_Q}{v_{\mathrm{rel}}} \stackrel{\mathrm{eq}}{=} \frac{1}{n_{\mathrm{He}}\sigma\sqrt{1 + m_Q/m_{\mathrm{He}}}},
\end{equation}
where $\nu$ is the $Q$--He collision frequency, $n_{\mathrm{He}}$ is the buffer gas number density, $\sigma$ the $Q$--He collision cross section, and $v_Q$ and $v_{\mathrm{rel}}$ the $Q$ and mean $Q$--He relative velocities, respectively. The last equality, labeled with ``eq'', holds in thermal equilibrium. 

Based on \cite{1986_Beneventi_He_molecule_collision_cross_sections, 1978_Slankas_300K_He_CH4_cross_section, 2013_Au_thesis_with_he3_cross_sections}, we estimate that SCS molecules have collision cross sections with He of order $10^{-14}$\,\si{cm^{2}} at $T = \SI{1.5}{K}$, and we take the $Q$--He collision cross section to therefore be $10^{-14}$\,\si{cm^2}.
With $n_{\mathrm{He}} = 10^{15}$\,\si{cm^{-3}}, $l$ is \SI{0.34}{mm} at \SI{1.5}{K}.

We simulate the thermalization dynamics, using the methods described in the Appendix of \cite{2021_Takahashi_Simulation_of_Cryogenic_Buffer_Gas_Beams}. $Q$ molecules are drawn by rejection sampling from an effusive initial speed distribution at $T_i$, directed out of a heated fill line at the origin in the $\mathbf{\hat{z}}$ direction. Each molecule undergoes billiard-ball collisions with He atoms in the buffer gas until it cools to the buffer gas temperature $T$.

\begin{figure}
        \centering
        \includegraphics[width=0.75\textwidth]{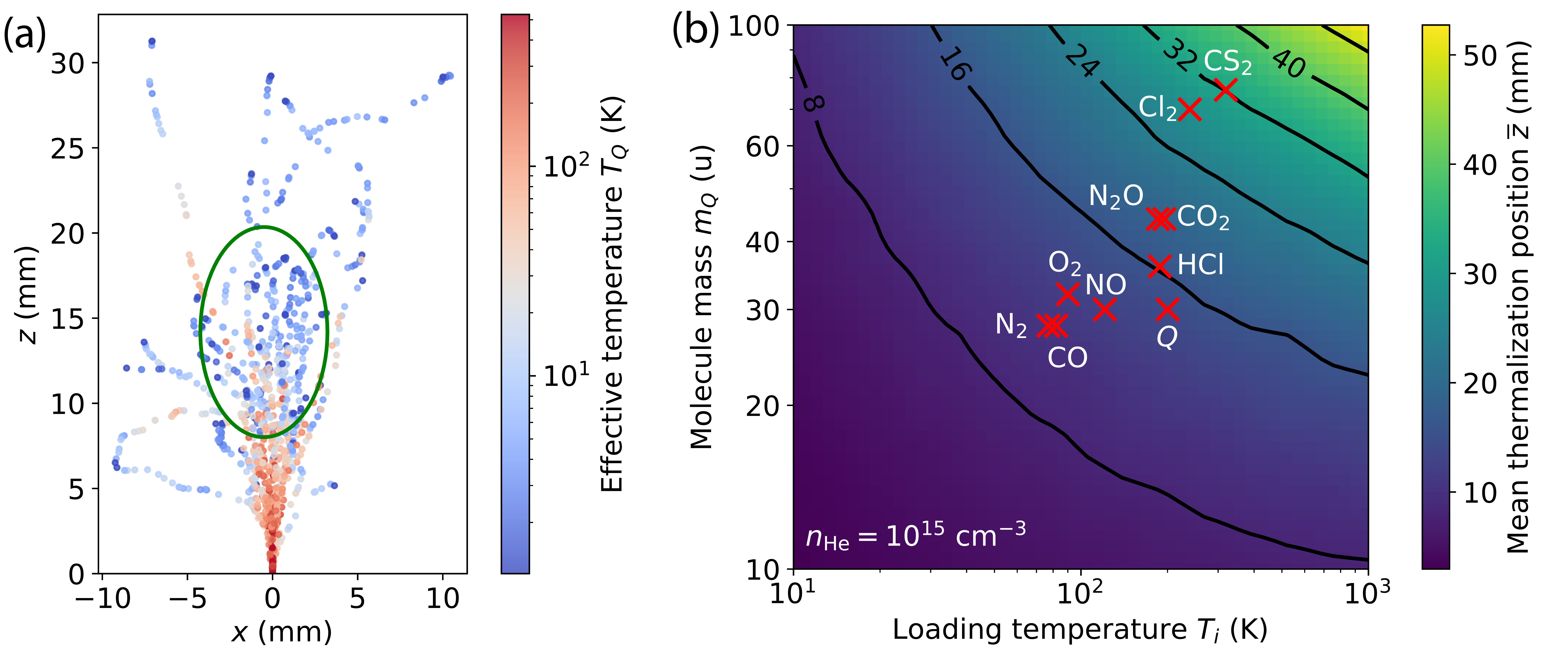}
        \caption{(a) 2-D ($xz$) projection of trajectories of 50 $Q$ molecules through a cold buffer gas. Each point shows a location of a collision, with the color axis representing the effective temperature $T_Q = m_Qv_Q^2/3\kB$ of the molecule at that point. Trajectories continue until the effective temperature reaches the 1.5-K buffer gas temperature. The green ellipse has semi-minor and -major axes equal to the $x$ and $z$ standard deviations respectively, showing the 1-$\sigma$ spread of the positions of molecules when thermalization occurs. (b) Mean $z$-distance traveled, $\overline z$, by 1000 simulated $Q$ molecules into the cold buffer gas as a function of their initial temperature $T_i$ and mass $m_Q$. Not shown are the $x$ and $z$ standard deviations, which are both less than $10$ mm at all $T_i, m_Q$ points. The coordinates corresponding to all molecules of \cref{tab: molecule list} (except H$_2$, which is outside the shown region) are marked with a red x.}
        \label{fig:buffer_gas_thermalization}
\end{figure}

A 2-D projection of 50 molecular trajectories from a fully 3-D simulation is shown in \cref{fig:buffer_gas_thermalization}\,(a). Each point marks a collision, and the color axis represents the effective temperature $T_Q = m_Qv_Q^2/3\kB T$ of $Q$ after the collision. The green ellipse shows the 1-$\sigma$ spread of positions where the molecules come to the buffer gas temperature of \SI{1.5}{K}. The $z$ coordinates of the centers of the green ellipses $\overline z$ for different values of $m_Q$ and $T_i$ are shown in \cref{fig:buffer_gas_thermalization}\,(b). Based on this figure, many SCS molecules will thermalize in a 30-mm-tall cell. 

\subsection{Buffer-gas loading}\label{subsec: Buffer Gas Loading}

With the molecules thermalized to \SI{1.5}{K}, we next discuss loading them into the optical trap. Since the dipole force is conservative, some supplementary dissipation in the trap volume is required for trapping \cite{2001_Metcalf_Book_Laser_cooling_and_Trapping}. Since here $Q$--He collisions are the only universal source of dissipation, buffer-gas cooling must occur inside the trap volume to achieve loading.

\subsubsection{Loss-free, equilibrium trapped molecule number}\label{subsec: loss-free equilibrium trapped molecule number}

The energy-level splittings associated with the trap frequencies are three orders of magnitude smaller than the buffer gas temperature, so the loading dynamics are semiclassical. Molecules arrive at the loading region after passing through one of the buffer-gas cells, so are already thermalized to \SI{1.5}{K}. In our closely-spaced, opposing cell geometry (\cref{fig: cell design}), the molecule and He number densities in the loading region are approximately the same as in the cells during loading, with $\nQcell = 10^{13}$\,\si{cm^{-3}}, and $n_{\mathrm{He}} = 10^{15}$\,\si{cm^{-3}}, respectively. If trap losses are negligible, the number density of trapped molecules $n(\mathbf{x})$ during trap loading must eventually follow a Boltzmann distribution \cite{2013_Landau_Lifshitz_Statistical_Physics}:
\begin{equation}\label{eq: Density Boltzmann Distribution}
    n(\mathbf{x}) = \nQcell e^{\eta(\mathbf{x})} P(\mathbf{x}),
\end{equation}
where $\eta(\mathbf{x}) = |U(r,z)|/\kB T$, and $P(\mathbf{x}) = \left[
\mathrm{erf}\left(\sqrt{\eta(\mathbf{x})}\right)- \left(2/\sqrt{\pi}\right)\sqrt{\eta(\mathbf{x})}e^{-\eta(\mathbf{x})}
\right]$ accounts for the truncation of the distribution at the trap depth. 

\begin{figure}
    \centering
    \includegraphics[width = 0.5\linewidth]{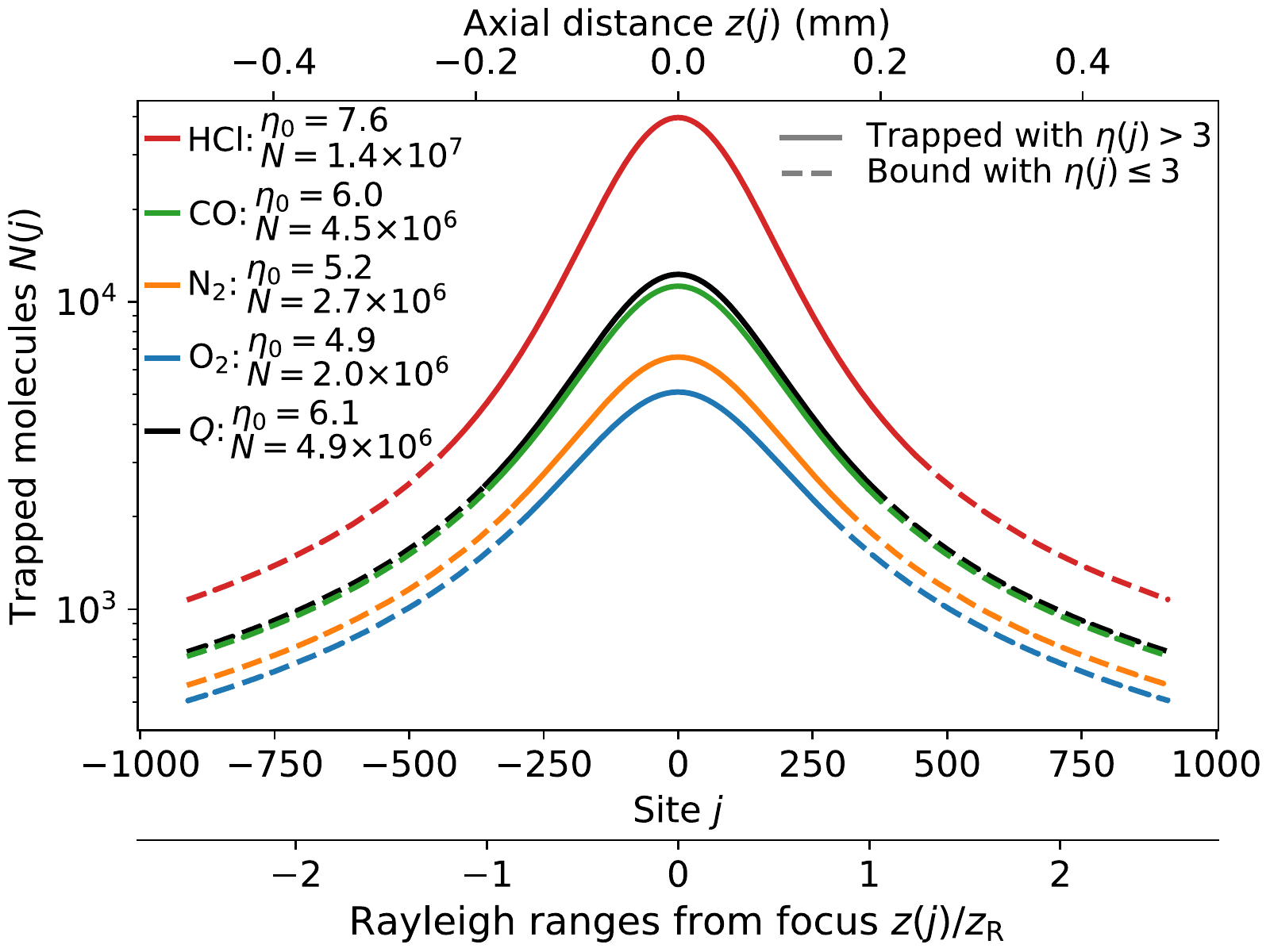}
    \caption{Number $N(j)$ of trapped molecules per lattice site $j$ during buffer-gas loading for different molecular species, based on \cref{eq: Density Boltzmann Distribution}, assuming no losses. Solid curves indicate molecules trapped in sites with site depth parameter $\eta(j) = \eta_0 w_0^2/w^2(z(j))>3$, dashed curves indicate molecules bound in the trap in weak lattice sites with $\eta(j)\leq 3$. The legend gives the total number of trapped molecules $N$, given by $N = \sum_j N(j)$, with the sum over all $j$ such that $\eta(j)>3$.
    }
    \label{fig: N_trapped_nonlinear_Boltzmann}
\end{figure}

We now integrate \cref{eq: Density Boltzmann Distribution} over space to determine the trapped molecule number. We will index distinct lattice sites with $j$, refer to their positions as $z(j) = j\lambda/2$, and describe the trap depth of each lattice site by the site depth parameter $\eta(j) = \eta_0 w_0^2/w^2(z(j))$ ($\eta(0) = \eta_0 = \trapdepth/T$). By integrating \cref{eq: Density Boltzmann Distribution} one site at a time, we can determine the total number of trapped molecules $N_Q(j)$ in each lattice site, as well as the total number of trapped molecules $N_Q = \sum_j N_Q(j)$. The sum to $j=\pm\infty$ diverges logarithmically, so we truncate the sum beyond sites where $\eta(j) = \eta_0 w_0^2/w^2(z(j)) \leq 3$. The results are shown in \cref{fig: N_trapped_nonlinear_Boltzmann}, with $N(j)$ plotted on the $y$ axis and the sum $N$ shown in the legend. The Boltzmann distribution predicts that several millions of molecules will be trapped for each species shown, at peak trapped densities of order $10^{15}$--$10^{16}$\,\si{cm^{-3}}. 

\cref{eq: Density Boltzmann Distribution} does not, however, indicate the timescale of equilibration. This makes it hard to determine the effect of losses on the number of trapped molecules per site, and it fails to estimate the number of trapped molecules remaining in the trap after the buffer gas and untrapped molecules are pumped out ($\nQcell\rightarrow0$). We therefore consider a microscopic approach to trap loading. 

\subsubsection{Microscopic, ergodic loading model}\label{subsubsec: microscopic ergodic loading model}

In our trap, the trap dimensions (set by $w_0 = \SI{8}{\micro\meter}$ and $\lambda = \SI{1064}{nm}$) are much smaller than the mean free path $l=\SI{340}{\micro\meter}$, so our loading model will differ from models used in buffer-gas-loaded magnetic trapping experiments, where $l$ is small compared to the trap dimensions \cite{2008_Doyle_buffer_gas_loading_review}.

Starting with an empty trap, any free molecule passing through the trap potential can only be loaded if a collision with a He atom reduces its energy to below the local trap depth. Of the free molecules that collide with a He atom in the trap, a substantial fraction $f_0$ will not lose enough energy in the collision to become bound. Since $l\gg w_0$, no other collision will occur in the trap volume for these molecules, so they are lost. The remaining molecules do lose enough energy to initially be bound in the trap, however not all of these molecules will go on to thermalize at the bottom of the trap potential. A fraction $f_1$ of molecules will be initially bound, but never fall to a total energy consistent with a truncated Boltzmann distribution that would indicate thermalization. Meanwhile, a fraction $f_2$ (with $f_0+f_1+f_2 = 1$) will be initially bound and also go on to thermalize. Molecules which thermalize will live in the trap until they eventually are ejected after a mean number of collisions $\kevap$ with the buffer gas.

The $Q$--He collision time from \cref{eq: mean free path} is \SI{10}{\micro\second}, but in the harmonic approximation the radial and axial orbital periods for a trapped molecule are \SI{600}{ns} and \SI{18}{ns}, respectively. Since the true trap potential \cref{eq: U trap} is highly nonlinear outside the harmonic approximation, the ergodic approximation can be taken for the motion of $Q$ in the trap. Hence, approximating collisions as equally likely to occur at each time interval, $Q$'s trajectory need not be integrated to determine its initial conditions for each collision, which are instead drawn randomly from the energy hypersurface. It then becomes computationally simple to track the energy of $Q$ after a number of hard-ball collisions with He atoms drawn from a Maxwell distribution at $T=\SI{1.5}{K}$. The details of these simulations are discussed in \cref{Appendix: Details of Ergodic Loading Simulations}.

\begin{figure}
    \centering
    \includegraphics[width = 0.5\linewidth]{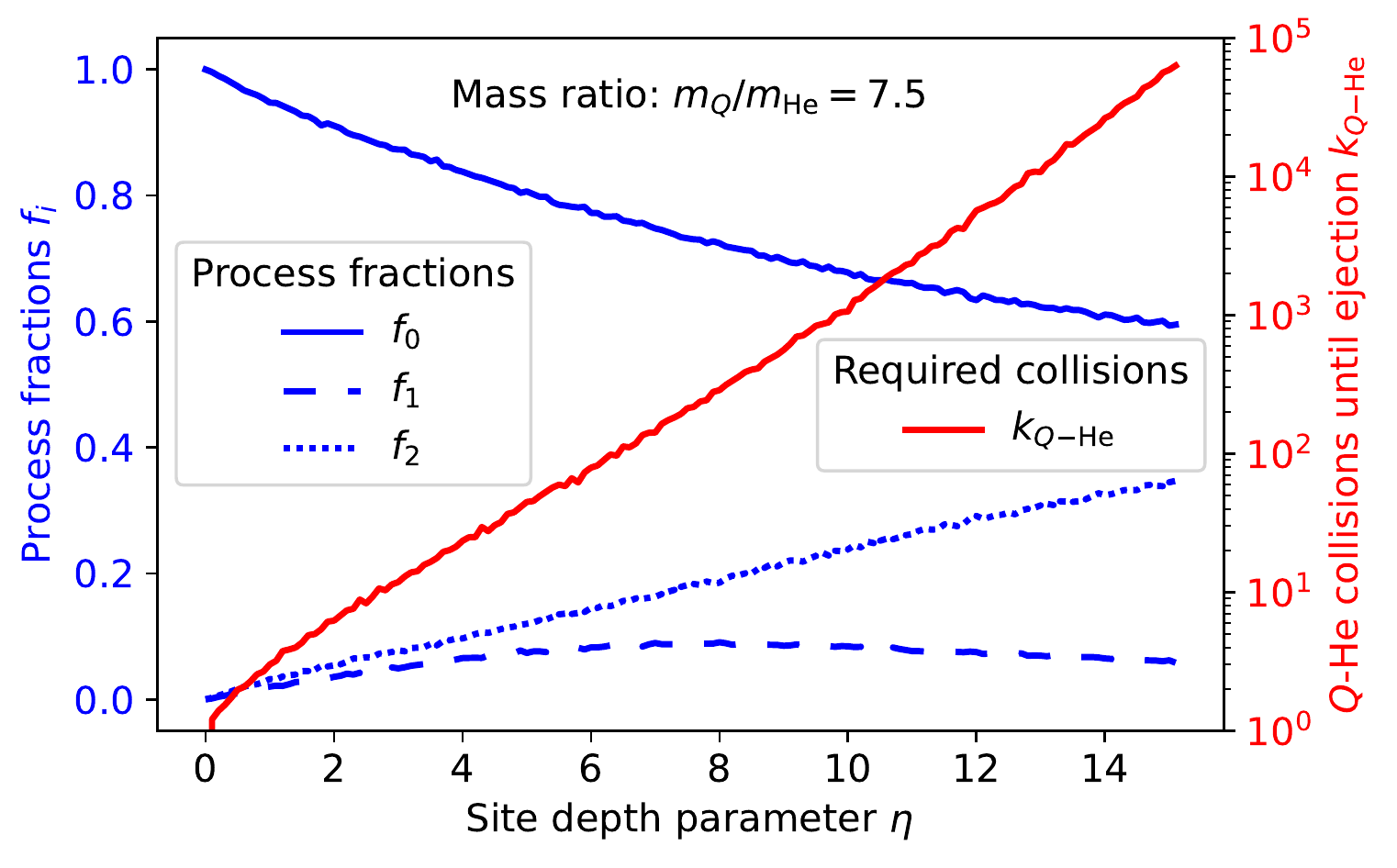}
    \caption{Ergodic loading simulations in a single lattice site. As a function of the site depth parameter $\eta$, we show the loading fractions of an unbound molecule undergoing a collision in the trap volume on the left axis. The fractions are: $f_0$ (solid blue curve): molecule undergoes a collision in the trap but does not lose enough energy and immediately escapes the trap; $f_1$ (dashed blue curve): molecule does lose enough energy to not immediately escape the trap, but in subsequent collisions never loses enough energy to fall to a thermal Boltzmann distribution; $f_2$ (dotted blue curve): molecule does lose enough energy to not immediately escape the trap, and in subsequent collisions loses enough energy to eventually fall below a Boltzmann distributed energy. For a molecule that thermalizes, the number of $Q$--He collisions needed for it to be ejected from the trap, $\kevap$ (red curve), is shown on the right axis.}
    \label{fig: ergodic loading simulations}
\end{figure}

The results of the ergodic loading simulations are shown in \cref{fig: ergodic loading simulations} as a function of the site depth parameter $\eta$ in one particular lattice site. The simulation results do not noticeably depend on the chosen site $j$ except through the site depth parameter $\eta(j)$, so to show their generality, we omit the index $j$ and write the site's depth as $\eta$. The loading fractions $f_i$ vary slowly with $\eta$. The number of $Q$--He collisions needed for a thermalized molecule to be ejected from the trap, $\kevap$ (right axis), however, grows exponentially with $\eta$. 

\subsubsection{Time-dependent, loss-inclusive loading model}\label{subsubsec: time dependent loss inclusive loading model}

Applying these results across all lattice sites, we can estimate the number of trapped molecules $N_Q(j)$ in a lattice site $j$ as well as its time dependence. The approximate model we develop will be limited mostly by three approximations.
\begin{enumerate}
    \item The loading fractions $f_i(j)$ and collision rates are computed assuming the trap is empty, which ignores the fact that the number density and energy distribution of molecules in the trap volume is affected by molecules in the trap and molecules recently ejected from the trap. Our approximate model will therefore only strictly be valid in the limit of high trap losses, and this effect leads to an underestimation of $N_Q(j)$ in thermal equilibrium with the buffer gas in the low-loss limit by a factor of about 3.
    \item The simulations are carried out in a truncated harmonic trap with total volume $V_\mathrm{e}(j) = w^2(z(j))\lambda/3$ as opposed to the true nonlinear trap potential (see \cref{Appendix: Details of Ergodic Loading Simulations}). This leads to an underestimation of $N_Q(j)$ by a factor of 1.5--2.
    \item The model will compute $N_Q(j)$ without explicitly determining the local density distribution in the trap, and will therefore not completely account for out-of-equilibrium loss rates.
\end{enumerate}

With these approximations in mind, we can write down an equation for $dN_Q(j)/dt$. Molecules will be loaded into an empty trap at a rate $\nQcell(t) V_\mathrm{e}(j) \nu(n_{\mathrm{He}}(t)) f_2(j)$, dependent on the number of free molecules in the trap volume $\nQcell(t) V_\mathrm{e}(j)$, the $Q$--He collision frequency $\nu(n_{\mathrm{He}}(t))$, and the fraction of collisions in $V_\mathrm{e}(j)$ that lead to loading $f_2(j)$. Molecules will be collisionally ejected from the trap at a rate $[\nu(n_{\mathrm{He}}(t))/\kevap(j)]N_Q(j)$, and two-body collisions between trapped molecules lead to an additional loss term $[\betatb/V_\mathrm{e}(j)]N^2_Q(j)$, where $\betatb$ is the $Q$--$Q$ two-body loss coefficient. Rotational Raman scattering (RRS) leads to an additional loss term $\Ramanlossrate$ dependent on the RRS rate $\rotRamanrate$. We neglect a non-collisional one-body loss term $-\Lambda N_Q(j)$, because ionization, dissociation, and recoil heating are small (see \cref{sec: Effects of High Intensity Light on Trapped Molecules}). Other sources of trap heating are also small, as discussed in \cref{subsec: One-Body Loss Rate}. Overall,
\begin{equation}\label{eq: dN/dt}
    \frac{dN_Q(j)}{dt} = \nQcell(t) V_\mathrm{e}(j) \nu(n_{\mathrm{He}}(t)) f_2(j)) - \frac{\nu(n_{\mathrm{He}}(t))}{\kevap(j)}N_Q(j)  - \frac{\betatb}{V_\mathrm{e}(j)} N^2_Q(j) - \Ramanlossrate.
\end{equation}
Note that in the two-body loss term, the effective volume is the full simulation volume $V_\mathrm{e}(j)$, as opposed to a commonly used $V_{\mathrm{eff}} = \sqrt{\frac \pi 2} w_0^2\lambda / \eta^{3/2}$ \cite{2021_Bause_universal_loss_debate}. This effective volume $V_{\mathrm{eff}}$ is only appropriate when the density distribution in the trap is close to a Boltzmann distribution. Our approximate loading model, on the other hand, is only strictly valid when losses are high ($\betatb\gtrsim 10^{-11}$\,\si{cm^3/s}), in which case the trap density distribution is pinned near a constant value $n_Q(j) \approx \sqrt{\nu(n_{\mathrm{He}}(t)) f_2(j) \nQcell(t)/\betatb}$. The molecules are therefore evenly distributed over the volume $V_\mathrm{e}(j)$ in the high-loss limit. In the low-loss limit, on the other hand, the choice of the effective two-body loss volume is not critical.

\cref{eq: dN/dt} can be integrated once $\betatb$ and $\Ramanlossrate$ are specified. In magnetic traps, two-body loss is usually caused by spin-changing inelastic collisions between trapped particles, and typical values of $\betatb$ range from $\SI{4e-11}{cm^3/s}$ \cite{2019_Segev_Magnetic_Trapping_O2} to $\SI{9e-13}{cm^3/s}$ and lower \cite{2011_Hummon_N_NH_Collisions_Magnetic_Trap}. Our trap, however, is insensitive to the molecules' internal state, so we are likely not limited by this kind of loss. In \cref{subsec: collision-induced absorption}, we show that collision-induced absorption, seen in O$_2$ gases, also does not cause significant two-body loss when trapping O$_2$ at the densities expected in our experiment. In optical trapping experiments on molecules, particularly bialkali molecules, a high two-body loss rate is observed. In this so-called ``universal loss'', a close to unity fraction of collisions between molecules in the trap are lossy. The mechanism is discussed in \cref{subsec: Universal Loss}, but in short, we believe universal loss is unlikely in our experiment due to the weak interactions between, and the high excitation energies of, SCS molecules, and consider universal loss only as a possible ``worst case scenario''. It is more likely that $\beta$ is small enough to be ignored, and the loss is dominated by the other loss terms in \cref{eq: dN/dt}.

The loss due to RRS, $\Ramanlossrate$, has a complicated form, which is detailed in \cref{Appendix: Trap Heating and Loss from RRS}. Nevertheless, the effect of $\Ramanlossrate$ is simple. During buffer-gas loading, rotational cooling is efficient \cite{2012_Hutzler_Buffer_Gas_Beam_Review}, and RRS does not lead to substantial loss. Once the buffer gas is removed, rotational heating causes exponential decay of the ground state trapped population at a rate $\rotRamanrate$.

 \subsubsection{Loading simulation results}\label{subsubsec: Loading Simulation Results}

\begin{figure}
        \centering
        \includegraphics[width = \linewidth]{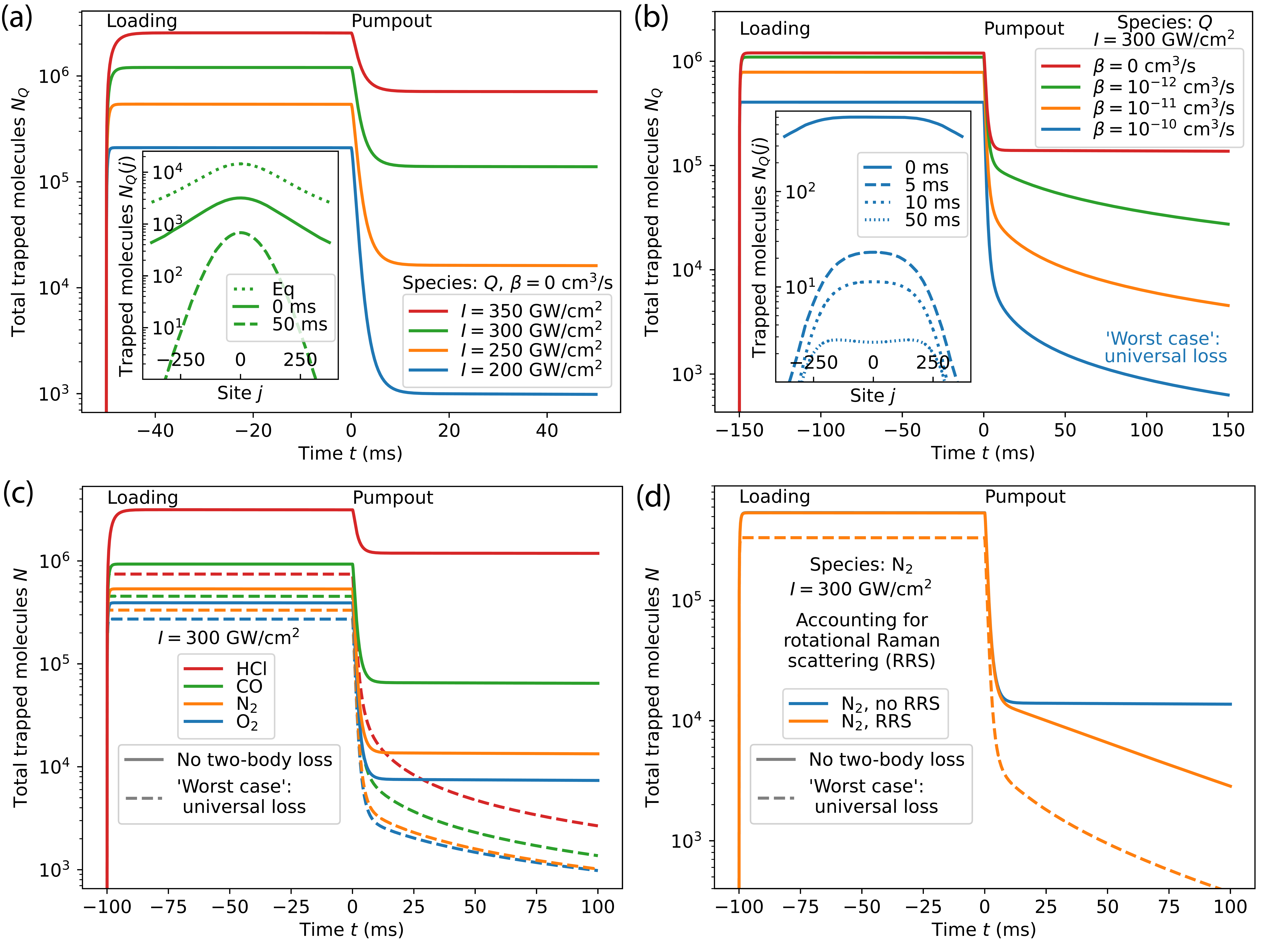}
        \caption{Number of trapped molecules $N_Q(t)$ from \cref{eq: dN/dt} for different scenarios. Loading occurs at negative $t$, with He density $n_{\mathrm{He}} = 10^{15}$\,\si{cm^{-3}} and Q density $\nQcell = 10^{13}$\,\si{cm^{-3}}. Pumpout to produce an isolated sample begins at $t=0$, after which $n_{\mathrm{He}}$ and $\nQcell$ exponentially decay with a timescale of 2 and 0.5 ms, respectively. $n_{\mathrm{He}}$ saturates at $10^{11}$\,\si{cm^{-3}} to model He film desorption dynamics. In (a), $Q$ is treated assuming no two-body loss for various optical intensities $I$. Inset shows the distribution of molecules across lattice sites at $t=\SI{0}{ms}$ (solid curve) and $t = \SI{50}{ms}$ (dashed curve). The $t=\SI{0}{ms}$ curve can be compared to the analytic result from \cref{eq: Density Boltzmann Distribution} (dotted curve), which shows that the approximate model \cref{eq: dN/dt} underestimates $N_Q(j)$ during loading in the low-loss limit by a factor of 6. In (b), $Q$ is treated with fixed intensity $I = \SI{300}{GW/cm^2}$, but varying two-body loss coefficients $\betatb$. The highest value of $\betatb=10^{10}$\,\si{cm^3/s} corresponds to the ``worst case'' of universal loss, which we believe to be unlikely in our trap (\cref{subsec: Universal Loss}). Inset shows the distribution of molecules across lattice sites at different times for the case of universal loss. In (c), we integrate \cref{eq: dN/dt} for N$_2$, CO, O$_2$, and HCl with $I = \SI{300}{GW/cm^2}$, ignoring rotational Raman scattering (RRS). For each molecule, we consider the case of no two-body loss (solid curves), as well as the unlikely ``worst case'' of universal loss (dashed curves). $\betatb$ coefficients for universal loss are computed as in \cref{subsec: Universal Loss}.
        For the case of no two-body loss, the peak trapped densities at $t = \SI{100}{ms}$ (averaged over the central lattice site volume $V_\mathrm{e}(0)$) are \SI{1.7e12}{cm^{-3}} (O$_2$), \SI{3.1e12}{cm^{-3}} (N$_2$),  \SI{1.4e13}{cm^{-3}} (CO), and \SI{2.3e14}{cm^{-3}} (HCl).
        In (d), we integrate \cref{eq: dN/dt} for N$_2$, comparing the cases of including and excluding RRS to exemplify its effect. The effect of RRS is modeled in \cref{Appendix: Trap Heating and Loss from RRS}.
        }
        \label{fig: trap_number_evolution}
\end{figure}

Integrating \cref{eq: dN/dt} to find approximate values for $N_Q(t)$ is done in four stages, and the results are shown in \cref{fig: trap_number_evolution}. We start in \cref{fig: trap_number_evolution}\,(a) with the simplest case of negligible two-body loss, and show the evolution of trapped $Q$ as a function of time for various optical intensities. We begin the simulation by turning on the trap at $t = \SI{-50}{ms}$ with $n_{\mathrm{He}} = 10^{15}$\,\si{cm^{-3}} and $\nQcell = 10^{13}$\,\si{cm^{-3}}. The figure shows the trapped molecule numbers coming to equilibrium within a few \si{ms}.

The pumpout begins at $t = \SI{0}{ms}$, when the cryogenic shutter actuates in \SI{1}{ms} to stop flow into the loading region. We conservatively assume molecules will cryopump to the surface of the shutter, rather than bounce off it, to underestimate the trap loading during the pumpout, and conservatively assume the He buffer gas will take longer to evacuate the loading region to overestimate the trap losses during pumpout. To this end, we set the He and molecule pumpout timescales from the loading region to be \SI{2}{ms} and \SI{0.5}{ms}, respectively. After $t=\SI{0}{ms}$, $n_{\mathrm{He}}$ and $\nQcell$ decay exponentially with these pumpout timescales. To model the effect of He film formation on the outside of the buffer-gas cells, we let $n_{\mathrm{He}}$ saturate at $10^{11}$\,\si{cm^{-3}} for the remainder of the simulation \cite{2004_Harris_cryogenic_films, 2004_Michniak_low_eta_magnetic_trapping}. The trapped densities all rapidly re-equilibrate to the new loading region densities, until after about 10 ms, when the loading region densities are so small that the trap is isolated and the trapped molecule number is constant.

With increasing intensity, the number of molecules trapped during the loading phase increases, and also the fraction of these molecules retained in the trap after pumpout increases. Our proposed experiment, with $I = \SI{300}{GW/cm^2}$, appears to trap about $10^5$ $Q$ molecules in our approximate model. The inset in \cref{fig: trap_number_evolution}\,(a) shows the evolution of the number of molecules over time in each site for $I = \SI{300}{GW/cm^2}$. Just before pumpout, at $t=\SI{0}{ms}$, the trapped molecule numbers predicted by \cref{eq: dN/dt} (solid curve) resemble a Boltzmann distribution (dotted curve), except that the approximations involved in \cref{eq: dN/dt} lead to an underestimation of the trapped molecule number by a factor of 6. After \SI{50}{ms}, when the trap is isolated from background-gas collisions, we see that molecules near the focus of the trap are retained more than molecules far from the focus, since the trap depth $\eta(j)$ is higher.

In \cref{fig: trap_number_evolution}\,(b), we consider the effect of two-body losses. The details of the pumpout are the same as in \cref{fig: trap_number_evolution}\,(a), but the integration time is increased so that the long-time behavior is visible. The blue curve, with $\betatb = 10^{-10}$\,\si{cm^3/s}, represents universal loss, while the red curve, with $\betatb = \SI{0}{cm^3/s}$, represents no two-body loss. We see that two-body loss leads to a reduction in the trapped molecule number during loading, the fraction of molecules that survive the pumpout, and the number of molecules in the long-time limit. Roughly speaking, in the high-$\betatb$ limit, each order-of-magnitude reduction in $\betatb$ leads to an order-of-magnitude increase in the trapped molecules \SI{200}{ms} after pumpout. 

The inset of \cref{fig: trap_number_evolution}\,(b) shows how molecules are distributed among lattice sites in the universal loss limit ($\betatb = 10^{-10}$\,\si{cm^3/s}). Pinning of the trapped molecule number near the trap focus is clearly visible, which retroactively justifies the use of $V_\mathrm{e}$ as the effective two-body loss volume instead of $V_\mathrm{eff}$. With universal loss, only a few molecules per lattice site survive after the buffer gas is pumped out, but because the molecules are spread over many lattice sites, the total number of trapped molecules remains large enough for sensitive detection schemes to detect them (see \cref{sec: Detection}).

Third, we consider some real molecules (N$_2$, CO, O$_2$, HCl) in \cref{fig: trap_number_evolution}\,(c). We ignore RRS in this figure for clarity, and treat it separately. For each molecule, we recompute the parameters $f_2(\eta(j))$ and $\kevap(\eta(j))$, which change with the molecular mass and $\eta_0$, and then separately integrate the case of no two-body loss (solid curves) and universal loss (dashed curves) to demonstrate a best and worst case scenario for the number of each molecule we can expect to trap. Even our worst case estimates based on universal loss suggest thousands of molecules will be trapped at peak densities of $10^{11}$\,\si{cm^{-3}} (at \SI{100}{ms}, averaged over the center lattice site volume $V_\mathrm{e}(0)$), which will be enough to demonstrate trapping. In the case of no universal loss, on the other hand, about $10^{4}$ O$_2$ and N$_2$ molecules, $10^{5}$ CO molecules, and $10^6$ HCl molecules can be trapped. These correspond to peak trapped densities (at \SI{100}{ms}, averaged over the central lattice site volume $V_\mathrm{e}(0)$) of \SI{1.7e12}{cm^{-3}} (O$_2$), \SI{3.1e12}{cm^{-3}} (N$_2$),  \SI{1.4e13}{cm^{-3}} (CO), and \SI{2.3e14}{cm^{-3}} (HCl).

Finally, we exemplify the effect of RRS in \cref{fig: trap_number_evolution}\,(d) by studying its impact on N$_2$ (see \cref{Appendix: Trap Heating and Loss from RRS}). During buffer-gas loading, the curves for the case without RRS (blue) and with RRS (orange) are indistinguishable, owing to efficient rotational cooling by the buffer gas. After the buffer-gas pumpout, exponential decay of the rotational-ground-state, trapped population is observed at the rate $\rotRamanrate$.

\subsubsection{Evaporative cooling and other considerations}\label{subsubsec: evaporative effects and other considerations}

In our models, for the case when $\betatb$ and $\rotRamanrate$ are small, we have so far ignored the evaporative effect of $Q$--$Q$ elastic collisions after the buffer-gas pumpout. They will lead to additional $Q$ loss, but also decrease the sample temperature and thus increase $\eta_0$ through evaporative cooling, making them fundamentally different in nature to the losses included in \cref{eq: dN/dt} \cite{1996_Ketterle_evaporative_cooling_review}. We compute the initial evaporation timescale at constant $\eta_0$ \cite{2013_Bourgain_Evaporative_Cooling_Small_Number_of_Molecules_Dipole_Trap, 2001_Ohara_Evaporative_Cooling_Time_Scaling_Theory} to be $N_Q(0)/\dot N_Q(0) = \tau_{\mathrm{ev}}=\SI{15}{ms}$ for the central lattice site of our trap, 1.5 times less than \cite{2013_Bourgain_Evaporative_Cooling_Small_Number_of_Molecules_Dipole_Trap}. Therefore, for species with $\rotRamanrate<\SI{70}{s^{-1}}$, at least some amount of evaporation can be achieved initially, allowing for a reduction of the trap depth which in turn proportionally reduces the RRS rates.

For molecules with low losses, we expect similar evaporative cooling dynamics to \cite{2013_Bourgain_Evaporative_Cooling_Small_Number_of_Molecules_Dipole_Trap}, where 800 optically trapped Rb atoms are evaporated to 40 atoms with $\eta\sim 5$, resulting in a 1000-fold reduction in temperature and increase in phase-space density. At the now reduced optical intensity required to maintain trapping of tens of molecules per lattice site at \SI{\sim1.5}{mK} (phase-space density $\sim\!10^{-4}$), the sites could be combined using a bichromatic light field \cite{2006_Lee_bichromatic_cavity_remove_standing_wave_pattern}. The resulting sample of thousands of molecules at \si{mK} temperatures could be evaporated further towards the ultracold regime.

We have also thus far ignored the effect of trapped He atoms. Although a small amount of sympathetic evaporative cooling can be expected from the rapid evaporation of trapped He \cite{2014_Edmunds_Selective_Heating_Med_Finesse_Cavity_Argon_Trap}, the number of He atoms that will survive the buffer-gas pumpout is negligible due to their low trap depth of $\eta_0=0.6$. 

One final consideration during He pumpout is the buffer-gas ``wind'' dragging molecules out of the trap, as observed in buffer-gas-loaded magnetic traps \cite{2004_Michniak_low_eta_magnetic_trapping}. In our trap, the $Q$--He collision frequency is slow compared to the trap frequencies, so the $Q$ position and velocity in the trap is randomized between collisions. The wind therefore does not provide a unidirectional drag force on trapped molecules, so need not be considered in our trap. 

\section{Detection}\label{sec: Detection}

To detect the molecules in the trap both during and after loading, a sensitive and background-free scheme is required. Absorption or fluorescence detection techniques, commonly used for cold and ultracold atoms, are challenging to implement for most SCS molecules due to the lack of optical cycling transitions.

Resonance-enhanced multiphoton ionization (REMPI) of molecules with an intense UV laser pulse, in combination with the detection of the charged products, is both sensitive and background-free \cite{2021_Boesl_REMPIReview}. The UV laser pulses can be derived from a frequency-converted, tunable dye laser, and a microchannel plate (MCP) can serve as a detector.
Due to its resonant nature, REMPI only ionizes a given species, but not any other species in the background gas. It can also resolve internal states, allowing a measurement of the rovibrational temperature of the molecules.
Furthermore, differential AC Stark shifts from the trap light will likely lead to the resonant frequencies for trapped and untrapped molecules to be different on the order of the trap depth of \SI{\sim200}{GHz} (\SI{\sim10}{K}). Thus, given a narrow enough resonance, REMPI can distinguish between trapped and untrapped molecules. For example, \cite{2012_Yamaguchi_REMPI_N2_rotational_temperature} has used REMPI to resolve different rotational levels in N$_2$ at \SI{15}{K}, corresponding to a resolution of better than \SI{120}{GHz}. This also opens up the intriguing possibility to selectively remove molecules with a certain kinetic energy from the trap, which could be of use in forced evaporative cooling schemes, or to forcibly remove rotationally excited molecules from the trap.

Although REMPI is only applicable to molecules which have selection-rule-allowed multiphoton transitions \cite{2017_Ashford_Multiphoton_Spectroscopy_Applications}, the technique is widely applicable. In addition to all of the molecules listed in \cref{tab: molecule list}, REMPI has been demonstrated on other symmetric tops like NH$_3$ \cite{2005_Nolde_REMPI_NH3} and benzene \cite{1976_Johnson_REMPI_Benzene}, asymmetric tops like SO$_2$ \cite{2000_Xue_REMPI_SO2}, H$_2$O \cite{1986_Meijer_REMPI_H2O}, methanol and ethanol \cite{2007_Philis_REMPIMethanolEthanol}, radicals like NH \cite{2017_Ashford_Multiphoton_Spectroscopy_Applications}, SF$_2$ \cite{2011_Dogariu_REMPI_NO_SF2} and OH \cite{1991_deBeer_REMPI_OH}, aromatics and organic compounds \cite{2014_Streibel_REMPI_TOFMS_review}, and a wide variety of other atomic and molecular species \cite{2021_Zhang_REMPI_review_microwave_scattering}.

An alternative scheme is nonresonant ionization of molecules in a tightly focused, ultrashort laser pulse, and subsequent characterization of the products using time-of-flight mass spectrometry (TOFMS) \cite{1971_Chin_multiphoton_ionisation_of_molecules, 1983_LHuillier_MPI_noble_gases_1064, 1983_Lhuillier_multiphoton_dissociation_ionisation_molecules}. This allows for the simultaneous detection of arbitrary molecular species in the trap and thus the monitoring of chemical populations as a function of time during cold chemical reactions. However, trapped and untrapped molecules cannot be easily distinguished with this technique, leading to a large background signal especially during the loading phase. Moreover, the long TOFMS path and ion optics required to distinguish different mass products will make the technique challenging in our experiment. We note that, in principle, an electron beam could be used for non-resonant ionization, but this will lead to a larger background signal compared to a laser beam as gas outside the focal volume will also be ionized.

Finally, the trap light scattered off the molecules could be used for detection. However, even for the high intensities assumed here, the Rayleigh scattering rate is only $\sim\!10^{3}$\,\si{s^{-1}} (see \cref{subsec: Rayleigh Scattering}), which will be difficult to distinguish from other sources of scattered light in a realistic apparatus. This could in principle be overcome by using a second resonant, but otherwise empty, cavity to enhance the Rayleigh scattering rate \cite{2010_Motsch_CavityRayleigh}.

\section{Summary and outlook}\label{sec: Outlook}

In this work, we have studied trapping small, chemically stable (SCS) molecules in a deep, very far-off-resonant, quasielectrostatic dipole trap formed by a tightly focused, high-intensity optical cavity.
We have analyzed the trapping and buffer-gas loading dynamics, and the potentially harmful effects of the high-intensity laser light on the molecules. For the examples of N$_2$, CO, O$_2$, and HCl, we conclude that on the order of a million molecules can be loaded into the trap, and a large fraction can be retained after removing the buffer gas to produce an isolated sample. Evaporative cooling can be used to further reduce the temperature of the trapped sample, possibly deep into the \si{mK} regime. Other molecules shown in \cref{tab: molecule list}, such as CO$_2$ and N$_2$O, may be similarly trapped, but might require a longer wavelength than the 1064-nm light proposed here to reduce their rotational Raman scattering or ionization rates. Likewise, H or H$_2$ could be trapped if the intensity can be further increased by a factor of two over the demonstrated value, or if the temperature of the buffer gas can be further reduced, e.g., with additional buffer-gas cells cooled below \SI{1}{K} with a $^3$He pot.

Although this work has focused on a sub-class of linear molecules, we see no obvious reasons why our proposed trap cannot be used on other classes of SCS molecules. For example, SO$_2$ ($\dcpolarizability=\SI{3.8}{\angstrom^3}$, $\ionienergy=\SI{12.3}{eV}$ \cite{CCCBDB}) and H$_2$O ($\dcpolarizability=\SI{1.5}{\angstrom^3}$, $\ionienergy = \SI{12.6}{eV}$ \cite{CCCBDB}) are small but non-linear atmospheric gases which may be trapped without significant modification to the trap design presented here. We have also focused on molecules which can be loaded into the buffer-gas cells from a heated fill line. Future renditions of our experiment could instead use laser ablation to seed molecules into the buffer gas, opening up the possibility of trapping heavy molecules and radicals that are of interest in atmospheric and interstellar chemistry \cite{2021_Heazlewood_Cold_Chemistry_Review}, and in tests of the Standard Model \cite{2011_Levshakov_CH3_fundamental_constants, 2018_ACME_EDM}.

Moreover, the literature on high-intensity, continuous-wave laser--molecule interactions is sparse, so although we have determined that molecules with $\leq 3$ atoms and $\ionienergy\geq\SI{12}{eV}$ are unlikely to be destroyed by the trap light, this does not necessarily imply that other molecules will be destroyed. We believe that many molecules outside these constraints will still be suitable for trapping. In the case of ionization, because of the approximations made in Popruzhenko's formula used here to estimate the ionization rate $\ionirate$, we here have placed a rather conservative constraint of $\ionirate<10^{-9}$\,\si{s^{-1}}. By measuring the ionization rates of molecules in our high-intensity cavity, without even the need for buffer-gas cooling, the robustness of molecules against ionization can be directly determined. As for dissociation, the arguments made above do not obviously rule out molecules like the planar BF$_3$ ($\dcpolarizability=\SI{2.4}{\angstrom^3}$, $\ionienergy = \SI{15.7}{eV}$ \cite{CCCBDB}) or the spherical top SF$_6$ ($\dcpolarizability=\SI{4.5}{\angstrom^3}$, $\ionienergy = \SI{15.3}{eV}$ \cite{CCCBDB}) as potential trap candidates. Although they have more than three atoms, initial excitation of vibrational modes may be suppressed because the 1064-nm trap light is so far from the fundamental vibrational modes in these molecules. Similarly, although room-temperature data for hydrogen-containing molecules with more than three atoms, like CH$_4$ ($\dcpolarizability = \SI{2.5}{\angstrom^3}$, $\ionienergy = \SI{12.6}{eV}$ \cite{CCCBDB}) and CH$_3$OH ($\dcpolarizability=\SI{3.2}{\angstrom^3}$, $\ionienergy = \SI{10.8}{eV}$ \cite{CCCBDB}), indicate some absorption at near-infrared frequencies \cite{1978_Giver_CH4_Rovib_Overtone_Spectroscopy, 2005_Rueda_Overtone_Spectroscopy_Methanol}, it is unclear that this will lead to infrared multiphoton dissociation (IRMPD) at cold temperatures. Deuteration and halogenation of these molecules may also reduce the risk of IRMPD by lowering their fundamental vibrational frequencies. Measurements of molecular ionization and dissociation rates in our cavity provide valuable information about the kinds of molecules we can trap in our proposed experiment, but also fill a gap in the literature surrounding the interaction of high-intensity, continuous-wave lasers with molecules.

Nevertheless, even the cold chemical reactions of very simple molecules such as those in \cref{tab: molecule list} are difficult to simulate, and therefore interesting to study \cite{2016_Balakrishnan_Controlled_Cold_Chemistry_Review, 2021_Heazlewood_Cold_Chemistry_Review}.
At cold and ultracold temperatures, accurate descriptions of chemical reactions require a fully quantum mechanical treatment \cite{2016_Balakrishnan_Controlled_Cold_Chemistry_Review, 2021_Heazlewood_Cold_Chemistry_Review,2019_Yang_Cold_Interstallar_Clouds,2013_Shannon_OH_CH3OH_Cold_Chemistry}. Despite this, state-of-the-art calculations, including that of the predissociative state lifetime of cold collisional complexes of diatomic bialkali molecules \cite{2019_Christianen_Karman_RRKM_DOS_Calculation}, still rely on semiclassical approximations, which are not valid for SCS molecules (see \cite{2013_Shannon_OH_CH3OH_Cold_Chemistry} and \cref{subsec: Universal Loss}). Our proposed trap is insensitive to most molecular properties and could be loaded with multiple different molecular species simultaneously, allowing for the study of a diverse range of cold chemical reactions. Our trap will therefore provide valuable information about the transition between classical and quantum mechanical descriptions of chemical reactions, and help benchmark new theoretical and numerical techniques to compute the dynamics of cold chemical reactions.

Spectroscopy on cold and ultracold, trapped molecules is another promising application of our trap.
For example, radio searches for interstellar organic molecules are partly limited by a lack of available experimental data to compare to observed spectra \cite{2005_Snyder_Interstellar_Glycine_Controversy, 2020_Puzzarini_Grand_Challenges_in_Astrochemistry}. In many cases, computational models are being used to augment experimental observations of molecular spectra, leading to their more frequent use in molecule searches \cite{2016_Tennyson_Comp_Chem_Rovibrational_Calculations_Better_Than_Measurement}. 
Laboratory measurements of cold molecular spectra in our trap will therefore not only directly assist interstellar molecule searches, but also provide valuable data to calibrate computational models and prove their general accuracy.
Likewise, astrophysical studies of the variation of the proton-to-electron mass ratio $\mu$ from the observation of molecular spectra, such as of CH$_3$OH, would benefit from improved laboratory measurements of the relevant transitions \cite{2018_Safronova_New_Physics_Atoms_Molecules}.
Atomic- or molecular-beam-based precision measurements \cite{2018_ACME_EDM,Beyer2017,Grinin2020} could be improved upon by trapping the atoms or molecules at \SI{1.5}{K} instead, thereby increasing the interaction time and averaging some systematic effects related to the particles' motion such as Doppler shifts.
The light shift from the trap light can be removed by releasing the cold molecules from the trap during measurements. Alternatively, a magic wavelength for the trap light \cite{2008_Ye_Magic_Wavelengths,2019_Kondov_Magic_Wavelength,2023_Leung_Zelevinski_Sr2_Clock_Magic_wavelength,2022_Jozwiak_Magic_Wavelength_Rovibrational_H2} can be chosen such that the differential light shift of a given transition is reduced (see \cref{Appendix: Rotational State Hybridization and Extra Trap Depth}).

\section*{Acknowledgments} 

This work was supported by the Gordon and Betty Moore Foundation (grant no.~9366), the U.S.~Department of Energy, Office of Science, National Quantum Information Science Research Centers, Quantum Systems Accelerator (QSA, no.~DE-AC02-05CH11231), the NASA Jet Propulsion Laboratory (JPL) (grant no.~1669913), the Chan Zuckerberg Initiative (award no.~2021-234606), and Thermo Fisher Scientific (award no.~AWD00004352). A.~S. acknowledges support from the Eleanor Sophia Wood Travelling Scholarship. L.~M. acknowledges support from the Alexander von Humboldt Foundation through a Feodor Lynen Fellowship. We would like to thank Ben Augenbraun, John Doyle and the members of his research group, Arthur Christianen, Bretislav Friedrich, Yair Segev, Tanya Zelevinsky, and Adrianne Zhong for fruitful discussions, and Howard Padmore for conducting mirror surface profile measurements.

\appendix

\section{Rotational state hybridization and extra trap depth}\label{Appendix: Rotational State Hybridization and Extra Trap Depth}

\cref{eq: trap depth} is an approximate expression for the trap depth. Here, two corrections are discussed: firstly, the mean dynamic polarizability $\dcpolarizability(\lambda)$ at $\lambda=\SI{1064}{nm}$ is usually slightly larger than the mean DC polarizability $\alpha_s$ used in \cref{eq: trap depth}. A correction of order $(\lambda_1/\lambda)^2\sim5\%$, where $\lambda_1$ is the wavelength of first electronic excitation of the given molecule, may be warranted for many of the molecules in \cref{tab: molecule list} \cite{1995_Takekoshi_QUEST_theory}. The correction is difficult to estimate accurately for most molecules, but is in any case of little consequence to the experiment.

The second correction to \cref{eq: trap depth} is a result of the hybridization of the rotational states of the molecule, and can have large consequences for the trapping of molecules with large polarizability anisotropies $\dcpolarizabilityani$. In general, symmetric top molecules have a different polarizability along their symmetry axis ($\alpha_\parallel$) and perpendicular to this axis ($\alpha_\perp$: $\dcpolarizabilityani = \alpha_\parallel-\alpha_\perp$, $\dcpolarizability=(\alpha_\parallel+2\alpha_\perp)/3$). In this analysis, these polarizabilities are assumed to be constant within any given rotational band. In a field-free setting, the rotational eigenstates of a molecule will be thermally populated such that there is no molecular alignment. However, optical fields of sufficient intensity will dress these rotational states and align molecules so that their maximally polarizable axis aligns with the optical polarization \cite{1995_Friedrich_Alignment_Trapping_Spheroidal_Wave_Eqn_Theory,2008_Boyd_Nonlinear_Optics}. Linear molecules are described by the Hamiltonian \cite{1995_Friedrich_Alignment_Trapping_Spheroidal_Wave_Eqn_Theory, 1995_Friedrich_Spheroidal_Wave_Eqn_Theory_Details}
\begin{equation}\label{eq: rotational hybridization hamiltonian}
    H = BJ^2 -\frac{E_0^2}{4}\left(\alpha_\parallel\cos^2\theta+\alpha_\perp\sin^2\theta\right).
\end{equation}
Here, $\theta$ is the angle between the molecule's symmetry axis and the electric field with amplitude $E_0$ in a molecule-fixed frame, $J$ is the rotational angular momentum of the molecule, and $B$ is the rotational constant. The time-independent Schr\"{o}dinger equation takes the form of an oblate spheroidal wave equation \cite{1995_Friedrich_Alignment_Trapping_Spheroidal_Wave_Eqn_Theory, 1995_Friedrich_Spheroidal_Wave_Eqn_Theory_Details}
\begin{equation}\label{eq: spheroidal wave equation}
    \left[-\frac{d}{dz}\left[(1-z^2)\frac{d}{dz}\right]+\frac{m^2}{1-z^2}  - \frac{\dcpolarizabilityani E_0^2}{4B}z^2\right] \psi_{\tilde J, m} = \left(\frac{u_{\tilde J, m}}{B} + \frac{\alpha_\perp E_0^2}{4B}\right)\psi_{\tilde J, m},
\end{equation}
where $z=\cos\theta$, $m$ is the projection of angular momentum onto the optical polarization axis, $\tilde J$ is the quantum number which adiabatically turns into $J$ in the limit of zero electric field, and $\psi_{\tilde J, m}e^{im\phi}$ and $u_{\tilde J, m}$ are the eigenfunction and energy of the state with quantum numbers $\tilde J, m$, respectively, where $\phi$ is the azimuthal angle about the optical polarization axis. The solutions for $\psi_{\tilde J, m}$ in \cref{eq: spheroidal wave equation} are the angular oblate spheroidal functions $S_{|m| \tilde J}$ as defined in section 21.6.4 of \cite{1965_Abramowitz_Stegun}. The eigenvalues of \cref{eq: spheroidal wave equation}, and hence the energies $u_{\tilde J, m}$, are readily computed using standard libraries (e.g., \texttt{scipy.special.obl\_cv} of \cite{Scipy}). For $\Delta\alpha E_0^2/4B\ll 1$, we can write a power series expansion for $u_{\tilde J, m}$ (see Section 21.8.1 in \cite{1965_Abramowitz_Stegun}), from which the trap potential $U_{\tilde J, m}$ is given by
\begin{align}\label{eq: trap potential}
\trappotJM =& \,u_{\tilde J, m} - B\tilde J(\tilde J+1) \\
=& - \frac{\dcpolarizability E_0^2}{4} - \frac 1 2 \left[\frac 1 3 - \frac{(2|m|-1)(2|m|+1)}{(2\tilde J-1)(2\tilde J + 3)}\right]\frac{\dcpolarizabilityani E_0^2}{4}\nonumber\\
&+ \left[\frac{-(\tilde J - |m| + 1)(\tilde J - |m| + 2)(\tilde J + |m| + 1)(\tilde J + |m| + 2)}{2(2\tilde J + 1)(2\tilde J + 3)^3(2\tilde J + 5)} \right.\nonumber\\
 &\hphantom{+}\left.+\frac{(\tilde J - |m| - 1)(\tilde J - |m|)(\tilde J + |m| - 1)(\tilde J + |m|)}{2(2\tilde J - 3)(2\tilde J - 1)^3(2\tilde J + 1)}\right]\frac{1}{B}\left(\frac{\dcpolarizabilityani E_0^2}{4}\right)^2 \nonumber\\&+ B\left(O\left(\frac{\dcpolarizabilityani E_0^2}{4B}\right)^3\right).
\end{align}
For the special case of $\tilde J = 0$ and $m = 0$, we have
\begin{equation}
    U_{0,0} = -\frac{\dcpolarizability I}{2\epsilon_0c}-\frac{\dcpolarizabilityani^2 I^2}{270 B \epsilon_0^2c^2} + B\left(O\left(\frac{\dcpolarizabilityani I}{2B\epsilon_0 c}\right)^3\right),
\end{equation}
where $I = \epsilon_0 c E_0^2/2$ is the optical intensity. In the main text, we use $U_{0,0}$ for the trap potential $U$, and define the trap depth as $\trapdepth = \mathrm{max}|U_{0,0}|/\kB$, and the extra trap depth resulting from molecular alignment as $\extratrapdepth = \trapdepth - \dcpolarizability I/2\epsilon_0c\kB$.

For some of the molecules in \cref{tab: molecule list}, including those treated in the detail in the main text, $\extratrapdepth$ is much smaller than \SI{1}{K}, and the degree of alignment (computed using the Hellmann-Feynman theorem as described in \cite{1995_Friedrich_Alignment_Trapping_Spheroidal_Wave_Eqn_Theory}) is small. These molecules are trapped in barely hybridized rotational ground states. However, some of the molecules on the list, such as CO$_2$, N$_2$O, Cl$_2$, and CS$_2$, see a substantial increase in the trap depth due to rotational alignment, which scales nonlinearly with the intensity around \SI{300}{GW/cm^2}. For these molecules, the character of the rotational ground state is highly aligned with the optical polarization (note that this effect is here ignored in the calculation of rotational Raman scattering rates).

The buffer gas collision frequencies ($\sim\!10^5$\,\si{s^{-1}}) and trap frequencies (\SI{<100}{MHz}) are small compared to the rotational constants of the molecules in \cref{tab: molecule list} (\SI{>3}{GHz}). Thus, trapped molecules adiabatically follow the dressed rotational ground state as they traverse the trap. We note in passing that this is unlike a previously proposed experiment to trap polar molecules in microwave fields \cite{2004_DeMille_Microwave_Trap_For_Cold_Polar_Molecules}. In particular, the polarizability at \SI{1064}{nm} is independent of the rotational state, so avoided crossings do not open up avenues for rotational state changes during trap traversal in our experiment, as opposed to \cite{2004_DeMille_Microwave_Trap_For_Cold_Polar_Molecules}.

In precision spectroscopy of atoms, magic wavelength schemes \cite{2008_Ye_Magic_Wavelengths} are used to cancel differential light shifts between two given states by making the dynamic polarizability $\dcpolarizability(\lambda)$, and thus $\trappotJM$, of the states equal by choice of a suitable (``magic'') trap light wavelength. To first order in intensity, a magic wavelength can still be found for molecules, although the wavelength must be tuned to not simply cancel the difference in each state's $\dcpolarizability(\lambda)$ (except in the special case of $\tilde J = 0$ \cite{2019_Kondov_Magic_Wavelength, 2023_Leung_Zelevinski_Sr2_Clock_Magic_wavelength}), but rather, to cancel the first-order term in \cref{eq: trap potential} \cite{2022_Jozwiak_Magic_Wavelength_Rovibrational_H2}. In general, more magic parameters are required to cancel higher-order terms in \cref{eq: trap potential}. 
Rotational state hybridization can represent a large additional contribution $\extratrapdepth$ to  $\trapdepth$ at $I=\SI{300}{GW/cm^2}$ (up to \SI{37}{\percent} of $\trapdepth$ in the case of CS$_2$) when compared to other corrections \cite{2008_Boyd_Nonlinear_Optics}, such as: nonlinear corrections due to electronic and vibrational contributions to molecular hyperpolarizabilities, which typically contribute a few mK in additional trap depth at $I=\SI{300}{GW/cm^2}$ \cite{1994_Archibong_hyperpolarizability_N2, 1995_Luo_hyperpolarizability_O2, 1996_Maroulis_hyperpolarizability_CO, 1998_Fernandez_hyperpolarizability_HCl_HBr, 2020_Fernandez_CS2_hyperpolarizability}, and; linear corrections due to higher order multipoles in the multipole expansion \cite{2022_Jozwiak_Magic_Wavelength_Rovibrational_H2, 2023_Leung_Zelevinski_Sr2_Clock_Magic_wavelength}, which typically are orders of magnitude smaller than the leading linear term \cite{1988_Maroulis_higher_order_multipole_polarizabilities_N2}. Small differences in the polarizability components $\dcpolarizability(\lambda)$ and $\dcpolarizabilityani(\lambda)$ within each rotational band \cite{2022_Jozwiak_Magic_Wavelength_Rovibrational_H2} additionally modify the analysis starting with \cref{eq: rotational hybridization hamiltonian}.

\section{Heat load from scattered trap light}\label{Appendix: Scattered Heat Load}

A major heat load on the cryogenic system in our experiment is scattering of the high-intensity cavity light. Our mirrors will be superpolished to \SI{\sim 1}{\angstrom} RMS (root-mean-square) surface roughness, and the resultant scattered light will be a few hundred \si{mW}. From white light interferometry measurements of our previous mirrors' surface profiles, we have determined the angular distribution of scattered light \cite{1953_Davies_Surface_Roughness, 1961_Bennet_Surface_Roughness, 1991_Steyerl_surface_roughness}. About half of the scattered light is diffusely scattered, and is managed by placing the mirrors more than \SI{10}{mm} from the apertures in the 50-K shields. The other half is scattered into a small cone around the cavity mode. From ray-tracing this scattered light through the near-concentric cavity, we know this light will be incident on either a buffer-gas cell, or the inside of one of the cryocooler's radiation shields, rather than the outside of the radiation shields. The buffer-gas cells will be polished and highly reflective, so light incident on them will be reflected towards the radiation shields (mostly the 4-K shield). We aim to absorb the scattered light that misses the buffer-gas cells on the 50-K shields rather than the 4-K shields, due to their larger cooling capacity. This is achieved by designing the 50-K shields with a smaller angular size than the 4-K shield apertures when viewed from the opposite mirror, so the scattered light will be absorbed on the non-reflective (NR) material on the 50-K shields shown in \cref{fig: cell design}. 

An alternative approach to handle the scattered light is to conically indent the radiation shields around the cavity mode, allowing the apertures to be closer to the cavity focus, and therefore smaller, so that almost no scattered light enters the cryogenic system in the first place. This approach has small consequences for the pumpout timescale, and has not been investigated thoroughly.

All other heat loads from scattering of the high-intensity light, including Rayleigh scattering from the buffer gas and trapped molecules (\si{nW}) and scattering of transmitted input light not matched into the cavity mode (\SI{<1}{mW}), are negligible for our experiment.

\section{Details of ergodic loading simulations}\label{Appendix: Details of Ergodic Loading Simulations} 

At their core, the ergodic loading simulations work by storing the energy $E$ of $Q$ molecules, and assigning initial conditions for $Q$--He collisions stochastically based on this energy within the ergodic approximation, before enacting a billiard-ball elastic collision and tracking the change to the energy after the collision, which is again stored, and so on. This is numerically efficient, since no trajectories need to be integrated to simulate the dynamics. The simulations rely on the ergodic approximation, valid when the orbits in the trap are fast compared to the mean $Q$--He collision time. The simulations also assume the $Q$ density is small compared to the He density, so $Q$--He collisions need to be considered but $Q$--$Q$ collisions can be mostly ignored. In all collisions, the small effect of the trap light on the buffer gas atoms is ignored. We also approximate all collisions as equally likely to occur at all times so that the ergodic theorem is relevant for drawing initial conditions, even though, as in \cref{subsec: Buffer Gas Cooling}, collisions are technically more likely to occur at times of $Q$'s motion where its speed is higher. 

In our simulations, we approximate the trap potential of a single lattice site $j$ as a truncated harmonic potential,
\begin{equation}\label{eq: harmonic trap potential approximation}
    U(\mathbf{x}; j)\approx \min\left[\left(- \eta(j)\kB T + \sum_{i=1}^3\frac 1 2 m_Q \omega_i^2(j)x_i^2\right), 0\right],
\end{equation}
in order to efficiently sample initial conditions for collisions from the energy hypersurfaces. Here, the $\omega_i$ represent the trap angular frequencies $\omega_x = \omega_y = \sqrt{4 \eta(j) k_B T/m_Qw(j)^2} = \omega_r$ and $\omega_z = \sqrt{8 \pi^2 \eta(j) k_B T / m_Q \lambda^2}$. We also ignore the constraint that orbits in the trap conserve the $z$ component of angular momentum. We stress that we are still making the ergodic approximation, we are simply taking initial conditions for collisions from a simpler physical system than the fully nonlinear trapping potential. The simulation volume for lattice site $j$, centered on the antinode at $z_j$, is defined as the region where the approximate trap potential in \cref{eq: harmonic trap potential approximation} is nonzero, and consists of an ellipsoid with volume $V_\mathrm{e}(z_j) = w^2(z_j)\lambda/3$, assuming the lattice site is near the trap center where wavefront curvature can be ignored.

In action-angle coordinates, sampling from the energy hypersurface amounts to drawing three random angle variables and three random actions $J_i$ under the constraint that $E +\eta k_B T = \sum_{i=1}^3\omega_iJ_i$. This is efficiently done by drawing two random numbers between 0 and 1 and using them as a partition of the interval into three randomly drawn pieces. These are translated back to regular phase-space coordinates to draw initial conditions for bound $Q$ molecules in collisions.

There are two forms of the ergodic loading simulations, namely, that which computes the loading fractions $f_i$, and that which computes the mean number of $Q$--He collisions needed to eject a trapped molecule, $\kevap$, which are discussed in \cref{subsubsec: microscopic ergodic loading model}. The results of both are required to construct the loading model in \cref{subsubsec: time dependent loss inclusive loading model}. For the simulation of the loading fractions $f_i$, an untrapped $Q$ molecule is first spawned randomly in the simulation volume. Its velocity is drawn from a Maxwell distribution at the buffer gas temperature $T = \SI{1.5}{K}$ and its position $\mathbf{x}$ is drawn uniformly within the simulation volume, before its kinetic energy is increased due to the local trap potential $U(\mathbf{x})$. Its speed is computed as $v = \sqrt{2(E-U)/m_Q}$, and its direction is randomized. The initial condition for a first billiard-ball collision is now set by drawing a buffer gas atom from a Maxwell distribution at $T$. If the molecule is bound by the collision, only its energy needs to be retained, since before every subsequent collision new initial conditions are drawn by randomly sampling the energy hypersurface within the ergodic approximation. There is no need to integrate the $Q$ trajectory. The simulation continues until $Q$'s energy either becomes positive and the molecule escapes, or it stays negative and falls below the energy of a molecule drawn from a Boltzmann distribution in the trap at $T$. 

For the simulation of the number of $Q$--He collisions needed to eject a trapped molecule, $\kevap$, a molecule is initially drawn with an energy from a Boltzmann distribution within the trap, and with phase-space coordinates drawn ergodically from the corresponding energy hypersurface. Billiard-ball collisions are repeatedly enacted with buffer gas atoms until the molecule is eventually ejected from the trap ($E>0$), giving $\kevap$.

\section{Trap heating caused by laser noise}\label{subsec: One-Body Loss Rate}

Laser noise can cause trap heating in two ways \cite{1997_Savard_heating_from_laser_noise, 2000_Gardiner_Savard_analysis_worse}. Firstly, laser intensity noise (RIN) at twice the trap angular frequencies $\omega_r = 2\pi\times \SI{2}{MHz}$ and $\omega_z = 2\pi\times\SI{68}{MHz}$ lead to parametric driving of trapped molecules, resulting in a heating rate $\Gamma^{\mathrm{RIN}}$ at which energy grows exponentially in the trap:
\begin{equation}
    \Gamma_i^{\mathrm{RIN}} = \frac{1}{4}\omega_{i}^2S_\epsilon(2\omega_i),
\end{equation}
where $S_{\epsilon}(\omega)$ is the one-sided power spectrum (in units of \si{dBc/Hz}) of the relative intensity noise $\Delta I/I$ at an angular frequency $\omega$ \cite{1997_Savard_heating_from_laser_noise}.

We have measured the RIN of our laser oscillator (NKT Koheras Adjustik Y10) to be \SI{-147}{dBc/Hz} (\SI{-154}{dBc/Hz}) at \SI{4}{MHz} (\SI{136}{MHz}), which corresponds to $\Gamma_r^{\mathrm{RIN}} = \SI{8e-2}{s^{-1}}$ ($\Gamma_z^{\mathrm{RIN}} = \SI{18}{s^{-1}}$). However, the cavity acts as a frequency filter with, for perfect spatial mode matching, the normalized transfer function $|G(\omega)|^2 = (\pi^2/2\mathcal{F}^2)/\left(1-\cos{(2\pi f/\mathrm{FSR})}\right)$ (for $\mathcal{F} \gg 1$ and $|f-N\,\mathrm{FSR}| \gg \Delta\nu$, $N = 0,1,\dots$), where $f$ is the frequency offset from the frequency resonant with the cavity, $\mathcal{F}$ is the cavity's finesse, and $L$ is its length, and $\mathrm{FSR} = c/2L$ and $\Delta\nu = \mathrm{FSR}/\mathcal{F}$ are the cavity's resulting free spectral range and linewidth, respectively. RIN at frequency $f$, which corresponds to amplitude modulation at $f$, is thus further suppressed. At $f = 2\omega_r/2\pi$ ($f = 2\omega_z/2\pi$), this results in a suppression by \SI{46}{dB} (\SI{76}{dB}), and a resulting $\Gamma_r^{\mathrm{RIN}} = \SI{\sim2e-6}{s^{-1}}$ ($\Gamma_z^{\mathrm{RIN}} = \SI{\sim5e-7}{s^{-1}}$), assuming the measured RIN values outside the cavity. We note that the value of $\Gamma_z^{\mathrm{RIN}}$ is within a factor of two of the shot-noise of the circulating power inside the cavity. We also note that imperfect spatial mode matching will reduce the cavity's suppression. In particular, $2 \omega_r = 2 \pi \times \SI{4}{MHz}$ is close to the difference in resonance frequencies of $\Delta f_{10} = \SI{3.6}{MHz}$ between the TEM$_{01}$/TEM$_{10}$ modes and the fundamental TEM$_{00}$ mode. Finally, we do not expect that the high-power fiber amplifier to be used in the experiment adds substantial RIN at the frequencies of interest.

Secondly, movement of the trap center by an amount $\boldsymbol{x}(t)$ results in a linear growth in average energy,
\begin{equation}
    \langle\dot E\rangle_i = \frac{\pi}{2}m_Q\omega_{i}^4S_{\boldsymbol{x}}(\omega_{i}),
\end{equation}
which now could lead to significant heating due to the quartic dependence on the trap frequency. Here, $S_{\mathbf{x}}$ is the one-sided power spectrum of position fluctuations in the trap center \cite{1997_Savard_heating_from_laser_noise}. However, we expect acoustic noise at MHz frequencies to be sufficiently small.

\section{Two-body loss models}\label{subsec: two-body loss models}

\subsection{Universal loss}\label{subsec: Universal Loss}

In optical trapping experiments on bialkali molecules, two-body loss is a result of so-called ``universal loss'', in which a large fraction of collisions between trapped molecules lead to loss. One argument suggests that our trap will not experience universal loss. In this argument, the loss mechanism involves strong perturbation of molecular energy levels by a partner, which leads to an excited state of the two-body complex located one photon energy of the trap light above the ground state \cite{2019_Christianen_Karman_Two-body_Loss_ab_initio_molpro}. For long enough lifetimes of the complex, the probability to reach this excited state, causing subsequent loss of the molecules, then approaches unity. However, the interactions between SCS molecules are often of order \SI{20}{meV} \cite{2008_Hellmann_CH4_CH4_interaction_potential, 2013_Hellmann_N2_N2_interaction_potential}, which is two orders of magnitude too small to perturb the excited states by several \si{eV} and unlock a single-photon excitation of the complex. Moreover, the complex lifetimes of SCS molecules (based on a semiclassical model presented in \cite{2019_Christianen_Karman_RRKM_DOS_Calculation}) are only of order picoseconds, so the notion of ``sticky molecular collisions'' seems inapplicable to our trap. Hence, one expects not to see universal loss for $Q$.

However, recent literature has cast some doubt over this explanation \cite{2021_Bause_universal_loss_debate}. Therefore, we consider a worst-case scenario in which universal loss occurs for all $Q$--$Q$ collisions in the trap. This worst-case scenario provides an upper bound for the two-body loss rates expected from general two-body loss processes that are difficult to estimate in the cold, high-intensity conditions of our trap. We do not, however, treat $Q$--He collisions as lossy, as the interactions are only $O(\SI{5}{meV})$ \cite{1978_Slankas_300K_He_CH4_cross_section}.

In universal loss, the functional form of the two-body loss coefficient $\betatb$ depends on whether or not an $s$-wave two-body interaction is allowed, which in turn depends on whether the molecules can be regarded as indistinguishable and Bosonic. Since the $s$-wave loss rates are higher than the p-wave loss rates, we must assume in a worst-case estimate that we will be limited by $s$-wave scattering, in which \cite{2009_Ospelkaus_universal_loss}
\begin{equation}
    \betatb = 4\pi\frac{\hbar}{\mu}a,
\end{equation}
where the Van der Waals length $a$ is here defined by the $C_6$ coefficient through $a = 0.4778 (2\mu C_6/\hbar^2)^{1/4} $ \cite{2009_Julienne_C6_def_and_inelastic_losses}, and $\mu$ is the reduced mass of the two bodies. 

For many small molecules, a typical value of $C_6$ is a few tens to hundreds of atomic units ($E_h a_0^6$), and the $C_6^{1/4}$ dependence of $\betatb$ means a good estimate is not critical. Using $C_6$ coefficients from an approximate formula in \cite{2016_Tao_C_6_coefficients_Simple_SFA_Model}, we obtain, including an extra factor of two as noted in \cite{2009_Ospelkaus_universal_loss}, values of $\betatb$ slightly less than $\SI{1e-10}{cm^3/s}$ for many molecules, and assign $\betatb=10^{-10}$\,\si{cm^3/s} for $Q$, which is consistent with typical values in \cite{2009_Ospelkaus_universal_loss}. In the main text, we keep in mind the strong likelihood that universal loss does not occur for $Q$, in which case two-body losses may not limit the trapped density. By experimentally observing the two-body loss rate for molecules in our trap, we will be able to provide useful insight into the nature of cold collisions and the mechanism of universal loss.

\subsection{Collision-induced absorption}\label{subsec: collision-induced absorption}

In high-density gases, optical transitions that are forbidden for single molecules can become allowed through collision-induced absorption (CIA) \cite{1994_Frommhold_Collision_Induced_Absorption_in_Gases}. To our knowledge, of the molecules in \cref{tab: molecule list}, only O$_2$ exhibits CIA at \SI{1064}{nm}. From Fig 1\,(b) of \cite{2018_Karman_O2_N2_collisional_absorption}, the absorption of this feature is $\alpha_{\mathrm{CIA}}=10^{-6}\,\si{cm^{-1} am^{-2}}=\SI{1.39e-45}{cm^5/molecule^2}$. Assuming every CIA event leads to loss, the resulting two-body loss coefficient is $\beta=\alpha_{\mathrm{CIA}}I/\hbar\omega = \SI{2.2e-15}{cm^3/s}$, which is small compared to other two-body loss coefficients considered in \cref{fig: trap_number_evolution}. It therefore appears that even in the worst case of CIA features centered at \SI{1064}{nm}, the densities required to induce absorption are high compared to those we expect to see in our experiment. CIA effects can therefore be ignored in our experiment.

\section{Trap heating and loss from rotational Raman scattering}\label{Appendix: Trap Heating and Loss from RRS}

Here, we compute the rotational Raman scattering (RSS) loss term $\Ramanlossrate$ of \cref{eq: dN/dt}. RRS from the $J=0$ rotational ground state to the $J=2$ excited state increases a molecule's internal energy by $\DeltaERRS = 6B$, which is \SI{\sim 15}{K}, and hence greater than $\trapdepth$, for many molecules in \cref{tab: molecule list}. Rotationally inelastic collisions typically occur once per $\sim10$ elastic collisions \cite{2012_Hutzler_Buffer_Gas_Beam_Review}, which is $10^{4}$\,\si{s^{-1}}, much faster than RRS rates in \cref{tab: molecule list}. We therefore assume molecules are only rotationally excited once before being immediately rotationally cooled to the ground state, whereby their rotational energy is converted to translational kinetic energy. 

In $Q(J=2)$--$Q(J=0)$ inelastic collisions, which occur at a rate $\QQinelasticrate$, on average $\DeltaERRS$ is split evenly between the molecules, so most likely a molecule will be ejected from the trap. On the other hand, in $Q(J=2)$--He inelastic collisions, which occur at a rate $\QHeinelasticrate$, only $\mu/m_Q = 1/8.5$ of $\DeltaERRS$ goes to the molecule's translational energy, the rest being carried away by the much lighter He atom. Therefore, while the fraction $\QQinelasticrate/(\QHeinelasticrate+\QQinelasticrate)$ of RRS events simply lead to loss, the remaining fraction $\QHeinelasticrate/(\QHeinelasticrate+\QQinelasticrate)$ do not. They instead lead to heating of the sample of $N$ trapped molecules, increasing the total kinetic and potential energy of the trapped sample at a rate $\Edoth = \frac{\mu}{m_Q}\DeltaERRS \rotRamanrate N$. 
This heating will lead to some molecule loss, which we can bound above by assuming, conservatively, that the buffer-gas cooling does not compensate for $\Edoth$ at all, and the heat is instead balanced exclusively by molecule loss. Every time a thermalized molecule is lost from the trapped sample, 
it carries away on average $(\eta+\kappa-3)k_BT$ of energy ($\kappa = (1-\Gamma(5, \eta))/(\eta\Gamma(3,\eta)-4\Gamma(4,\eta))$, where $\Gamma$ is the incomplete Gamma function \cite{2013_Bourgain_Evaporative_Cooling_Small_Number_of_Molecules_Dipole_Trap}). Balancing $\Edoth$ with this loss and combining with the $Q(J=2)$--$Q(J=0)$ case, an upper bound for the overall loss rate due to RRS is
\begin{equation}\label{eq: Ramanlossrate}
    \Ramanlossrate = -\frac{\QHeinelasticrate}{\QHeinelasticrate+\QQinelasticrate}\frac{(\mu/m_Q)\DeltaERRS\rotRamanrate}{(\eta+\kappa-3)\kB T}N - \frac{\QQinelasticrate}{\QHeinelasticrate+\QQinelasticrate}\rotRamanrate N.
\end{equation}
This loss term has a nonlinear dependence on $N$ as $\QQinelasticrate$ depends on the trapped density.

To evaluate \cref{eq: Ramanlossrate}, inelastic collision cross sections are required. For N$_2$ (\cref{fig: trap_number_evolution} (d)), we approximate He--N$_2$ inelastic collision cross sections using the elastic cross section from \cite{1986_Beneventi_He_molecule_collision_cross_sections} and the inelastic-to-elastic collision ratio from \cite{1995_Belikov_N2_He_Inelastic_Rate}, and we take the N$_2$--N$_2$ inelastic collision cross sections from \cite{1967_Miller_N2_N2_inelastic_cross_section, 1988_Belikov_N2_N2_inelastic_cross_section}. We estimate the inelastic N$_2$--N$_2$ and N$_2$--He cross sections at \SI{1.5}{K} to be \SI{100}{\angstrom^2} and \SI{40}{\angstrom^2}, respectively.

\section{Laser-induced breakdown}\label{appendix: Laser Induced Breakdown}

Laser-induced breakdown occurs when an electron created in the laser beam, for instance by multiphoton ionization (MPI), is heated by inverse bremsstrahlung in collisions with neighboring molecules to above the ionization energy of a molecule, so it can free new electrons by impact ionization. This can lead to exponential growth in the electron number. The process is well understood in high-intensity, pulsed lasers and at high gas densities \cite{1975_Morgan_Breakdown_ionisation_review, 1983_Ali_breakdown_threshold, 1987_Rosen_breakdown_N2_noble_gases, 1989_Radziemski_laser_induced_plasmas_and_applications, 2016_Isaacs_breakdown_thresholds, 2020_Woodbury_PRL_AbsoluteMeasuremntLaserIonization}. However, the treatment of breakdown in continuous-wave (CW) laser beams is often left as an afterthought, because CW intensities are rarely high enough to induce breakdown.

Fortunately, the theory of breakdown in CW laser beams is straightforward. In this case, breakdown can be treated as an electron diffusion problem: breakdown occurs when an electron spawns in the beam and causes more than one ionization event before it leaves. Once outside the beam, the electrons will cool down rather than be collisionally heated, and eventually recombine with ions, and so will no longer contribute to further ionization events. We do not consider the ion motion as contributing to breakdown, because the collisional heating rate of the ions is suppressed by their much higher mass.

At the relatively low buffer gas densities we are interested in, macroscopic treatments of breakdown \cite{1987_Rosen_breakdown_N2_noble_gases} are not required, and it is instead numerically tractable to simulate entire electron trajectories through the laser beam. Electrons are spawned at the center of the beam in the radial direction, and within a Rayleigh range of the focus in the axial direction. Based on \cite{1975_Morgan_Breakdown_ionisation_review}, their initial energies are drawn from a Boltzmann distribution with mean energy $\ionienergy/3$, $\ionienergy$ being the He ionization energy. For self-consistency, we have confirmed in our simulations that electrons formed from impact ionization events at intensities of order \SI{100}{GW/cm^2} have roughly this energy. The electron motion is modeled step-wise, with step sizes drawn from an exponential distribution of mean size $l_e$, where $l_e$ is the electron mean free path, computed as 
\begin{equation}
    \frac{1}{l_e} = \sum_{\mathrm{gas}\, i}\sum_{\mathrm{process\, j}} n_i\sigma_{ij},
\end{equation}
where $i$ indexes between $Q$ and He, $n_i$ represent the gas number densities, and $\sigma_{ij}$ is the cross section for electron collisions with gas $i$ that lead to process $j$. The gas densities are taken to be consistent with \cref{eq: dN/dt} during the loading phase, and we assume there is no ion density, as electron multiplication begins before there is a large number of ions in the beam. The processes we consider are elastic collisions, electronically exciting collisions, and ionizing collisions. Exciting collisions are counted as ionizing collisions as the trap light would likely ionize any electronically excited molecule. In each case, we treat the gas particles as stationary targets due to their much lower velocity, and model the cross sections as spherically symmetric for simplicity.

In each collision, a number of photons may be absorbed due to inverse bremsstrahlung, which modifies the cross sections. For an exciting or elastic process with a laser-free cross section $\sigma_{E_1\rightarrow E_2}^{(0)}(E_1)$, the cross section is modified to \cite{1971_Brehme_multiphoton_inv_bms, 1972_Seely_Inverse_Bremsstrahlung_Full, 1980_Cavaliere_light_assisted_e_impact_ionisation, Zarcone_1983_light_assisted_impact_ionisation_He}
\begin{align}\label{eq:light assisted electron cross sections}
    &\sigma_{E_1\rightarrow E_2}^{(0)}(E_1)\rightarrow\sum_{n=-\infty}^{\infty}\sigma_{E_1\rightarrow E_2}^{(n)} \nonumber\\&= \sum_{n=-\infty}^{\infty}\sigma_{E_1+n\hbar\omega\rightarrow E_2}^{(0)}(E_1+n\hbar\omega)\sqrt{1 + \frac{n\hbar\omega}{E_1}}J_n^2\left(\frac{eE_0}{\electronmass\omega^2}|\Delta k_x|\right)\bigg{|}_{E_2=E_1+n\hbar\omega>0},
\end{align}
where $n$ represents the number of photons absorbed in the process, $e$ the electron charge, $E_0$ the electric field amplitude, $\electronmass$ the electron mass, $\omega$ the optical angular frequency, $J_n$ the $n^{\mathrm{th}}$-order Bessel function of the first kind, and $\Delta k_x$ the change (along the axis of optical polarization, here the $x$ axis) in the electron's $\mathbf{k}$ vector in the collision. We do not modify cross sections in which a third body is created in this way.

In our simulations, we use a particular target, CH$_4$, to represent $Q$, because electron-impact cross section data is difficult to choose for $Q$. The energy-dependent, laser-free cross sections for helium are compiled from \cite{1992_Brunger_He_e_elastic_cross_section} (elastic), \cite{2000_Ralchenko_electron_helium_cross_sections} (exciting), and \cite{1984_Montague_He_e_ionisation} (ionizing). CH$_4$ cross sections are taken from Song \cite{2014_Song_Cross_Sections_for_electron_CH4_collisions}, except for dissociation cross sections (which we take to be inelastic but not ionizing) which are found in \cite{2010_Fuss_e_impact_CH4_cross_sections} and multiplied by 2 to fit with Song's sparsely populated but likely more accurate data. The binary encounter dipole model (\cite{2000_Kim_BED_ionisation_cross_section} for helium and \cite{1996_Hwang_BED_ionisation_cross_section} for methane) is used to draw electron velocities after impact ionization events to continue their trajectories. Other molecules, like N$_2$ and O$_2$, have similar orders of magnitude for their electron-impact cross sections \cite{2006_Itikawa_e_N2_cross_sections, 2008_Itikawa_e_O2_cross_sections}, so we believe that our conclusions for breakdown in a mostly helium He--CH$_4$ mixture will generalize to other species we would like to trap.

One challenge with using these cross sections is that $l_e$ depends on the local light intensity and gas density. One would need to update $l_e$ depending on where the electron moves in one step, which is itself determined by $l_e$. We overcome this using a ``feedforward'' method, in which $l_e$ is determined by the local intensity and densities at the previous step. We expect this to only cause small errors, and this approximation is needed to make the problem tractable.

Armed with the relevant cross sections, at each step of mean distance $l_e$ and random direction, we stochastically decide which process occurs and how many photons are absorbed with relative weights $n_i\sigma_{ij}^{(n)}$. From the chosen process, we update the electron energy, and count whether an ionization event has occurred. We also update the electron energy by the ponderomotive potential, which causes minimal change to the simulation results. We stop an electron's simulation when it has moved three waists from the beam's central axis. We also track the total time taken for the electron to diffuse, and we can add electrons spawned from multiphoton ionization (MPI) during this time to compute the breakdown condition when MPI is included. We count how many electrons are created from each electron we spawned, and if this number is greater than one, we decide that breakdown has occurred.

\begin{figure}
    \centering
    \includegraphics[width = 0.5\linewidth]{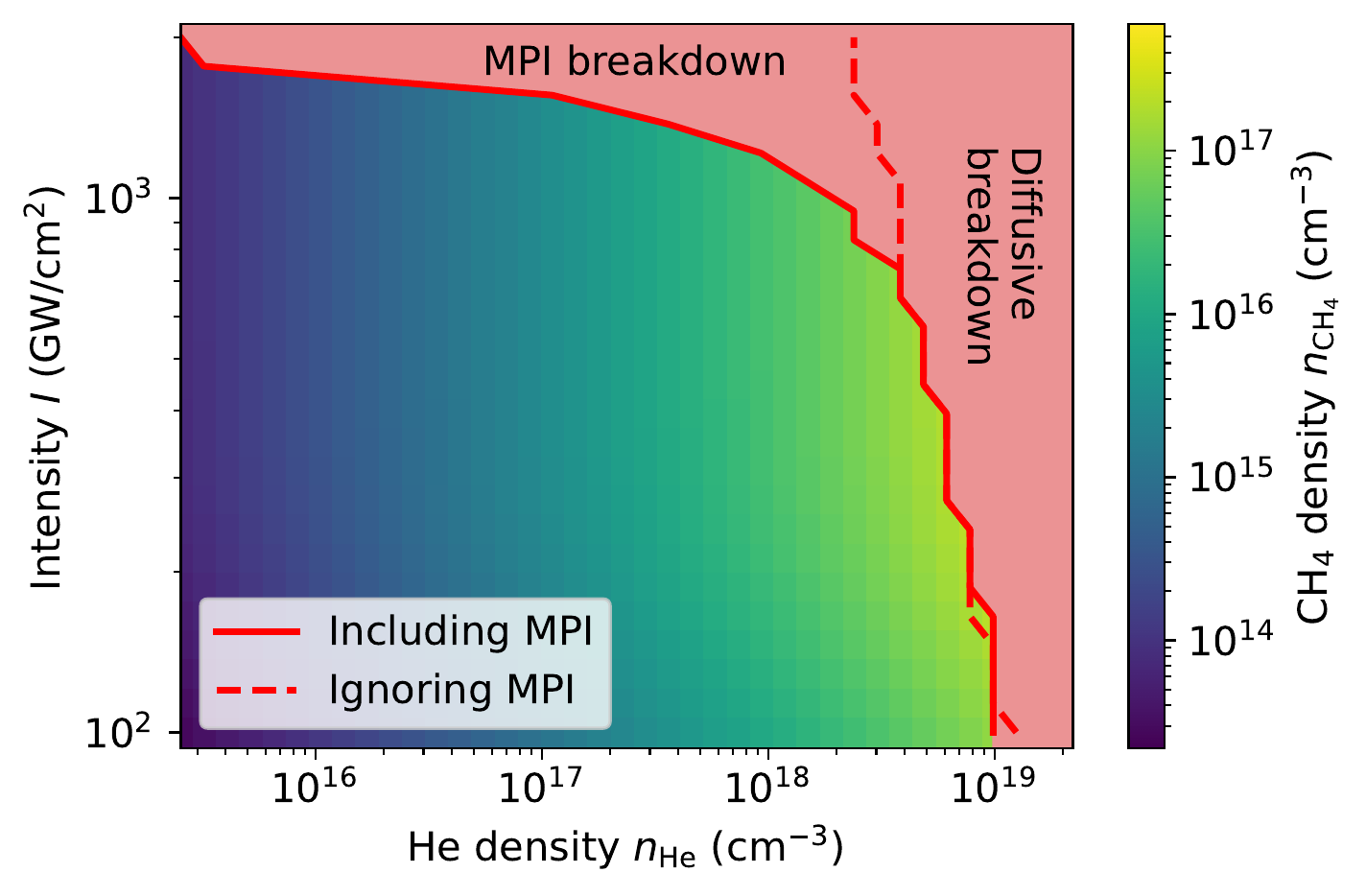}
    \caption{Laser-induced breakdown simulations. Simulations determine when one electron spawned in the beam creates more than one electron in the laser beam before it diffuses out. The solid curve includes electrons created independently through multiphoton ionization (MPI) before the electron diffusion out of the beam. The dashed curve ignores these electrons and purely counts electrons formed by impact ionization. The red-filled region is that bounded below by the solid curve, indicating the region where breakdown occurs. Inset are the CH$_4$ densities used in the simulation, which are taken from the steady state of \cref{eq: dN/dt}, despite the fact that the model is not strictly valid at the comparatively high densities used in this simulation.
    Here, we fix $n_{\mathrm{CH}_4,\mathrm{LR}}/n_{\mathrm{He}}=1/100$ and assume universal loss in \cref{eq: dN/dt}. Our conclusions on breakdown are not sensitive to these choices.        
    }
    \label{fig:breakdown figure}
\end{figure}

The results are shown in \cref{fig:breakdown figure}. It is known that in the diffusive breakdown regime, the intensity threshold $I_{\mathrm{th}}$ depends on the gas pressure $p$ as \cite{1989_Radziemski_laser_induced_plasmas_and_applications}
\begin{equation}
    I_{\mathrm{th}} \sim p^{-2/m'},
\end{equation}
where $m'$ is a constant near 1. For breakdown dominated by MPI instead of electron diffusion, this constant $m'$ becomes quite large, and the pressure dependence becomes small. We can see both of these regimes in \cref{fig:breakdown figure}. When MPI electrons are ignored, there is a strong dependence on pressure, but when they are added, $I_{\mathrm{th}}$ becomes a slow-varying function of pressure at high enough intensities. Breakdown appears to occur at densities and intensities much higher than we will require, so it will not limit our experiment.


\bibliography{bibliography}

\begin{thebibliography}{204}%
\makeatletter
\providecommand \@ifxundefined [1]{%
 \@ifx{#1\undefined}
}%
\providecommand \@ifnum [1]{%
 \ifnum #1\expandafter \@firstoftwo
 \else \expandafter \@secondoftwo
 \fi
}%
\providecommand \@ifx [1]{%
 \ifx #1\expandafter \@firstoftwo
 \else \expandafter \@secondoftwo
 \fi
}%
\providecommand \natexlab [1]{#1}%
\providecommand \enquote  [1]{``#1''}%
\providecommand \bibnamefont  [1]{#1}%
\providecommand \bibfnamefont [1]{#1}%
\providecommand \citenamefont [1]{#1}%
\providecommand \href@noop [0]{\@secondoftwo}%
\providecommand \href [0]{\begingroup \@sanitize@url \@href}%
\providecommand \@href[1]{\@@startlink{#1}\@@href}%
\providecommand \@@href[1]{\endgroup#1\@@endlink}%
\providecommand \@sanitize@url [0]{\catcode `\\12\catcode `\$12\catcode
  `\&12\catcode `\#12\catcode `\^12\catcode `\_12\catcode `\%12\relax}%
\providecommand \@@startlink[1]{}%
\providecommand \@@endlink[0]{}%
\providecommand \url  [0]{\begingroup\@sanitize@url \@url }%
\providecommand \@url [1]{\endgroup\@href {#1}{\urlprefix }}%
\providecommand \urlprefix  [0]{URL }%
\providecommand \Eprint [0]{\href }%
\providecommand \doibase [0]{https://doi.org/}%
\providecommand \selectlanguage [0]{\@gobble}%
\providecommand \bibinfo  [0]{\@secondoftwo}%
\providecommand \bibfield  [0]{\@secondoftwo}%
\providecommand \translation [1]{[#1]}%
\providecommand \BibitemOpen [0]{}%
\providecommand \bibitemStop [0]{}%
\providecommand \bibitemNoStop [0]{.\EOS\space}%
\providecommand \EOS [0]{\spacefactor3000\relax}%
\providecommand \BibitemShut  [1]{\csname bibitem#1\endcsname}%
\let\auto@bib@innerbib\@empty
\bibitem [{\citenamefont {Metcalf}\ and\ \citenamefont {van~der
  Straten}(2001)}]{2001_Metcalf_Book_Laser_cooling_and_Trapping}%
  \BibitemOpen
  \bibfield  {author} {\bibinfo {author} {\bibfnamefont {H.}~\bibnamefont
  {Metcalf}}\ and\ \bibinfo {author} {\bibfnamefont {P.}~\bibnamefont {van~der
  Straten}},\ }\href {https://books.google.com.au/books?id=i-40VaXqrj0C} {\emph
  {\bibinfo {title} {Laser Cooling and Trapping}}},\ Graduate Texts in
  Contemporary Physics\ (\bibinfo  {publisher} {Springer New York},\ \bibinfo
  {year} {2001})\BibitemShut {NoStop}%
\bibitem [{\citenamefont {Postler}\ \emph {et~al.}(2022)\citenamefont
  {Postler}, \citenamefont {Heu{\ss}en}, \citenamefont {Pogorelov},
  \citenamefont {Rispler}, \citenamefont {Feldker}, \citenamefont {Meth},
  \citenamefont {Marciniak}, \citenamefont {Stricker}, \citenamefont
  {Ringbauer}, \citenamefont {Blatt}, \citenamefont {Schindler}, \citenamefont
  {Müller},\ and\ \citenamefont
  {Monz}}]{2021_Postler_Universal_Quantum_Gates}%
  \BibitemOpen
  \bibfield  {author} {\bibinfo {author} {\bibfnamefont {L.}~\bibnamefont
  {Postler}}, \bibinfo {author} {\bibfnamefont {S.}~\bibnamefont {Heu{\ss}en}},
  \bibinfo {author} {\bibfnamefont {I.}~\bibnamefont {Pogorelov}}, \bibinfo
  {author} {\bibfnamefont {M.}~\bibnamefont {Rispler}}, \bibinfo {author}
  {\bibfnamefont {T.}~\bibnamefont {Feldker}}, \bibinfo {author} {\bibfnamefont
  {M.}~\bibnamefont {Meth}}, \bibinfo {author} {\bibfnamefont {C.~D.}\
  \bibnamefont {Marciniak}}, \bibinfo {author} {\bibfnamefont {R.}~\bibnamefont
  {Stricker}}, \bibinfo {author} {\bibfnamefont {M.}~\bibnamefont {Ringbauer}},
  \bibinfo {author} {\bibfnamefont {R.}~\bibnamefont {Blatt}}, \bibinfo
  {author} {\bibfnamefont {P.}~\bibnamefont {Schindler}}, \bibinfo {author}
  {\bibfnamefont {M.}~\bibnamefont {Müller}},\ and\ \bibinfo {author}
  {\bibfnamefont {T.}~\bibnamefont {Monz}},\ }\bibfield  {title} {\bibinfo
  {title} {Demonstration of fault-tolerant universal quantum gate operations},\
  }\href {https://doi.org/10.1038/s41586-022-04721-1} {\bibfield  {journal}
  {\bibinfo  {journal} {Nature}\ }\textbf {\bibinfo {volume} {605}},\ \bibinfo
  {pages} {675} (\bibinfo {year} {2022})}\BibitemShut {NoStop}%
\bibitem [{\citenamefont {Bloch}\ \emph {et~al.}(2012)\citenamefont {Bloch},
  \citenamefont {Dalibard},\ and\ \citenamefont
  {Nascimb{\`e}ne}}]{2012_Bloch_Quantum_Simulations_with_ultracold_Quantum_Gases}%
  \BibitemOpen
  \bibfield  {author} {\bibinfo {author} {\bibfnamefont {I.}~\bibnamefont
  {Bloch}}, \bibinfo {author} {\bibfnamefont {J.}~\bibnamefont {Dalibard}},\
  and\ \bibinfo {author} {\bibfnamefont {S.}~\bibnamefont {Nascimb{\`e}ne}},\
  }\bibfield  {title} {\bibinfo {title} {Quantum simulations with ultracold
  quantum gases},\ }\href {https://doi.org/10.1038/nphys2259} {\bibfield
  {journal} {\bibinfo  {journal} {Nature Physics}\ }\textbf {\bibinfo {volume}
  {8}},\ \bibinfo {pages} {267} (\bibinfo {year} {2012})}\BibitemShut {NoStop}%
\bibitem [{\citenamefont {Lewenstein}\ \emph {et~al.}(2012)\citenamefont
  {Lewenstein}, \citenamefont {Sanpera},\ and\ \citenamefont
  {Ahufinger}}]{2012_Lewenstein_Ultracold_Atoms_in_Optical_Lattices}%
  \BibitemOpen
  \bibfield  {author} {\bibinfo {author} {\bibfnamefont {M.}~\bibnamefont
  {Lewenstein}}, \bibinfo {author} {\bibfnamefont {A.}~\bibnamefont
  {Sanpera}},\ and\ \bibinfo {author} {\bibfnamefont {V.}~\bibnamefont
  {Ahufinger}},\ }\href@noop {} {\emph {\bibinfo {title} {Ultracold Atoms in
  Optical Lattices: {S}imulating quantum many-body systems}}}\ (\bibinfo
  {publisher} {OUP Oxford},\ \bibinfo {year} {2012})\BibitemShut {NoStop}%
\bibitem [{\citenamefont {Ludlow}\ \emph {et~al.}(2015)\citenamefont {Ludlow},
  \citenamefont {Boyd}, \citenamefont {Ye}, \citenamefont {Peik},\ and\
  \citenamefont {Schmidt}}]{2015_Ludlow_Optical_Atomic_Clocks}%
  \BibitemOpen
  \bibfield  {author} {\bibinfo {author} {\bibfnamefont {A.~D.}\ \bibnamefont
  {Ludlow}}, \bibinfo {author} {\bibfnamefont {M.~M.}\ \bibnamefont {Boyd}},
  \bibinfo {author} {\bibfnamefont {J.}~\bibnamefont {Ye}}, \bibinfo {author}
  {\bibfnamefont {E.}~\bibnamefont {Peik}},\ and\ \bibinfo {author}
  {\bibfnamefont {P.~O.}\ \bibnamefont {Schmidt}},\ }\bibfield  {title}
  {\bibinfo {title} {Optical atomic clocks},\ }\href
  {https://doi.org/10.1103/RevModPhys.87.637} {\bibfield  {journal} {\bibinfo
  {journal} {Reviews of Modern Physics}\ }\textbf {\bibinfo {volume} {87}},\
  \bibinfo {pages} {637} (\bibinfo {year} {2015})}\BibitemShut {NoStop}%
\bibitem [{\citenamefont {Parker}\ \emph {et~al.}(2018)\citenamefont {Parker},
  \citenamefont {Yu}, \citenamefont {Zhong}, \citenamefont {Estey},\ and\
  \citenamefont {M{\"u}ller}}]{2018_Muller_alpha}%
  \BibitemOpen
  \bibfield  {author} {\bibinfo {author} {\bibfnamefont {R.~H.}\ \bibnamefont
  {Parker}}, \bibinfo {author} {\bibfnamefont {C.}~\bibnamefont {Yu}}, \bibinfo
  {author} {\bibfnamefont {W.}~\bibnamefont {Zhong}}, \bibinfo {author}
  {\bibfnamefont {B.}~\bibnamefont {Estey}},\ and\ \bibinfo {author}
  {\bibfnamefont {H.}~\bibnamefont {M{\"u}ller}},\ }\bibfield  {title}
  {\bibinfo {title} {Measurement of the fine-structure constant as a test of
  the {S}tandard {M}odel},\ }\href {https://doi.org/10.1126/science.aap7706}
  {\bibfield  {journal} {\bibinfo  {journal} {Science}\ }\textbf {\bibinfo
  {volume} {360}},\ \bibinfo {pages} {191} (\bibinfo {year}
  {2018})}\BibitemShut {NoStop}%
\bibitem [{\citenamefont {Krems}\ \emph {et~al.}(2009)\citenamefont {Krems},
  \citenamefont {Stwalley},\ and\ \citenamefont
  {Friedrich}}]{2009_Krems_Cold_Molecules_Theory_Experiment_Applications}%
  \BibitemOpen
  \bibfield  {author} {\bibinfo {author} {\bibfnamefont {R.}~\bibnamefont
  {Krems}}, \bibinfo {author} {\bibfnamefont {W.~C.}\ \bibnamefont
  {Stwalley}},\ and\ \bibinfo {author} {\bibfnamefont {B.}~\bibnamefont
  {Friedrich}},\ }\href@noop {} {\emph {\bibinfo {title} {Cold molecules:
  {T}heory, experiment, applications}}}\ (\bibinfo  {publisher} {CRC Press},\
  \bibinfo {address} {Boca Raton},\ \bibinfo {year} {2009})\BibitemShut
  {NoStop}%
\bibitem [{\citenamefont {Friedrich}\ and\ \citenamefont
  {Doyle}(2009)}]{2009_Friedrich_Why_Are_Cold_Molecules_So_Hot}%
  \BibitemOpen
  \bibfield  {author} {\bibinfo {author} {\bibfnamefont {B.}~\bibnamefont
  {Friedrich}}\ and\ \bibinfo {author} {\bibfnamefont {J.~M.}\ \bibnamefont
  {Doyle}},\ }\bibfield  {title} {\bibinfo {title} {Why are cold molecules so
  hot?},\ }\href {https://doi.org/https://doi.org/10.1002/cphc.200800577}
  {\bibfield  {journal} {\bibinfo  {journal} {ChemPhysChem}\ }\textbf {\bibinfo
  {volume} {10}},\ \bibinfo {pages} {604} (\bibinfo {year} {2009})}\BibitemShut
  {NoStop}%
\bibitem [{\citenamefont
  {DeMille}(2002)}]{2002_DeMille_QC_Polar_Molecules_Theory}%
  \BibitemOpen
  \bibfield  {author} {\bibinfo {author} {\bibfnamefont {D.}~\bibnamefont
  {DeMille}},\ }\bibfield  {title} {\bibinfo {title} {Quantum computation with
  trapped polar molecules},\ }\href
  {https://doi.org/10.1103/PhysRevLett.88.067901} {\bibfield  {journal}
  {\bibinfo  {journal} {Physical Review Letters}\ }\textbf {\bibinfo {volume}
  {88}},\ \bibinfo {pages} {067901} (\bibinfo {year} {2002})}\BibitemShut
  {NoStop}%
\bibitem [{\citenamefont {Albert}\ \emph {et~al.}(2020)\citenamefont {Albert},
  \citenamefont {Covey},\ and\ \citenamefont
  {Preskill}}]{2020_Albert_Robust_Encoding_of_a_Qubit_in_a_Molecule}%
  \BibitemOpen
  \bibfield  {author} {\bibinfo {author} {\bibfnamefont {V.~V.}\ \bibnamefont
  {Albert}}, \bibinfo {author} {\bibfnamefont {J.~P.}\ \bibnamefont {Covey}},\
  and\ \bibinfo {author} {\bibfnamefont {J.}~\bibnamefont {Preskill}},\
  }\bibfield  {title} {\bibinfo {title} {Robust encoding of a qubit in a
  molecule},\ }\href {https://doi.org/10.1103/PhysRevX.10.031050} {\bibfield
  {journal} {\bibinfo  {journal} {Physical Review X}\ }\textbf {\bibinfo
  {volume} {10}},\ \bibinfo {pages} {031050} (\bibinfo {year}
  {2020})}\BibitemShut {NoStop}%
\bibitem [{\citenamefont {Bohn}\ \emph {et~al.}(2017)\citenamefont {Bohn},
  \citenamefont {Rey},\ and\ \citenamefont
  {Ye}}]{2017_Bohn_Cold_Molecules_Chemistry_Materials_Review}%
  \BibitemOpen
  \bibfield  {author} {\bibinfo {author} {\bibfnamefont {J.~L.}\ \bibnamefont
  {Bohn}}, \bibinfo {author} {\bibfnamefont {A.~M.}\ \bibnamefont {Rey}},\ and\
  \bibinfo {author} {\bibfnamefont {J.}~\bibnamefont {Ye}},\ }\bibfield
  {title} {\bibinfo {title} {Cold molecules: {P}rogress in quantum engineering
  of chemistry and quantum matter},\ }\href
  {https://doi.org/10.1126/science.aam6299} {\bibfield  {journal} {\bibinfo
  {journal} {Science}\ }\textbf {\bibinfo {volume} {357}},\ \bibinfo {pages}
  {1002} (\bibinfo {year} {2017})}\BibitemShut {NoStop}%
\bibitem [{\citenamefont {Santos}\ \emph {et~al.}(2000)\citenamefont {Santos},
  \citenamefont {Shlyapnikov}, \citenamefont {Zoller},\ and\ \citenamefont
  {Lewenstein}}]{2000_Lewenstein_BEC_Polar_Molecules_Theory}%
  \BibitemOpen
  \bibfield  {author} {\bibinfo {author} {\bibfnamefont {L.}~\bibnamefont
  {Santos}}, \bibinfo {author} {\bibfnamefont {G.~V.}\ \bibnamefont
  {Shlyapnikov}}, \bibinfo {author} {\bibfnamefont {P.}~\bibnamefont
  {Zoller}},\ and\ \bibinfo {author} {\bibfnamefont {M.}~\bibnamefont
  {Lewenstein}},\ }\bibfield  {title} {\bibinfo {title} {Bose-{E}instein
  condensation in trapped dipolar gases},\ }\href
  {https://doi.org/10.1103/PhysRevLett.85.1791} {\bibfield  {journal} {\bibinfo
   {journal} {Physical Review Letters}\ }\textbf {\bibinfo {volume} {85}},\
  \bibinfo {pages} {1791} (\bibinfo {year} {2000})}\BibitemShut {NoStop}%
\bibitem [{\citenamefont {{De Marco}}\ \emph {et~al.}(2019)\citenamefont {{De
  Marco}}, \citenamefont {Valtolina}, \citenamefont {Matsuda}, \citenamefont
  {Tobias}, \citenamefont {Covey},\ and\ \citenamefont
  {Ye}}]{2020_Ye_KRb_Degenerate_Gas}%
  \BibitemOpen
  \bibfield  {author} {\bibinfo {author} {\bibfnamefont {L.}~\bibnamefont {{De
  Marco}}}, \bibinfo {author} {\bibfnamefont {G.}~\bibnamefont {Valtolina}},
  \bibinfo {author} {\bibfnamefont {K.}~\bibnamefont {Matsuda}}, \bibinfo
  {author} {\bibfnamefont {W.~G.}\ \bibnamefont {Tobias}}, \bibinfo {author}
  {\bibfnamefont {J.~P.}\ \bibnamefont {Covey}},\ and\ \bibinfo {author}
  {\bibfnamefont {J.}~\bibnamefont {Ye}},\ }\bibfield  {title} {\bibinfo
  {title} {A degenerate {F}ermi gas of polar molecules},\ }\href
  {https://doi.org/10.1126/science.aau7230} {\bibfield  {journal} {\bibinfo
  {journal} {Science}\ }\textbf {\bibinfo {volume} {363}},\ \bibinfo {pages}
  {853} (\bibinfo {year} {2019})}\BibitemShut {NoStop}%
\bibitem [{\citenamefont {Safronova}\ \emph {et~al.}(2018)\citenamefont
  {Safronova}, \citenamefont {Budker}, \citenamefont {DeMille}, \citenamefont
  {Kimball}, \citenamefont {Derevianko},\ and\ \citenamefont
  {Clark}}]{2018_Safronova_New_Physics_Atoms_Molecules}%
  \BibitemOpen
  \bibfield  {author} {\bibinfo {author} {\bibfnamefont {M.~S.}\ \bibnamefont
  {Safronova}}, \bibinfo {author} {\bibfnamefont {D.}~\bibnamefont {Budker}},
  \bibinfo {author} {\bibfnamefont {D.}~\bibnamefont {DeMille}}, \bibinfo
  {author} {\bibfnamefont {D.~F.~J.}\ \bibnamefont {Kimball}}, \bibinfo
  {author} {\bibfnamefont {A.}~\bibnamefont {Derevianko}},\ and\ \bibinfo
  {author} {\bibfnamefont {C.~W.}\ \bibnamefont {Clark}},\ }\bibfield  {title}
  {\bibinfo {title} {Search for new physics with atoms and molecules},\ }\href
  {https://doi.org/10.1103/RevModPhys.90.025008} {\bibfield  {journal}
  {\bibinfo  {journal} {Rev. Mod. Phys.}\ }\textbf {\bibinfo {volume} {90}},\
  \bibinfo {pages} {025008} (\bibinfo {year} {2018})}\BibitemShut {NoStop}%
\bibitem [{\citenamefont {{ACME Collaboration}}(2018)}]{2018_ACME_EDM}%
  \BibitemOpen
  \bibfield  {author} {\bibinfo {author} {\bibnamefont {{ACME
  Collaboration}}},\ }\bibfield  {title} {\bibinfo {title} {Improved limit on
  the electric dipole moment of the electron},\ }\href
  {https://doi.org/10.1038/s41586-018-0599-8} {\bibfield  {journal} {\bibinfo
  {journal} {Nature}\ }\textbf {\bibinfo {volume} {562}},\ \bibinfo {pages}
  {355} (\bibinfo {year} {2018})}\BibitemShut {NoStop}%
\bibitem [{\citenamefont {Cairncross}\ and\ \citenamefont
  {Ye}(2019)}]{2019_Ye_EDM_Review}%
  \BibitemOpen
  \bibfield  {author} {\bibinfo {author} {\bibfnamefont {W.~B.}\ \bibnamefont
  {Cairncross}}\ and\ \bibinfo {author} {\bibfnamefont {J.}~\bibnamefont
  {Ye}},\ }\bibfield  {title} {\bibinfo {title} {Atoms and molecules in the
  search for time-reversal symmetry violation},\ }\href
  {https://doi.org/10.1038/s42254-019-0080-0} {\bibfield  {journal} {\bibinfo
  {journal} {Nature Reviews Physics}\ }\textbf {\bibinfo {volume} {1}},\
  \bibinfo {pages} {510} (\bibinfo {year} {2019})}\BibitemShut {NoStop}%
\bibitem [{\citenamefont
  {Balakrishnan}(2016)}]{2016_Balakrishnan_Controlled_Cold_Chemistry_Review}%
  \BibitemOpen
  \bibfield  {author} {\bibinfo {author} {\bibfnamefont {N.}~\bibnamefont
  {Balakrishnan}},\ }\bibfield  {title} {\bibinfo {title} {Perspective:
  {U}ltracold molecules and the dawn of cold controlled chemistry},\ }\href
  {https://doi.org/10.1063/1.4964096} {\bibfield  {journal} {\bibinfo
  {journal} {The Journal of Chemical Physics}\ }\textbf {\bibinfo {volume}
  {145}},\ \bibinfo {pages} {150901} (\bibinfo {year} {2016})}\BibitemShut
  {NoStop}%
\bibitem [{\citenamefont {Heazlewood}\ and\ \citenamefont
  {Softley}(2021)}]{2021_Heazlewood_Cold_Chemistry_Review}%
  \BibitemOpen
  \bibfield  {author} {\bibinfo {author} {\bibfnamefont {B.~R.}\ \bibnamefont
  {Heazlewood}}\ and\ \bibinfo {author} {\bibfnamefont {T.~P.}\ \bibnamefont
  {Softley}},\ }\bibfield  {title} {\bibinfo {title} {Towards chemistry at
  absolute zero},\ }\href {https://doi.org/10.1038/s41570-020-00239-0}
  {\bibfield  {journal} {\bibinfo  {journal} {Nature Reviews Chemistry}\
  }\textbf {\bibinfo {volume} {5}},\ \bibinfo {pages} {125} (\bibinfo {year}
  {2021})}\BibitemShut {NoStop}%
\bibitem [{\citenamefont
  {Puzzarini}(2020)}]{2020_Puzzarini_Grand_Challenges_in_Astrochemistry}%
  \BibitemOpen
  \bibfield  {author} {\bibinfo {author} {\bibfnamefont {C.}~\bibnamefont
  {Puzzarini}},\ }\bibfield  {title} {\bibinfo {title} {Grand challenges in
  astrochemistry},\ }\bibfield  {journal} {\bibinfo  {journal} {Frontiers in
  Astronomy and Space Sciences}\ }\textbf {\bibinfo {volume} {7}},\ \href
  {https://doi.org/10.3389/fspas.2020.00019} {10.3389/fspas.2020.00019}
  (\bibinfo {year} {2020})\BibitemShut {NoStop}%
\bibitem [{\citenamefont {Herschbach}(1987)}]{1987_Herschbach_1987_Nobel}%
  \BibitemOpen
  \bibfield  {author} {\bibinfo {author} {\bibfnamefont {D.~R.}\ \bibnamefont
  {Herschbach}},\ }\bibfield  {title} {\bibinfo {title} {Molecular dynamics of
  elementary chemical reactions ({N}obel lecture)},\ }\href
  {https://doi.org/10.1002/anie.198712211} {\bibfield  {journal} {\bibinfo
  {journal} {Angewandte Chemie International Edition in English}\ }\textbf
  {\bibinfo {volume} {26}},\ \bibinfo {pages} {1221} (\bibinfo {year}
  {1987})}\BibitemShut {NoStop}%
\bibitem [{\citenamefont {Lee}(1987)}]{1987_Lee_1987_Nobel}%
  \BibitemOpen
  \bibfield  {author} {\bibinfo {author} {\bibfnamefont {Y.~T.}\ \bibnamefont
  {Lee}},\ }\bibfield  {title} {\bibinfo {title} {Molecular beam studies of
  elementary chemical processes ({N}obel lecture)},\ }\href
  {https://doi.org/10.1002/anie.198709393} {\bibfield  {journal} {\bibinfo
  {journal} {Angewandte Chemie International Edition in English}\ }\textbf
  {\bibinfo {volume} {26}},\ \bibinfo {pages} {939} (\bibinfo {year}
  {1987})}\BibitemShut {NoStop}%
\bibitem [{\citenamefont {Phaneuf}\ \emph {et~al.}(1999)\citenamefont
  {Phaneuf}, \citenamefont {Havener}, \citenamefont {Dunn},\ and\ \citenamefont
  {M\"{u}ller}}]{1999_Phaneuf_Merged_beams}%
  \BibitemOpen
  \bibfield  {author} {\bibinfo {author} {\bibfnamefont {R.~A.}\ \bibnamefont
  {Phaneuf}}, \bibinfo {author} {\bibfnamefont {C.~C.}\ \bibnamefont
  {Havener}}, \bibinfo {author} {\bibfnamefont {G.~H.}\ \bibnamefont {Dunn}},\
  and\ \bibinfo {author} {\bibfnamefont {A.}~\bibnamefont {M\"{u}ller}},\
  }\bibfield  {title} {\bibinfo {title} {Merged-beams experiments in atomic and
  molecular physics},\ }\href {https://doi.org/10.1088/0034-4885/62/7/202}
  {\bibfield  {journal} {\bibinfo  {journal} {Reports on Progress in Physics}\
  }\textbf {\bibinfo {volume} {62}},\ \bibinfo {pages} {1143} (\bibinfo {year}
  {1999})}\BibitemShut {NoStop}%
\bibitem [{\citenamefont {Campbell}\ and\ \citenamefont
  {Doyle}(2008)}]{2008_Doyle_buffer_gas_loading_review}%
  \BibitemOpen
  \bibfield  {author} {\bibinfo {author} {\bibfnamefont {W.~C.}\ \bibnamefont
  {Campbell}}\ and\ \bibinfo {author} {\bibfnamefont {J.~M.}\ \bibnamefont
  {Doyle}},\ }\bibfield  {title} {\bibinfo {title} {Cooling, trap loading, and
  beam production using a cryogenic helium buffer gas},\ }\href
  {http://www.doylegroup.harvard.edu/wiki/images/8/87/BGCoolingReview27.pdf}
  {\bibfield  {journal} {\bibinfo  {journal} {Cold Molecules: {T}heory,
  Experiment, Applications}\ ,\ \bibinfo {pages} {1}} (\bibinfo {year}
  {2008})}\BibitemShut {NoStop}%
\bibitem [{\citenamefont {Spaun}\ \emph {et~al.}(2016)\citenamefont {Spaun},
  \citenamefont {Changala}, \citenamefont {Patterson}, \citenamefont {Bjork},
  \citenamefont {Heckl}, \citenamefont {Doyle},\ and\ \citenamefont
  {Ye}}]{2016_Spaun_Buffer_Gas_Molecule_Spectroscopy}%
  \BibitemOpen
  \bibfield  {author} {\bibinfo {author} {\bibfnamefont {B.}~\bibnamefont
  {Spaun}}, \bibinfo {author} {\bibfnamefont {P.~B.}\ \bibnamefont {Changala}},
  \bibinfo {author} {\bibfnamefont {D.}~\bibnamefont {Patterson}}, \bibinfo
  {author} {\bibfnamefont {B.~J.}\ \bibnamefont {Bjork}}, \bibinfo {author}
  {\bibfnamefont {O.~H.}\ \bibnamefont {Heckl}}, \bibinfo {author}
  {\bibfnamefont {J.~M.}\ \bibnamefont {Doyle}},\ and\ \bibinfo {author}
  {\bibfnamefont {J.}~\bibnamefont {Ye}},\ }\bibfield  {title} {\bibinfo
  {title} {Continuous probing of cold complex molecules with infrared frequency
  comb spectroscopy},\ }\href {https://doi.org/10.1038/nature17440} {\bibfield
  {journal} {\bibinfo  {journal} {Nature}\ }\textbf {\bibinfo {volume} {533}},\
  \bibinfo {pages} {517} (\bibinfo {year} {2016})}\BibitemShut {NoStop}%
\bibitem [{\citenamefont {Changala}\ \emph {et~al.}(2016)\citenamefont
  {Changala}, \citenamefont {Spaun}, \citenamefont {Patterson}, \citenamefont
  {Doyle},\ and\ \citenamefont
  {Ye}}]{2016_Changala_Buffer_Gas_Molecule_Spectroscopy_Details_Fast_Shutter}%
  \BibitemOpen
  \bibfield  {author} {\bibinfo {author} {\bibfnamefont {P.~B.}\ \bibnamefont
  {Changala}}, \bibinfo {author} {\bibfnamefont {B.}~\bibnamefont {Spaun}},
  \bibinfo {author} {\bibfnamefont {D.}~\bibnamefont {Patterson}}, \bibinfo
  {author} {\bibfnamefont {J.~M.}\ \bibnamefont {Doyle}},\ and\ \bibinfo
  {author} {\bibfnamefont {J.}~\bibnamefont {Ye}},\ }\bibfield  {title}
  {\bibinfo {title} {Sensitivity and resolution in frequency comb spectroscopy
  of buffer gas cooled polyatomic molecules},\ }\href
  {https://doi.org/10.1007/s00340-016-6569-7} {\bibfield  {journal} {\bibinfo
  {journal} {Applied Physics B}\ }\textbf {\bibinfo {volume} {122}},\ \bibinfo
  {pages} {292} (\bibinfo {year} {2016})}\BibitemShut {NoStop}%
\bibitem [{\citenamefont {Weinstein}\ \emph {et~al.}(1998)\citenamefont
  {Weinstein}, \citenamefont {deCarvalho}, \citenamefont {Guillet},
  \citenamefont {Friedrich},\ and\ \citenamefont
  {Doyle}}]{1998_Doyle_CaH_Magnetic_Trap}%
  \BibitemOpen
  \bibfield  {author} {\bibinfo {author} {\bibfnamefont {J.~D.}\ \bibnamefont
  {Weinstein}}, \bibinfo {author} {\bibfnamefont {R.}~\bibnamefont
  {deCarvalho}}, \bibinfo {author} {\bibfnamefont {T.}~\bibnamefont {Guillet}},
  \bibinfo {author} {\bibfnamefont {B.}~\bibnamefont {Friedrich}},\ and\
  \bibinfo {author} {\bibfnamefont {J.~M.}\ \bibnamefont {Doyle}},\ }\bibfield
  {title} {\bibinfo {title} {Magnetic trapping of calcium monohydride molecules
  at millikelvin temperatures},\ }\href {https://doi.org/10.1038/25949}
  {\bibfield  {journal} {\bibinfo  {journal} {Nature}\ }\textbf {\bibinfo
  {volume} {395}},\ \bibinfo {pages} {148} (\bibinfo {year}
  {1998})}\BibitemShut {NoStop}%
\bibitem [{\citenamefont {Campbell}\ \emph {et~al.}(2007)\citenamefont
  {Campbell}, \citenamefont {Tsikata}, \citenamefont {Lu}, \citenamefont {van
  Buuren},\ and\ \citenamefont {Doyle}}]{2007_Doyle_Magnetic_Trap_NH}%
  \BibitemOpen
  \bibfield  {author} {\bibinfo {author} {\bibfnamefont {W.~C.}\ \bibnamefont
  {Campbell}}, \bibinfo {author} {\bibfnamefont {E.}~\bibnamefont {Tsikata}},
  \bibinfo {author} {\bibfnamefont {{\relax Hsin{-}I}.}~\bibnamefont {Lu}},
  \bibinfo {author} {\bibfnamefont {L.~D.}\ \bibnamefont {van Buuren}},\ and\
  \bibinfo {author} {\bibfnamefont {J.~M.}\ \bibnamefont {Doyle}},\ }\bibfield
  {title} {\bibinfo {title} {Magnetic trapping and {Z}eeman relaxation of {NH}
  $x^3\sigma^-$)},\ }\href {https://doi.org/10.1103/PhysRevLett.98.213001}
  {\bibfield  {journal} {\bibinfo  {journal} {Physical Review Letters}\
  }\textbf {\bibinfo {volume} {98}},\ \bibinfo {pages} {213001} (\bibinfo
  {year} {2007})}\BibitemShut {NoStop}%
\bibitem [{\citenamefont {Segev}\ \emph {et~al.}(2019)\citenamefont {Segev},
  \citenamefont {Pitzer}, \citenamefont {Karpov}, \citenamefont {Akerman},
  \citenamefont {Narevicius},\ and\ \citenamefont
  {Narevicius}}]{2019_Segev_Magnetic_Trapping_O2}%
  \BibitemOpen
  \bibfield  {author} {\bibinfo {author} {\bibfnamefont {Y.}~\bibnamefont
  {Segev}}, \bibinfo {author} {\bibfnamefont {M.}~\bibnamefont {Pitzer}},
  \bibinfo {author} {\bibfnamefont {M.}~\bibnamefont {Karpov}}, \bibinfo
  {author} {\bibfnamefont {N.}~\bibnamefont {Akerman}}, \bibinfo {author}
  {\bibfnamefont {J.}~\bibnamefont {Narevicius}},\ and\ \bibinfo {author}
  {\bibfnamefont {E.}~\bibnamefont {Narevicius}},\ }\bibfield  {title}
  {\bibinfo {title} {Collisions between cold molecules in a superconducting
  magnetic trap},\ }\href {https://doi.org/10.1038/s41586-019-1446-2}
  {\bibfield  {journal} {\bibinfo  {journal} {Nature}\ }\textbf {\bibinfo
  {volume} {572}},\ \bibinfo {pages} {189} (\bibinfo {year}
  {2019})}\BibitemShut {NoStop}%
\bibitem [{\citenamefont {Wild}\ \emph {et~al.}(2021)\citenamefont {Wild},
  \citenamefont {N{\"o}tzold}, \citenamefont {Lochmann},\ and\ \citenamefont
  {Wester}}]{2021_Wester_Cl-H2_Complex_Studies}%
  \BibitemOpen
  \bibfield  {author} {\bibinfo {author} {\bibfnamefont {R.}~\bibnamefont
  {Wild}}, \bibinfo {author} {\bibfnamefont {M.}~\bibnamefont {N{\"o}tzold}},
  \bibinfo {author} {\bibfnamefont {C.}~\bibnamefont {Lochmann}},\ and\
  \bibinfo {author} {\bibfnamefont {R.}~\bibnamefont {Wester}},\ }\bibfield
  {title} {\bibinfo {title} {Complex formation in three-body reactions of
  {C}l$^-$ with {H}$_2$},\ }\href {https://doi.org/10.1021/acs.jpca.1c05458}
  {\bibfield  {journal} {\bibinfo  {journal} {The Journal of Physical Chemistry
  A}\ }\textbf {\bibinfo {volume} {125}},\ \bibinfo {pages} {8581} (\bibinfo
  {year} {2021})}\BibitemShut {NoStop}%
\bibitem [{\citenamefont {Gianturco}\ \emph {et~al.}(2021)\citenamefont
  {Gianturco}, \citenamefont {Giri}, \citenamefont {Gonzalez-Sanchez},
  \citenamefont {Yurtsever}, \citenamefont {Sathyamurthy},\ and\ \citenamefont
  {Wester}}]{2021_Wester_HeH+}%
  \BibitemOpen
  \bibfield  {author} {\bibinfo {author} {\bibfnamefont {F.~A.}\ \bibnamefont
  {Gianturco}}, \bibinfo {author} {\bibfnamefont {K.}~\bibnamefont {Giri}},
  \bibinfo {author} {\bibfnamefont {L.}~\bibnamefont {Gonzalez-Sanchez}},
  \bibinfo {author} {\bibfnamefont {E.}~\bibnamefont {Yurtsever}}, \bibinfo
  {author} {\bibfnamefont {N.}~\bibnamefont {Sathyamurthy}},\ and\ \bibinfo
  {author} {\bibfnamefont {R.}~\bibnamefont {Wester}},\ }\bibfield  {title}
  {\bibinfo {title} {Efficiency of rovibrational cooling of {H}e{H}$^+$ by
  collisions with {H}e: {C}ross sections and rate coefficients from quantum
  dynamics},\ }\href {https://doi.org/10.1063/5.0062147} {\bibfield  {journal}
  {\bibinfo  {journal} {The Journal of Chemical Physics}\ }\textbf {\bibinfo
  {volume} {155}},\ \bibinfo {pages} {154301} (\bibinfo {year}
  {2021})}\BibitemShut {NoStop}%
\bibitem [{\citenamefont {Bethlem}\ \emph {et~al.}(2000)\citenamefont
  {Bethlem}, \citenamefont {Berden}, \citenamefont {Crompvoets}, \citenamefont
  {Jongma}, \citenamefont {van Roij},\ and\ \citenamefont
  {Meijer}}]{2000_Bethlem_Meijer_Electrostatic_Trapping_Ammonia}%
  \BibitemOpen
  \bibfield  {author} {\bibinfo {author} {\bibfnamefont {H.~L.}\ \bibnamefont
  {Bethlem}}, \bibinfo {author} {\bibfnamefont {G.}~\bibnamefont {Berden}},
  \bibinfo {author} {\bibfnamefont {F.~M.~H.}\ \bibnamefont {Crompvoets}},
  \bibinfo {author} {\bibfnamefont {R.~T.}\ \bibnamefont {Jongma}}, \bibinfo
  {author} {\bibfnamefont {A.~J.~A.}\ \bibnamefont {van Roij}},\ and\ \bibinfo
  {author} {\bibfnamefont {G.}~\bibnamefont {Meijer}},\ }\bibfield  {title}
  {\bibinfo {title} {Electrostatic trapping of ammonia molecules},\ }\href
  {https://doi.org/10.1038/35020030} {\bibfield  {journal} {\bibinfo  {journal}
  {Nature}\ }\textbf {\bibinfo {volume} {406}},\ \bibinfo {pages} {491}
  (\bibinfo {year} {2000})}\BibitemShut {NoStop}%
\bibitem [{\citenamefont {van~de Meerakker}\ \emph {et~al.}(2005)\citenamefont
  {van~de Meerakker}, \citenamefont {Smeets}, \citenamefont {Vanhaecke},
  \citenamefont {Jongma},\ and\ \citenamefont
  {Meijer}}]{2005_van_de_Meerakker_Meijer_Electrostatic_Trapping_OH}%
  \BibitemOpen
  \bibfield  {author} {\bibinfo {author} {\bibfnamefont {S.~Y.~T.}\
  \bibnamefont {van~de Meerakker}}, \bibinfo {author} {\bibfnamefont
  {P.~H.~M.}\ \bibnamefont {Smeets}}, \bibinfo {author} {\bibfnamefont
  {N.}~\bibnamefont {Vanhaecke}}, \bibinfo {author} {\bibfnamefont {R.~T.}\
  \bibnamefont {Jongma}},\ and\ \bibinfo {author} {\bibfnamefont
  {G.}~\bibnamefont {Meijer}},\ }\bibfield  {title} {\bibinfo {title}
  {Deceleration and electrostatic trapping of {OH} radicals},\ }\href
  {https://doi.org/10.1103/PhysRevLett.94.023004} {\bibfield  {journal}
  {\bibinfo  {journal} {Physical Review Letters}\ }\textbf {\bibinfo {volume}
  {94}},\ \bibinfo {pages} {023004} (\bibinfo {year} {2005})}\BibitemShut
  {NoStop}%
\bibitem [{\citenamefont {Prehn}\ \emph {et~al.}(2016)\citenamefont {Prehn},
  \citenamefont {Ibr\"ugger}, \citenamefont {Gl\"ockner}, \citenamefont
  {Rempe},\ and\ \citenamefont
  {Zeppenfeld}}]{2016_Prehn_Rempe_Electrostatic_Trapping_Opeoelectrical_Cooling_CH2O}%
  \BibitemOpen
  \bibfield  {author} {\bibinfo {author} {\bibfnamefont {A.}~\bibnamefont
  {Prehn}}, \bibinfo {author} {\bibfnamefont {M.}~\bibnamefont {Ibr\"ugger}},
  \bibinfo {author} {\bibfnamefont {R.}~\bibnamefont {Gl\"ockner}}, \bibinfo
  {author} {\bibfnamefont {G.}~\bibnamefont {Rempe}},\ and\ \bibinfo {author}
  {\bibfnamefont {M.}~\bibnamefont {Zeppenfeld}},\ }\bibfield  {title}
  {\bibinfo {title} {Optoelectrical cooling of polar molecules to
  submillikelvin temperatures},\ }\href
  {https://doi.org/10.1103/PhysRevLett.116.063005} {\bibfield  {journal}
  {\bibinfo  {journal} {Physical Review Letters}\ }\textbf {\bibinfo {volume}
  {116}},\ \bibinfo {pages} {063005} (\bibinfo {year} {2016})}\BibitemShut
  {NoStop}%
\bibitem [{\citenamefont {DeMille}\ \emph {et~al.}(2004)\citenamefont
  {DeMille}, \citenamefont {Glenn},\ and\ \citenamefont
  {Petricka}}]{2004_DeMille_Microwave_Trap_For_Cold_Polar_Molecules}%
  \BibitemOpen
  \bibfield  {author} {\bibinfo {author} {\bibfnamefont {D.}~\bibnamefont
  {DeMille}}, \bibinfo {author} {\bibfnamefont {D.~R.}\ \bibnamefont {Glenn}},\
  and\ \bibinfo {author} {\bibfnamefont {J.}~\bibnamefont {Petricka}},\
  }\bibfield  {title} {\bibinfo {title} {Microwave traps for cold polar
  molecules},\ }\href {https://doi.org/10.1140/epjd/e2004-00163-6} {\bibfield
  {journal} {\bibinfo  {journal} {The European Physical Journal D - Atomic,
  Molecular, Optical and Plasma Physics}\ }\textbf {\bibinfo {volume} {31}},\
  \bibinfo {pages} {375} (\bibinfo {year} {2004})}\BibitemShut {NoStop}%
\bibitem [{\citenamefont {Wright}\ \emph {et~al.}(2019)\citenamefont {Wright},
  \citenamefont {Wall},\ and\ \citenamefont
  {Tarbutt}}]{2019_Wright_Tarbutt_Microwave_Trap_for_Atoms_and_Molecules}%
  \BibitemOpen
  \bibfield  {author} {\bibinfo {author} {\bibfnamefont {S.~C.}\ \bibnamefont
  {Wright}}, \bibinfo {author} {\bibfnamefont {T.~E.}\ \bibnamefont {Wall}},\
  and\ \bibinfo {author} {\bibfnamefont {M.~R.}\ \bibnamefont {Tarbutt}},\
  }\bibfield  {title} {\bibinfo {title} {Microwave trap for atoms and
  molecules},\ }\href {https://doi.org/10.1103/PhysRevResearch.1.033035}
  {\bibfield  {journal} {\bibinfo  {journal} {Physical Review Research}\
  }\textbf {\bibinfo {volume} {1}},\ \bibinfo {pages} {033035} (\bibinfo {year}
  {2019})}\BibitemShut {NoStop}%
\bibitem [{\citenamefont {Moses}\ \emph {et~al.}(2017)\citenamefont {Moses},
  \citenamefont {Covey}, \citenamefont {Miecnikowski}, \citenamefont {Jin},\
  and\ \citenamefont {Ye}}]{2016_Moses_Bialkali_Review}%
  \BibitemOpen
  \bibfield  {author} {\bibinfo {author} {\bibfnamefont {S.}~\bibnamefont
  {Moses}}, \bibinfo {author} {\bibfnamefont {J.}~\bibnamefont {Covey}},
  \bibinfo {author} {\bibfnamefont {M.}~\bibnamefont {Miecnikowski}}, \bibinfo
  {author} {\bibfnamefont {D.}~\bibnamefont {Jin}},\ and\ \bibinfo {author}
  {\bibfnamefont {J.}~\bibnamefont {Ye}},\ }\bibfield  {title} {\bibinfo
  {title} {New frontiers for quantum gases of polar molecules},\ }\href
  {https://doi.org/10.1038/nphys3985} {\bibfield  {journal} {\bibinfo
  {journal} {Nature Physics}\ }\textbf {\bibinfo {volume} {13}},\ \bibinfo
  {pages} {13} (\bibinfo {year} {2017})}\BibitemShut {NoStop}%
\bibitem [{\citenamefont {Liu}\ \emph {et~al.}(2020)\citenamefont {Liu},
  \citenamefont {Hu}, \citenamefont {Nichols}, \citenamefont {Grimes},
  \citenamefont {Karman}, \citenamefont {Guo},\ and\ \citenamefont
  {Ni}}]{2020_Liu_Kang-kuen-ni_light_assisted_pathways}%
  \BibitemOpen
  \bibfield  {author} {\bibinfo {author} {\bibfnamefont {Y.}~\bibnamefont
  {Liu}}, \bibinfo {author} {\bibfnamefont {M.-G.}\ \bibnamefont {Hu}},
  \bibinfo {author} {\bibfnamefont {M.~A.}\ \bibnamefont {Nichols}}, \bibinfo
  {author} {\bibfnamefont {D.~D.}\ \bibnamefont {Grimes}}, \bibinfo {author}
  {\bibfnamefont {T.}~\bibnamefont {Karman}}, \bibinfo {author} {\bibfnamefont
  {H.}~\bibnamefont {Guo}},\ and\ \bibinfo {author} {\bibfnamefont {K.-K.}\
  \bibnamefont {Ni}},\ }\bibfield  {title} {\bibinfo {title} {Photo-excitation
  of long-lived transient intermediates in ultracold reactions},\ }\href
  {https://doi.org/10.1038/s41567-020-0968-8} {\bibfield  {journal} {\bibinfo
  {journal} {Nature Physics}\ }\textbf {\bibinfo {volume} {16}},\ \bibinfo
  {pages} {1132} (\bibinfo {year} {2020})}\BibitemShut {NoStop}%
\bibitem [{\citenamefont {Gregory}\ \emph {et~al.}(2021)\citenamefont
  {Gregory}, \citenamefont {Blackmore}, \citenamefont {Bromley}, \citenamefont
  {Hutson},\ and\ \citenamefont {Cornish}}]{2021_Cornish_RbCs_Qubit}%
  \BibitemOpen
  \bibfield  {author} {\bibinfo {author} {\bibfnamefont {P.~D.}\ \bibnamefont
  {Gregory}}, \bibinfo {author} {\bibfnamefont {J.~A.}\ \bibnamefont
  {Blackmore}}, \bibinfo {author} {\bibfnamefont {S.~L.}\ \bibnamefont
  {Bromley}}, \bibinfo {author} {\bibfnamefont {J.~M.}\ \bibnamefont
  {Hutson}},\ and\ \bibinfo {author} {\bibfnamefont {S.~L.}\ \bibnamefont
  {Cornish}},\ }\bibfield  {title} {\bibinfo {title} {Robust storage qubits in
  ultracold polar molecules},\ }\href
  {https://doi.org/10.1038/s41567-021-01328-7} {\bibfield  {journal} {\bibinfo
  {journal} {Nature Physics}\ }\textbf {\bibinfo {volume} {17}},\ \bibinfo
  {pages} {1149} (\bibinfo {year} {2021})}\BibitemShut {NoStop}%
\bibitem [{\citenamefont {Mitra}\ \emph {et~al.}(2020)\citenamefont {Mitra},
  \citenamefont {Vilas}, \citenamefont {Hallas}, \citenamefont {Anderegg},
  \citenamefont {Augenbraun}, \citenamefont {Baum}, \citenamefont {Miller},
  \citenamefont {Raval},\ and\ \citenamefont
  {Doyle}}]{2020_Mitra_Laser_Cooling_CaCOH3}%
  \BibitemOpen
  \bibfield  {author} {\bibinfo {author} {\bibfnamefont {D.}~\bibnamefont
  {Mitra}}, \bibinfo {author} {\bibfnamefont {N.~B.}\ \bibnamefont {Vilas}},
  \bibinfo {author} {\bibfnamefont {C.}~\bibnamefont {Hallas}}, \bibinfo
  {author} {\bibfnamefont {L.}~\bibnamefont {Anderegg}}, \bibinfo {author}
  {\bibfnamefont {B.~L.}\ \bibnamefont {Augenbraun}}, \bibinfo {author}
  {\bibfnamefont {L.}~\bibnamefont {Baum}}, \bibinfo {author} {\bibfnamefont
  {C.}~\bibnamefont {Miller}}, \bibinfo {author} {\bibfnamefont
  {S.}~\bibnamefont {Raval}},\ and\ \bibinfo {author} {\bibfnamefont {J.~M.}\
  \bibnamefont {Doyle}},\ }\bibfield  {title} {\bibinfo {title} {Direct laser
  cooling of a symmetric top molecule},\ }\href
  {https://doi.org/10.1126/science.abc5357} {\bibfield  {journal} {\bibinfo
  {journal} {Science}\ }\textbf {\bibinfo {volume} {369}},\ \bibinfo {pages}
  {1366} (\bibinfo {year} {2020})}\BibitemShut {NoStop}%
\bibitem [{\citenamefont {Shuman}\ \emph {et~al.}(2010)\citenamefont {Shuman},
  \citenamefont {Barry},\ and\ \citenamefont
  {DeMille}}]{2010_Shuman_DeMille_Laser_Cooling_SrF}%
  \BibitemOpen
  \bibfield  {author} {\bibinfo {author} {\bibfnamefont {E.~S.}\ \bibnamefont
  {Shuman}}, \bibinfo {author} {\bibfnamefont {J.~F.}\ \bibnamefont {Barry}},\
  and\ \bibinfo {author} {\bibfnamefont {D.}~\bibnamefont {DeMille}},\
  }\bibfield  {title} {\bibinfo {title} {Laser cooling of a diatomic
  molecule},\ }\href {https://doi.org/10.1038/nature09443} {\bibfield
  {journal} {\bibinfo  {journal} {Nature}\ }\textbf {\bibinfo {volume} {467}},\
  \bibinfo {pages} {820} (\bibinfo {year} {2010})}\BibitemShut {NoStop}%
\bibitem [{\citenamefont {Anderegg}\ \emph {et~al.}(2018)\citenamefont
  {Anderegg}, \citenamefont {Augenbraun}, \citenamefont {Bao}, \citenamefont
  {Burchesky}, \citenamefont {Cheuk}, \citenamefont {Ketterle},\ and\
  \citenamefont {Doyle}}]{2018_Anderegg_Laser_Cooling_Trapping_CaF}%
  \BibitemOpen
  \bibfield  {author} {\bibinfo {author} {\bibfnamefont {L.}~\bibnamefont
  {Anderegg}}, \bibinfo {author} {\bibfnamefont {B.~L.}\ \bibnamefont
  {Augenbraun}}, \bibinfo {author} {\bibfnamefont {Y.}~\bibnamefont {Bao}},
  \bibinfo {author} {\bibfnamefont {S.}~\bibnamefont {Burchesky}}, \bibinfo
  {author} {\bibfnamefont {L.~W.}\ \bibnamefont {Cheuk}}, \bibinfo {author}
  {\bibfnamefont {W.}~\bibnamefont {Ketterle}},\ and\ \bibinfo {author}
  {\bibfnamefont {J.~M.}\ \bibnamefont {Doyle}},\ }\bibfield  {title} {\bibinfo
  {title} {Laser cooling of optically trapped molecules},\ }\href
  {https://doi.org/10.1038/s41567-018-0191-z} {\bibfield  {journal} {\bibinfo
  {journal} {Nature Physics}\ }\textbf {\bibinfo {volume} {14}},\ \bibinfo
  {pages} {890} (\bibinfo {year} {2018})}\BibitemShut {NoStop}%
\bibitem [{\citenamefont {Baum}\ \emph {et~al.}(2020)\citenamefont {Baum},
  \citenamefont {Vilas}, \citenamefont {Hallas}, \citenamefont {Augenbraun},
  \citenamefont {Raval}, \citenamefont {Mitra},\ and\ \citenamefont
  {Doyle}}]{2020_Doyle_CaOH_Laser_Cooling}%
  \BibitemOpen
  \bibfield  {author} {\bibinfo {author} {\bibfnamefont {L.}~\bibnamefont
  {Baum}}, \bibinfo {author} {\bibfnamefont {N.~B.}\ \bibnamefont {Vilas}},
  \bibinfo {author} {\bibfnamefont {C.}~\bibnamefont {Hallas}}, \bibinfo
  {author} {\bibfnamefont {B.~L.}\ \bibnamefont {Augenbraun}}, \bibinfo
  {author} {\bibfnamefont {S.}~\bibnamefont {Raval}}, \bibinfo {author}
  {\bibfnamefont {D.}~\bibnamefont {Mitra}},\ and\ \bibinfo {author}
  {\bibfnamefont {J.~M.}\ \bibnamefont {Doyle}},\ }\bibfield  {title} {\bibinfo
  {title} {1{D} magneto-optical trap of polyatomic molecules},\ }\href
  {https://doi.org/10.1103/PhysRevLett.124.133201} {\bibfield  {journal}
  {\bibinfo  {journal} {Physical Review Letters}\ }\textbf {\bibinfo {volume}
  {124}},\ \bibinfo {pages} {133201} (\bibinfo {year} {2020})}\BibitemShut
  {NoStop}%
\bibitem [{\citenamefont {Ding}\ \emph {et~al.}(2020)\citenamefont {Ding},
  \citenamefont {Wu}, \citenamefont {Finneran}, \citenamefont {Burau},\ and\
  \citenamefont {Ye}}]{2020_Ye_Sub-Doppler_Laser_Cooling_YO}%
  \BibitemOpen
  \bibfield  {author} {\bibinfo {author} {\bibfnamefont {S.}~\bibnamefont
  {Ding}}, \bibinfo {author} {\bibfnamefont {Y.}~\bibnamefont {Wu}}, \bibinfo
  {author} {\bibfnamefont {I.~A.}\ \bibnamefont {Finneran}}, \bibinfo {author}
  {\bibfnamefont {J.~J.}\ \bibnamefont {Burau}},\ and\ \bibinfo {author}
  {\bibfnamefont {J.}~\bibnamefont {Ye}},\ }\bibfield  {title} {\bibinfo
  {title} {Sub-{D}oppler cooling and compressed trapping of {YO} molecules at
  \si{\micro\kelvin} temperatures},\ }\href
  {https://doi.org/10.1103/PhysRevX.10.021049} {\bibfield  {journal} {\bibinfo
  {journal} {Physical Review X}\ }\textbf {\bibinfo {volume} {10}},\ \bibinfo
  {pages} {021049} (\bibinfo {year} {2020})}\BibitemShut {NoStop}%
\bibitem [{\citenamefont {Lim}\ \emph {et~al.}(2018)\citenamefont {Lim},
  \citenamefont {Almond}, \citenamefont {Trigatzis}, \citenamefont {Devlin},
  \citenamefont {Fitch}, \citenamefont {Sauer}, \citenamefont {Tarbutt},\ and\
  \citenamefont {Hinds}}]{2018_Lim_Laser_Cool_YbF}%
  \BibitemOpen
  \bibfield  {author} {\bibinfo {author} {\bibfnamefont {J.}~\bibnamefont
  {Lim}}, \bibinfo {author} {\bibfnamefont {J.~R.}\ \bibnamefont {Almond}},
  \bibinfo {author} {\bibfnamefont {M.~A.}\ \bibnamefont {Trigatzis}}, \bibinfo
  {author} {\bibfnamefont {J.~A.}\ \bibnamefont {Devlin}}, \bibinfo {author}
  {\bibfnamefont {N.~J.}\ \bibnamefont {Fitch}}, \bibinfo {author}
  {\bibfnamefont {B.~E.}\ \bibnamefont {Sauer}}, \bibinfo {author}
  {\bibfnamefont {M.~R.}\ \bibnamefont {Tarbutt}},\ and\ \bibinfo {author}
  {\bibfnamefont {E.~A.}\ \bibnamefont {Hinds}},\ }\bibfield  {title} {\bibinfo
  {title} {Laser cooled {Y}b{F} molecules for measuring the electron's electric
  dipole moment},\ }\href {https://doi.org/10.1103/PhysRevLett.120.123201}
  {\bibfield  {journal} {\bibinfo  {journal} {Physical Review Letters}\
  }\textbf {\bibinfo {volume} {120}},\ \bibinfo {pages} {123201} (\bibinfo
  {year} {2018})}\BibitemShut {NoStop}%
\bibitem [{\citenamefont {Kozyryev}\ \emph {et~al.}(2017)\citenamefont
  {Kozyryev}, \citenamefont {Baum}, \citenamefont {Matsuda}, \citenamefont
  {Augenbraun}, \citenamefont {Anderegg}, \citenamefont {Sedlack},\ and\
  \citenamefont {Doyle}}]{2017_Doyle_Laser_cooling_SrOH}%
  \BibitemOpen
  \bibfield  {author} {\bibinfo {author} {\bibfnamefont {I.}~\bibnamefont
  {Kozyryev}}, \bibinfo {author} {\bibfnamefont {L.}~\bibnamefont {Baum}},
  \bibinfo {author} {\bibfnamefont {K.}~\bibnamefont {Matsuda}}, \bibinfo
  {author} {\bibfnamefont {B.~L.}\ \bibnamefont {Augenbraun}}, \bibinfo
  {author} {\bibfnamefont {L.}~\bibnamefont {Anderegg}}, \bibinfo {author}
  {\bibfnamefont {A.~P.}\ \bibnamefont {Sedlack}},\ and\ \bibinfo {author}
  {\bibfnamefont {J.~M.}\ \bibnamefont {Doyle}},\ }\bibfield  {title} {\bibinfo
  {title} {Sisyphus laser cooling of a polyatomic molecule},\ }\href
  {https://doi.org/10.1103/PhysRevLett.118.173201} {\bibfield  {journal}
  {\bibinfo  {journal} {Physical Review Letters}\ }\textbf {\bibinfo {volume}
  {118}},\ \bibinfo {pages} {173201} (\bibinfo {year} {2017})}\BibitemShut
  {NoStop}%
\bibitem [{\citenamefont {Augenbraun}\ \emph {et~al.}(2020)\citenamefont
  {Augenbraun}, \citenamefont {Lasner}, \citenamefont {Frenett}, \citenamefont
  {Sawaoka}, \citenamefont {Miller}, \citenamefont {Steimle},\ and\
  \citenamefont {Doyle}}]{2020_Augenbraun_Laser_Cooling_YbOH}%
  \BibitemOpen
  \bibfield  {author} {\bibinfo {author} {\bibfnamefont {B.~L.}\ \bibnamefont
  {Augenbraun}}, \bibinfo {author} {\bibfnamefont {Z.~D.}\ \bibnamefont
  {Lasner}}, \bibinfo {author} {\bibfnamefont {A.}~\bibnamefont {Frenett}},
  \bibinfo {author} {\bibfnamefont {H.}~\bibnamefont {Sawaoka}}, \bibinfo
  {author} {\bibfnamefont {C.}~\bibnamefont {Miller}}, \bibinfo {author}
  {\bibfnamefont {T.~C.}\ \bibnamefont {Steimle}},\ and\ \bibinfo {author}
  {\bibfnamefont {J.~M.}\ \bibnamefont {Doyle}},\ }\bibfield  {title} {\bibinfo
  {title} {Laser-cooled polyatomic molecules for improved electron electric
  dipole moment searches},\ }\href {https://doi.org/10.1088/1367-2630/ab687b}
  {\bibfield  {journal} {\bibinfo  {journal} {New Journal of Physics}\ }\textbf
  {\bibinfo {volume} {22}},\ \bibinfo {pages} {022003} (\bibinfo {year}
  {2020})}\BibitemShut {NoStop}%
\bibitem [{\citenamefont {McNally}\ \emph {et~al.}(2020)\citenamefont
  {McNally}, \citenamefont {Kozyryev}, \citenamefont {Vazquez-Carson},
  \citenamefont {Wenz}, \citenamefont {Wang},\ and\ \citenamefont
  {Zelevinsky}}]{2020_Zelevinsky_Laser_Cooling_BaH}%
  \BibitemOpen
  \bibfield  {author} {\bibinfo {author} {\bibfnamefont {R.~L.}\ \bibnamefont
  {McNally}}, \bibinfo {author} {\bibfnamefont {I.}~\bibnamefont {Kozyryev}},
  \bibinfo {author} {\bibfnamefont {S.}~\bibnamefont {Vazquez-Carson}},
  \bibinfo {author} {\bibfnamefont {K.}~\bibnamefont {Wenz}}, \bibinfo {author}
  {\bibfnamefont {T.}~\bibnamefont {Wang}},\ and\ \bibinfo {author}
  {\bibfnamefont {T.}~\bibnamefont {Zelevinsky}},\ }\bibfield  {title}
  {\bibinfo {title} {Optical cycling, radiative deflection and laser cooling of
  barium monohydride ($^{138}${B}a$^1${H})},\ }\href
  {https://doi.org/10.1088/1367-2630/aba3e9} {\bibfield  {journal} {\bibinfo
  {journal} {New Journal of Physics}\ }\textbf {\bibinfo {volume} {22}},\
  \bibinfo {pages} {083047} (\bibinfo {year} {2020})}\BibitemShut {NoStop}%
\bibitem [{\citenamefont {Fitch}\ and\ \citenamefont
  {Tarbutt}(2021)}]{2021_Fitch_Tarbutt_Laser_Cooled_Molecules_Review}%
  \BibitemOpen
  \bibfield  {author} {\bibinfo {author} {\bibfnamefont {N.}~\bibnamefont
  {Fitch}}\ and\ \bibinfo {author} {\bibfnamefont {M.}~\bibnamefont
  {Tarbutt}},\ }\bibfield  {title} {\bibinfo {title} {Laser-cooled molecules},\
  }in\ \href {https://doi.org/https://doi.org/10.1016/bs.aamop.2021.04.003}
  {\emph {\bibinfo {booktitle} {Advances In Atomic, Molecular, and Optical
  Physics}}},\ Vol.~\bibinfo {volume} {70}\ (\bibinfo  {publisher} {Academic
  Press},\ \bibinfo {year} {2021})\ pp.\ \bibinfo {pages}
  {157--262}\BibitemShut {NoStop}%
\bibitem [{\citenamefont {Friedrich}\ and\ \citenamefont
  {Herschbach}(1995{\natexlab{a}})}]{1995_Friedrich_Alignment_Trapping_Spheroidal_Wave_Eqn_Theory}%
  \BibitemOpen
  \bibfield  {author} {\bibinfo {author} {\bibfnamefont {B.}~\bibnamefont
  {Friedrich}}\ and\ \bibinfo {author} {\bibfnamefont {D.}~\bibnamefont
  {Herschbach}},\ }\bibfield  {title} {\bibinfo {title} {Alignment and trapping
  of molecules in intense laser fields},\ }\href
  {https://doi.org/10.1103/PhysRevLett.74.4623} {\bibfield  {journal} {\bibinfo
   {journal} {Physical Review Letters}\ }\textbf {\bibinfo {volume} {74}},\
  \bibinfo {pages} {4623} (\bibinfo {year} {1995}{\natexlab{a}})}\BibitemShut
  {NoStop}%
\bibitem [{\citenamefont {Turnbaugh}\ \emph {et~al.}(2021)\citenamefont
  {Turnbaugh}, \citenamefont {Axelrod}, \citenamefont {Campbell}, \citenamefont
  {Dioquino}, \citenamefont {Petrov}, \citenamefont {Remis}, \citenamefont
  {Schwartz}, \citenamefont {Yu}, \citenamefont {Cheng}, \citenamefont
  {Glaeser},\ and\ \citenamefont {Mueller}}]{2020_Turnbaugh_microscope_paper}%
  \BibitemOpen
  \bibfield  {author} {\bibinfo {author} {\bibfnamefont {C.}~\bibnamefont
  {Turnbaugh}}, \bibinfo {author} {\bibfnamefont {J.~J.}\ \bibnamefont
  {Axelrod}}, \bibinfo {author} {\bibfnamefont {S.~L.}\ \bibnamefont
  {Campbell}}, \bibinfo {author} {\bibfnamefont {J.~Y.}\ \bibnamefont
  {Dioquino}}, \bibinfo {author} {\bibfnamefont {P.~N.}\ \bibnamefont
  {Petrov}}, \bibinfo {author} {\bibfnamefont {J.}~\bibnamefont {Remis}},
  \bibinfo {author} {\bibfnamefont {O.}~\bibnamefont {Schwartz}}, \bibinfo
  {author} {\bibfnamefont {Z.}~\bibnamefont {Yu}}, \bibinfo {author}
  {\bibfnamefont {Y.}~\bibnamefont {Cheng}}, \bibinfo {author} {\bibfnamefont
  {R.~M.}\ \bibnamefont {Glaeser}},\ and\ \bibinfo {author} {\bibfnamefont
  {H.}~\bibnamefont {Mueller}},\ }\bibfield  {title} {\bibinfo {title}
  {High-power near-concentric {F}abry-{P}\'{e}rot cavity for phase contrast
  electron microscopy},\ }\href {https://doi.org/10.1063/5.0045496} {\bibfield
  {journal} {\bibinfo  {journal} {Review of Scientific Instruments}\ }\textbf
  {\bibinfo {volume} {92}},\ \bibinfo {pages} {053005} (\bibinfo {year}
  {2021})}\BibitemShut {NoStop}%
\bibitem [{\citenamefont {Schwartz}\ \emph {et~al.}(2019)\citenamefont
  {Schwartz}, \citenamefont {Axelrod}, \citenamefont {Campbell}, \citenamefont
  {Turnbaugh}, \citenamefont {Glaeser},\ and\ \citenamefont
  {M{\"u}ller}}]{2019_Schwartz_LPP}%
  \BibitemOpen
  \bibfield  {author} {\bibinfo {author} {\bibfnamefont {O.}~\bibnamefont
  {Schwartz}}, \bibinfo {author} {\bibfnamefont {J.~J.}\ \bibnamefont
  {Axelrod}}, \bibinfo {author} {\bibfnamefont {S.~L.}\ \bibnamefont
  {Campbell}}, \bibinfo {author} {\bibfnamefont {C.}~\bibnamefont {Turnbaugh}},
  \bibinfo {author} {\bibfnamefont {R.~M.}\ \bibnamefont {Glaeser}},\ and\
  \bibinfo {author} {\bibfnamefont {H.}~\bibnamefont {M{\"u}ller}},\ }\bibfield
   {title} {\bibinfo {title} {Laser phase plate for transmission electron
  microscopy},\ }\href {https://doi.org/10.1038/s41592-019-0552-2} {\bibfield
  {journal} {\bibinfo  {journal} {Nature Methods}\ }\textbf {\bibinfo {volume}
  {16}},\ \bibinfo {pages} {1016} (\bibinfo {year} {2019})}\BibitemShut
  {NoStop}%
\bibitem [{\citenamefont {Schwartz}\ \emph {et~al.}(2017)\citenamefont
  {Schwartz}, \citenamefont {Axelrod}, \citenamefont {Tuthill}, \citenamefont
  {Haslinger}, \citenamefont {Ophus}, \citenamefont {Glaeser},\ and\
  \citenamefont {M\"{u}ller}}]{2017_Schwartz_LPP}%
  \BibitemOpen
  \bibfield  {author} {\bibinfo {author} {\bibfnamefont {O.}~\bibnamefont
  {Schwartz}}, \bibinfo {author} {\bibfnamefont {J.}~\bibnamefont {Axelrod}},
  \bibinfo {author} {\bibfnamefont {D.~R.}\ \bibnamefont {Tuthill}}, \bibinfo
  {author} {\bibfnamefont {P.}~\bibnamefont {Haslinger}}, \bibinfo {author}
  {\bibfnamefont {C.}~\bibnamefont {Ophus}}, \bibinfo {author} {\bibfnamefont
  {R.}~\bibnamefont {Glaeser}},\ and\ \bibinfo {author} {\bibfnamefont
  {H.}~\bibnamefont {M\"{u}ller}},\ }\bibfield  {title} {\bibinfo {title}
  {Near-concentric {F}abry-{P}\'{e}rot cavity for continuous-wave laser control
  of electron waves},\ }\href {https://doi.org/10.1364/OE.25.014453} {\bibfield
   {journal} {\bibinfo  {journal} {Optics Express}\ }\textbf {\bibinfo {volume}
  {25}},\ \bibinfo {pages} {14453} (\bibinfo {year} {2017})}\BibitemShut
  {NoStop}%
\bibitem [{CCC()}]{CCCBDB}%
  \BibitemOpen
  \href@noop {} {}\bibinfo {note} {NIST Computational Chemistry Comparison and
  Benchmark Database, NIST Standard Reference Database Number 101. Release 22,
  May 2022, Editor: Russell D. Johnson III}\BibitemShut {NoStop}%
\bibitem [{NIS()}]{NIST_Webbook}%
  \BibitemOpen
  \href@noop {} {}\bibinfo {note} {NIST Chemistry WebBook: NIST Standard
  Reference Database Number 69 (2021)}\BibitemShut {NoStop}%
\bibitem [{HIT()}]{HITRAN}%
  \BibitemOpen
  \href@noop {} {}\bibinfo {note} {The HIgh-resolution TRANsmission (HITRAN)
  database (2021). Atomic and Molecular Physics Division, Harvard-Smithsonian
  Center for Astrophysics.}\BibitemShut {Stop}%
\bibitem [{\citenamefont {Keller-Rudek}\ \emph {et~al.}(2013)\citenamefont
  {Keller-Rudek}, \citenamefont {Moortgat}, \citenamefont {Sander},\ and\
  \citenamefont {S\"orensen}}]{MPI_Spectral_Atlas_Database}%
  \BibitemOpen
  \bibfield  {author} {\bibinfo {author} {\bibfnamefont {H.}~\bibnamefont
  {Keller-Rudek}}, \bibinfo {author} {\bibfnamefont {G.~K.}\ \bibnamefont
  {Moortgat}}, \bibinfo {author} {\bibfnamefont {R.}~\bibnamefont {Sander}},\
  and\ \bibinfo {author} {\bibfnamefont {R.}~\bibnamefont {S\"orensen}},\
  }\bibfield  {title} {\bibinfo {title} {The {MPI-Mainz} {UV/VIS} spectral
  atlas of gaseous molecules of atmospheric interest},\ }\href
  {https://doi.org/10.5194/essd-5-365-2013} {\bibfield  {journal} {\bibinfo
  {journal} {Earth System Science Data}\ }\textbf {\bibinfo {volume} {5}},\
  \bibinfo {pages} {365} (\bibinfo {year} {2013})}\BibitemShut {NoStop}%
\bibitem [{\citenamefont {Bridge}\ \emph {et~al.}(1966)\citenamefont {Bridge},
  \citenamefont {Buckingham},\ and\ \citenamefont {Linnett}}]{Bridge1966}%
  \BibitemOpen
  \bibfield  {author} {\bibinfo {author} {\bibfnamefont {N.~J.}\ \bibnamefont
  {Bridge}}, \bibinfo {author} {\bibfnamefont {A.~D.}\ \bibnamefont
  {Buckingham}},\ and\ \bibinfo {author} {\bibfnamefont {J.~W.}\ \bibnamefont
  {Linnett}},\ }\bibfield  {title} {\bibinfo {title} {The polarization of laser
  light scattered by gases},\ }\href {https://doi.org/10.1098/rspa.1966.0244}
  {\bibfield  {journal} {\bibinfo  {journal} {Proceedings of the Royal Society
  of London. Series A. Mathematical and Physical Sciences}\ }\textbf {\bibinfo
  {volume} {295}},\ \bibinfo {pages} {334} (\bibinfo {year}
  {1966})}\BibitemShut {NoStop}%
\bibitem [{\citenamefont {Bogaard}\ \emph {et~al.}(1978)\citenamefont
  {Bogaard}, \citenamefont {Buckingham}, \citenamefont {Pierens},\ and\
  \citenamefont {White}}]{1978_Bogaard_Polarizability_Anisotropies}%
  \BibitemOpen
  \bibfield  {author} {\bibinfo {author} {\bibfnamefont {M.~P.}\ \bibnamefont
  {Bogaard}}, \bibinfo {author} {\bibfnamefont {A.~D.}\ \bibnamefont
  {Buckingham}}, \bibinfo {author} {\bibfnamefont {R.~K.}\ \bibnamefont
  {Pierens}},\ and\ \bibinfo {author} {\bibfnamefont {A.~H.}\ \bibnamefont
  {White}},\ }\bibfield  {title} {\bibinfo {title} {{R}ayleigh scattering
  depolarization ratio and molecular polarizability anisotropy for gases},\
  }\href {https://doi.org/10.1039/F19787403008} {\bibfield  {journal} {\bibinfo
   {journal} {Journal of the Chemical Society, Faraday Transactions 1: Physical
  Chemistry in Condensed Phases}\ }\textbf {\bibinfo {volume} {74}},\ \bibinfo
  {pages} {3008} (\bibinfo {year} {1978})}\BibitemShut {NoStop}%
\bibitem [{\citenamefont {McGuire}\ and\ \citenamefont
  {Miles}(2014)}]{2014_McGuire_REMPI_N2_2_2}%
  \BibitemOpen
  \bibfield  {author} {\bibinfo {author} {\bibfnamefont {S.}~\bibnamefont
  {McGuire}}\ and\ \bibinfo {author} {\bibfnamefont {R.}~\bibnamefont
  {Miles}},\ }\bibfield  {title} {\bibinfo {title} {Collision induced
  ultraviolet structure in nitrogen radar {REMPI} spectra},\ }\href
  {https://doi.org/10.1063/1.4904261} {\bibfield  {journal} {\bibinfo
  {journal} {The Journal of Chemical Physics}\ }\textbf {\bibinfo {volume}
  {141}},\ \bibinfo {pages} {244301} (\bibinfo {year} {2014})}\BibitemShut
  {NoStop}%
\bibitem [{\citenamefont {Peng}\ \emph {et~al.}(1997)\citenamefont {Peng},
  \citenamefont {Ledingham},\ and\ \citenamefont
  {Singhal}}]{1997_Peng_REMPI_CO}%
  \BibitemOpen
  \bibfield  {author} {\bibinfo {author} {\bibfnamefont {W.~X.}\ \bibnamefont
  {Peng}}, \bibinfo {author} {\bibfnamefont {K.~W.~D.}\ \bibnamefont
  {Ledingham}},\ and\ \bibinfo {author} {\bibfnamefont {R.~P.}\ \bibnamefont
  {Singhal}},\ }\bibfield  {title} {\bibinfo {title} {Trace {CO} detection by
  {REMPI} at 230 nm},\ }\href {https://doi.org/10.1063/1.52187} {\bibfield
  {journal} {\bibinfo  {journal} {AIP Conference Proceedings}\ }\textbf
  {\bibinfo {volume} {388}},\ \bibinfo {pages} {219} (\bibinfo {year}
  {1997})}\BibitemShut {NoStop}%
\bibitem [{\citenamefont {Yokelson}\ \emph {et~al.}(1992)\citenamefont
  {Yokelson}, \citenamefont {Lipert},\ and\ \citenamefont
  {Chupka}}]{1992_Yokelson_REMPI_O2}%
  \BibitemOpen
  \bibfield  {author} {\bibinfo {author} {\bibfnamefont {R.~J.}\ \bibnamefont
  {Yokelson}}, \bibinfo {author} {\bibfnamefont {R.~J.}\ \bibnamefont
  {Lipert}},\ and\ \bibinfo {author} {\bibfnamefont {W.~A.}\ \bibnamefont
  {Chupka}},\ }\bibfield  {title} {\bibinfo {title} {Identification of the
  ns$\sigma$ and nd$\lambda$ {R}ydberg states of {O}$_2$ for n=3-5},\ }\href
  {https://doi.org/10.1063/1.463724} {\bibfield  {journal} {\bibinfo  {journal}
  {The Journal of Chemical Physics}\ }\textbf {\bibinfo {volume} {97}},\
  \bibinfo {pages} {6153} (\bibinfo {year} {1992})}\BibitemShut {NoStop}%
\bibitem [{\citenamefont {Green}\ \emph {et~al.}(1991)\citenamefont {Green},
  \citenamefont {Bickel},\ and\ \citenamefont
  {Wallace}}]{1991_Green_REMPI_HCl}%
  \BibitemOpen
  \bibfield  {author} {\bibinfo {author} {\bibfnamefont {D.~S.}\ \bibnamefont
  {Green}}, \bibinfo {author} {\bibfnamefont {G.~A.}\ \bibnamefont {Bickel}},\
  and\ \bibinfo {author} {\bibfnamefont {S.~C.}\ \bibnamefont {Wallace}},\
  }\bibfield  {title} {\bibinfo {title} {(2 + 1) resonance enhanced multiphoton
  ionization of hydrogen chloride in a pulsed supersonic jet: {S}pectroscopic
  survey},\ }\href
  {https://doi.org/https://doi.org/10.1016/0022-2852(91)90238-6} {\bibfield
  {journal} {\bibinfo  {journal} {Journal of Molecular Spectroscopy}\ }\textbf
  {\bibinfo {volume} {150}},\ \bibinfo {pages} {303} (\bibinfo {year}
  {1991})}\BibitemShut {NoStop}%
\bibitem [{\citenamefont {Galea}\ \emph {et~al.}(2020)\citenamefont {Galea},
  \citenamefont {Shneider}, \citenamefont {Gragston},\ and\ \citenamefont
  {Zhang}}]{2020_Galea_REMPI_Xe}%
  \BibitemOpen
  \bibfield  {author} {\bibinfo {author} {\bibfnamefont {C.~A.}\ \bibnamefont
  {Galea}}, \bibinfo {author} {\bibfnamefont {M.~N.}\ \bibnamefont {Shneider}},
  \bibinfo {author} {\bibfnamefont {M.}~\bibnamefont {Gragston}},\ and\
  \bibinfo {author} {\bibfnamefont {Z.}~\bibnamefont {Zhang}},\ }\bibfield
  {title} {\bibinfo {title} {Coherent microwave scattering from xenon
  resonance-enhanced multiphoton ionization-initiated plasma in air},\ }\href
  {https://doi.org/10.1063/1.5135316} {\bibfield  {journal} {\bibinfo
  {journal} {Journal of Applied Physics}\ }\textbf {\bibinfo {volume} {127}},\
  \bibinfo {pages} {053301} (\bibinfo {year} {2020})}\BibitemShut {NoStop}%
\bibitem [{\citenamefont {Zernik}(1964)}]{1964_Zernik_REMPI_H}%
  \BibitemOpen
  \bibfield  {author} {\bibinfo {author} {\bibfnamefont {W.}~\bibnamefont
  {Zernik}},\ }\bibfield  {title} {\bibinfo {title} {Two-photon ionization of
  atomic hydrogen},\ }\href {https://doi.org/10.1103/PhysRev.135.A51}
  {\bibfield  {journal} {\bibinfo  {journal} {Physical Review}\ }\textbf
  {\bibinfo {volume} {135}},\ \bibinfo {pages} {A51} (\bibinfo {year}
  {1964})}\BibitemShut {NoStop}%
\bibitem [{\citenamefont {Marinero}\ \emph {et~al.}(1982)\citenamefont
  {Marinero}, \citenamefont {Rettner},\ and\ \citenamefont
  {Zare}}]{1982_Marinero_REMPI_H2}%
  \BibitemOpen
  \bibfield  {author} {\bibinfo {author} {\bibfnamefont {E.~E.}\ \bibnamefont
  {Marinero}}, \bibinfo {author} {\bibfnamefont {C.~T.}\ \bibnamefont
  {Rettner}},\ and\ \bibinfo {author} {\bibfnamefont {R.~N.}\ \bibnamefont
  {Zare}},\ }\bibfield  {title} {\bibinfo {title} {Quantum-state-specific
  detection of molecular hydrogen by three-photon ionization},\ }\href
  {https://doi.org/10.1103/PhysRevLett.48.1323} {\bibfield  {journal} {\bibinfo
   {journal} {Physical Review Letters}\ }\textbf {\bibinfo {volume} {48}},\
  \bibinfo {pages} {1323} (\bibinfo {year} {1982})}\BibitemShut {NoStop}%
\bibitem [{\citenamefont {Taylor}\ and\ \citenamefont
  {Johnson}(1993)}]{1993_Taylor_REMPI_CO2}%
  \BibitemOpen
  \bibfield  {author} {\bibinfo {author} {\bibfnamefont {D.~P.}\ \bibnamefont
  {Taylor}}\ and\ \bibinfo {author} {\bibfnamefont {P.~M.}\ \bibnamefont
  {Johnson}},\ }\bibfield  {title} {\bibinfo {title} {Resonance enhanced
  multiphoton ionization photoelectron spectra of {CO}$_2$. {III}.
  {A}utoionization dominates direct ionization},\ }\href
  {https://doi.org/10.1063/1.464215} {\bibfield  {journal} {\bibinfo  {journal}
  {The Journal of Chemical Physics}\ }\textbf {\bibinfo {volume} {98}},\
  \bibinfo {pages} {1810} (\bibinfo {year} {1993})}\BibitemShut {NoStop}%
\bibitem [{\citenamefont {Hanisco}\ and\ \citenamefont
  {Kummel}(1993)}]{1993_Hanisco_REMPI_N2O}%
  \BibitemOpen
  \bibfield  {author} {\bibinfo {author} {\bibfnamefont {T.~F.}\ \bibnamefont
  {Hanisco}}\ and\ \bibinfo {author} {\bibfnamefont {A.~C.}\ \bibnamefont
  {Kummel}},\ }\bibfield  {title} {\bibinfo {title} {State-resolved
  photodissociation of nitrous oxide},\ }\href
  {https://doi.org/10.1021/j100130a020} {\bibfield  {journal} {\bibinfo
  {journal} {Journal of Physical Chemistry}\ }\textbf {\bibinfo {volume}
  {97}},\ \bibinfo {pages} {7242} (\bibinfo {year} {1993})}\BibitemShut
  {NoStop}%
\bibitem [{\citenamefont {Al-Kahali}\ \emph {et~al.}(1996)\citenamefont
  {Al-Kahali}, \citenamefont {Donovan}, \citenamefont {Lawley},\ and\
  \citenamefont {Ridley}}]{1996_Al-Kahali_REMPI_Cl2}%
  \BibitemOpen
  \bibfield  {author} {\bibinfo {author} {\bibfnamefont {M.~S.~N.}\
  \bibnamefont {Al-Kahali}}, \bibinfo {author} {\bibfnamefont {R.~J.}\
  \bibnamefont {Donovan}}, \bibinfo {author} {\bibfnamefont {K.~P.}\
  \bibnamefont {Lawley}},\ and\ \bibinfo {author} {\bibfnamefont
  {T.}~\bibnamefont {Ridley}},\ }\bibfield  {title} {\bibinfo {title}
  {Mass-resolved multiphoton ionization spectroscopy of jet-cooled {C}l$_2$.
  {II}. {T}he (2+1) {REMPI} spectrum between 76000 and 90000 cm$^{-1}$},\
  }\href {https://doi.org/10.1063/1.470980} {\bibfield  {journal} {\bibinfo
  {journal} {The Journal of Chemical Physics}\ }\textbf {\bibinfo {volume}
  {104}},\ \bibinfo {pages} {1833} (\bibinfo {year} {1996})}\BibitemShut
  {NoStop}%
\bibitem [{\citenamefont {Hu}\ \emph {et~al.}(2008)\citenamefont {Hu},
  \citenamefont {Lee}, \citenamefont {Zhang}, \citenamefont {Wei},\ and\
  \citenamefont {Lin}}]{2008_Hu_REMPI_CS2}%
  \BibitemOpen
  \bibfield  {author} {\bibinfo {author} {\bibfnamefont {Z.}~\bibnamefont
  {Hu}}, \bibinfo {author} {\bibfnamefont {W.-B.}\ \bibnamefont {Lee}},
  \bibinfo {author} {\bibfnamefont {X.-P.}\ \bibnamefont {Zhang}}, \bibinfo
  {author} {\bibfnamefont {P.-Y.}\ \bibnamefont {Wei}},\ and\ \bibinfo {author}
  {\bibfnamefont {K.-C.}\ \bibnamefont {Lin}},\ }\bibfield  {title} {\bibinfo
  {title} {(1+1) resonance-enhanced multiphoton ionization and
  photodissociation study of {{CS}}{\textsubscript{2}} via the
  {\textsuperscript{2}}{{B}}{\textsubscript{2}} state},\ }\href
  {https://doi.org/10.1002/cphc.200700620} {\bibfield  {journal} {\bibinfo
  {journal} {ChemPhysChem}\ }\textbf {\bibinfo {volume} {9}},\ \bibinfo {pages}
  {422} (\bibinfo {year} {2008})}\BibitemShut {NoStop}%
\bibitem [{\citenamefont {Fulton}\ \emph {et~al.}(2006)\citenamefont {Fulton},
  \citenamefont {Bishop}, \citenamefont {Shneider},\ and\ \citenamefont
  {Barker}}]{2006_Fulton_REMPI_NO}%
  \BibitemOpen
  \bibfield  {author} {\bibinfo {author} {\bibfnamefont {R.}~\bibnamefont
  {Fulton}}, \bibinfo {author} {\bibfnamefont {A.~I.}\ \bibnamefont {Bishop}},
  \bibinfo {author} {\bibfnamefont {M.~N.}\ \bibnamefont {Shneider}},\ and\
  \bibinfo {author} {\bibfnamefont {P.~F.}\ \bibnamefont {Barker}},\ }\bibfield
   {title} {\bibinfo {title} {Controlling the motion of cold molecules with
  deep periodic optical potentials},\ }\href {https://doi.org/10.1038/nphys339}
  {\bibfield  {journal} {\bibinfo  {journal} {Nature Physics}\ }\textbf
  {\bibinfo {volume} {2}},\ \bibinfo {pages} {465} (\bibinfo {year}
  {2006})}\BibitemShut {NoStop}%
\bibitem [{\citenamefont {Salumbides}\ \emph {et~al.}(2009)\citenamefont
  {Salumbides}, \citenamefont {Khramov},\ and\ \citenamefont
  {Ubachs}}]{2009_Salumbides_REMPI_N2}%
  \BibitemOpen
  \bibfield  {author} {\bibinfo {author} {\bibfnamefont {E.~J.}\ \bibnamefont
  {Salumbides}}, \bibinfo {author} {\bibfnamefont {A.}~\bibnamefont
  {Khramov}},\ and\ \bibinfo {author} {\bibfnamefont {W.}~\bibnamefont
  {Ubachs}},\ }\bibfield  {title} {\bibinfo {title} {High-{{Resolution}} 2 + 1
  {{REMPI Study}} of the
  a$^{\prime\prime}${\textsuperscript{1}}{{$\Sigma$}}{\textsubscript{g}}+
  {{State}} in {{N}}{\textsubscript{2}}},\ }\href
  {https://doi.org/10.1021/jp808698u} {\bibfield  {journal} {\bibinfo
  {journal} {Journal of Physical Chemistry A}\ }\textbf {\bibinfo {volume}
  {113}},\ \bibinfo {pages} {2383} (\bibinfo {year} {2009})}\BibitemShut
  {NoStop}%
\bibitem [{\citenamefont {Axelrod}\ \emph {et~al.}(2020)\citenamefont
  {Axelrod}, \citenamefont {Campbell}, \citenamefont {Schwartz}, \citenamefont
  {Turnbaugh}, \citenamefont {Glaeser},\ and\ \citenamefont
  {M\"uller}}]{2020_Axelrod_reversal_of_ponderomotive_potential}%
  \BibitemOpen
  \bibfield  {author} {\bibinfo {author} {\bibfnamefont {J.~J.}\ \bibnamefont
  {Axelrod}}, \bibinfo {author} {\bibfnamefont {S.~L.}\ \bibnamefont
  {Campbell}}, \bibinfo {author} {\bibfnamefont {O.}~\bibnamefont {Schwartz}},
  \bibinfo {author} {\bibfnamefont {C.}~\bibnamefont {Turnbaugh}}, \bibinfo
  {author} {\bibfnamefont {R.~M.}\ \bibnamefont {Glaeser}},\ and\ \bibinfo
  {author} {\bibfnamefont {H.}~\bibnamefont {M\"uller}},\ }\bibfield  {title}
  {\bibinfo {title} {Observation of the relativistic reversal of the
  ponderomotive potential},\ }\href
  {https://doi.org/10.1103/PhysRevLett.124.174801} {\bibfield  {journal}
  {\bibinfo  {journal} {Physical Review Letters}\ }\textbf {\bibinfo {volume}
  {124}},\ \bibinfo {pages} {174801} (\bibinfo {year} {2020})}\BibitemShut
  {NoStop}%
\bibitem [{\citenamefont {Truppe}\ \emph {et~al.}(2018)\citenamefont {Truppe},
  \citenamefont {Hambach}, \citenamefont {Skoff}, \citenamefont {Bulleid},
  \citenamefont {Bumby}, \citenamefont {Hendricks}, \citenamefont {Hinds},
  \citenamefont {Sauer},\ and\ \citenamefont
  {Tarbutt}}]{2018_Truppe_A_buffer_gas_beam_source_for_short_intense_and_slow_molecular_pulses}%
  \BibitemOpen
  \bibfield  {author} {\bibinfo {author} {\bibfnamefont {S.}~\bibnamefont
  {Truppe}}, \bibinfo {author} {\bibfnamefont {M.}~\bibnamefont {Hambach}},
  \bibinfo {author} {\bibfnamefont {S.~M.}\ \bibnamefont {Skoff}}, \bibinfo
  {author} {\bibfnamefont {N.~E.}\ \bibnamefont {Bulleid}}, \bibinfo {author}
  {\bibfnamefont {J.~S.}\ \bibnamefont {Bumby}}, \bibinfo {author}
  {\bibfnamefont {R.~J.}\ \bibnamefont {Hendricks}}, \bibinfo {author}
  {\bibfnamefont {E.~A.}\ \bibnamefont {Hinds}}, \bibinfo {author}
  {\bibfnamefont {B.~E.}\ \bibnamefont {Sauer}},\ and\ \bibinfo {author}
  {\bibfnamefont {M.~R.}\ \bibnamefont {Tarbutt}},\ }\bibfield  {title}
  {\bibinfo {title} {A buffer gas beam source for short, intense and slow
  molecular pulses},\ }\href {https://doi.org/10.1080/09500340.2017.1384516}
  {\bibfield  {journal} {\bibinfo  {journal} {Journal of Modern Optics}\
  }\textbf {\bibinfo {volume} {65}},\ \bibinfo {pages} {648} (\bibinfo {year}
  {2018})}\BibitemShut {NoStop}%
\bibitem [{\citenamefont
  {Anderegg}(2019)}]{2019_Anderegg_thesis_ultracold_molecules}%
  \BibitemOpen
  \bibfield  {author} {\bibinfo {author} {\bibfnamefont {L.~G.}\ \bibnamefont
  {Anderegg}},\ }\emph {\bibinfo {title} {Ultracold molecules in optical
  arrays: {F}rom laser cooling to molecular collisions}},\ \href@noop {} {Ph.D.
  thesis},\ \bibinfo  {school} {Harvard University} (\bibinfo {year}
  {2019})\BibitemShut {NoStop}%
\bibitem [{\citenamefont
  {Baum}(2020)}]{2020_Baum_CaOH_MOT_cryogenic_techniques}%
  \BibitemOpen
  \bibfield  {author} {\bibinfo {author} {\bibfnamefont {L.~W.}\ \bibnamefont
  {Baum}},\ }\emph {\bibinfo {title} {Laser Cooling and 1{D} Magneto-Optical
  Trapping of Calcium Monohydroxide}},\ \href@noop {} {Ph.D. thesis},\ \bibinfo
   {school} {Harvard University} (\bibinfo {year} {2020})\BibitemShut {NoStop}%
\bibitem [{\citenamefont {Takahashi}\ \emph {et~al.}(2021)\citenamefont
  {Takahashi}, \citenamefont {Shlivko}, \citenamefont {Woolls},\ and\
  \citenamefont
  {Hutzler}}]{2021_Takahashi_Simulation_of_Cryogenic_Buffer_Gas_Beams}%
  \BibitemOpen
  \bibfield  {author} {\bibinfo {author} {\bibfnamefont {Y.}~\bibnamefont
  {Takahashi}}, \bibinfo {author} {\bibfnamefont {D.}~\bibnamefont {Shlivko}},
  \bibinfo {author} {\bibfnamefont {G.}~\bibnamefont {Woolls}},\ and\ \bibinfo
  {author} {\bibfnamefont {N.~R.}\ \bibnamefont {Hutzler}},\ }\bibfield
  {title} {\bibinfo {title} {Simulation of cryogenic buffer gas beams},\ }\href
  {https://doi.org/10.1103/PhysRevResearch.3.023018} {\bibfield  {journal}
  {\bibinfo  {journal} {Physical Review Research}\ }\textbf {\bibinfo {volume}
  {3}},\ \bibinfo {pages} {023018} (\bibinfo {year} {2021})}\BibitemShut
  {NoStop}%
\bibitem [{\citenamefont
  {Day}(2007)}]{2007_Day_Basics_and_Applications_of_Cryopumps}%
  \BibitemOpen
  \bibfield  {author} {\bibinfo {author} {\bibfnamefont {C.}~\bibnamefont
  {Day}},\ }\bibfield  {title} {\bibinfo {title} {{Basics and applications of
  cryopumps}},\ }\href {https://doi.org/10.5170/CERN-2007-003.241} {\bibfield
  {journal} {\bibinfo  {journal} {CAS - CERN Accelerator School: Vacuum in
  Accelerators}\ ,\ \bibinfo {pages} {241}} (\bibinfo {year}
  {2007})}\BibitemShut {NoStop}%
\bibitem [{\citenamefont {Vilas}\ \emph {et~al.}(2022)\citenamefont {Vilas},
  \citenamefont {Hallas}, \citenamefont {Anderegg}, \citenamefont {Robichaud},
  \citenamefont {Winnicki}, \citenamefont {Mitra},\ and\ \citenamefont
  {Doyle}}]{2022_Doyle_CaOH_MOT}%
  \BibitemOpen
  \bibfield  {author} {\bibinfo {author} {\bibfnamefont {N.~B.}\ \bibnamefont
  {Vilas}}, \bibinfo {author} {\bibfnamefont {C.}~\bibnamefont {Hallas}},
  \bibinfo {author} {\bibfnamefont {L.}~\bibnamefont {Anderegg}}, \bibinfo
  {author} {\bibfnamefont {P.}~\bibnamefont {Robichaud}}, \bibinfo {author}
  {\bibfnamefont {A.}~\bibnamefont {Winnicki}}, \bibinfo {author}
  {\bibfnamefont {D.}~\bibnamefont {Mitra}},\ and\ \bibinfo {author}
  {\bibfnamefont {J.~M.}\ \bibnamefont {Doyle}},\ }\bibfield  {title} {\bibinfo
  {title} {Magneto-optical trapping and sub-{Doppler} cooling of a polyatomic
  molecule},\ }\href {https://doi.org/10.1038/s41586-022-04620-5} {\bibfield
  {journal} {\bibinfo  {journal} {Nature}\ }\textbf {\bibinfo {volume} {606}},\
  \bibinfo {pages} {70} (\bibinfo {year} {2022})}\BibitemShut {NoStop}%
\bibitem [{\citenamefont {Takekoshi}\ \emph {et~al.}(1995)\citenamefont
  {Takekoshi}, \citenamefont {Yeh},\ and\ \citenamefont
  {Knize}}]{1995_Takekoshi_QUEST_theory}%
  \BibitemOpen
  \bibfield  {author} {\bibinfo {author} {\bibfnamefont {T.}~\bibnamefont
  {Takekoshi}}, \bibinfo {author} {\bibfnamefont {J.}~\bibnamefont {Yeh}},\
  and\ \bibinfo {author} {\bibfnamefont {R.}~\bibnamefont {Knize}},\ }\bibfield
   {title} {\bibinfo {title} {Quasi-electrostatic trap for neutral atoms},\
  }\href {https://doi.org/https://doi.org/10.1016/0030-4018(94)00638-B}
  {\bibfield  {journal} {\bibinfo  {journal} {Optics Communications}\ }\textbf
  {\bibinfo {volume} {114}},\ \bibinfo {pages} {421 } (\bibinfo {year}
  {1995})}\BibitemShut {NoStop}%
\bibitem [{\citenamefont {Takekoshi}\ and\ \citenamefont
  {Knize}(1996)}]{1996_Takekoshi_QUEST}%
  \BibitemOpen
  \bibfield  {author} {\bibinfo {author} {\bibfnamefont {T.}~\bibnamefont
  {Takekoshi}}\ and\ \bibinfo {author} {\bibfnamefont {R.~J.}\ \bibnamefont
  {Knize}},\ }\bibfield  {title} {\bibinfo {title} {{CO}$_2$ laser trap for
  cesium atoms},\ }\href {http://ol.osa.org/abstract.cfm?URI=ol-21-1-77}
  {\bibfield  {journal} {\bibinfo  {journal} {Optics Letters}\ }\textbf
  {\bibinfo {volume} {21}},\ \bibinfo {pages} {77} (\bibinfo {year}
  {1996})}\BibitemShut {NoStop}%
\bibitem [{\citenamefont {Engler}(2000)}]{2000_Hans_Engler_QUEST_thesis}%
  \BibitemOpen
  \bibfield  {author} {\bibinfo {author} {\bibfnamefont {H.}~\bibnamefont
  {Engler}},\ }\emph {\bibinfo {title} {A quasi-electrostatic trap for neutral
  atoms}},\ \href@noop {} {Ph.D. thesis},\ \bibinfo  {school} {University of
  Heidelberg} (\bibinfo {year} {2000})\BibitemShut {NoStop}%
\bibitem [{\citenamefont {Grimm}\ \emph {et~al.}(2000)\citenamefont {Grimm},
  \citenamefont {Weidem\"{u}ller},\ and\ \citenamefont
  {Ovchinnikov}}]{2000_Grimm_optical_trapping_review}%
  \BibitemOpen
  \bibfield  {author} {\bibinfo {author} {\bibfnamefont {R.}~\bibnamefont
  {Grimm}}, \bibinfo {author} {\bibfnamefont {M.}~\bibnamefont
  {Weidem\"{u}ller}},\ and\ \bibinfo {author} {\bibfnamefont {Y.~B.}\
  \bibnamefont {Ovchinnikov}},\ }\bibfield  {title} {\bibinfo {title} {Optical
  dipole traps for neutral atoms},\ }in\ \href
  {https://doi.org/https://doi.org/10.1016/S1049-250X(08)60186-X} {\emph
  {\bibinfo {booktitle} {Advances In Atomic, Molecular, and Optical
  Physics}}},\ Vol.~\bibinfo {volume} {42},\ \bibinfo {editor} {edited by\
  \bibinfo {editor} {\bibfnamefont {B.}~\bibnamefont {Bederson}}\ and\ \bibinfo
  {editor} {\bibfnamefont {H.}~\bibnamefont {Walther}}}\ (\bibinfo  {publisher}
  {Academic Press},\ \bibinfo {year} {2000})\ pp.\ \bibinfo {pages} {95 --
  170}\BibitemShut {NoStop}%
\bibitem [{\citenamefont {Le~Kien}\ \emph {et~al.}(2013)\citenamefont
  {Le~Kien}, \citenamefont {Schneeweiss},\ and\ \citenamefont
  {Rauschenbeutel}}]{2013_LeKien_dynaic_polarizability_tutorial_derivation}%
  \BibitemOpen
  \bibfield  {author} {\bibinfo {author} {\bibfnamefont {F.}~\bibnamefont
  {Le~Kien}}, \bibinfo {author} {\bibfnamefont {P.}~\bibnamefont
  {Schneeweiss}},\ and\ \bibinfo {author} {\bibfnamefont {A.}~\bibnamefont
  {Rauschenbeutel}},\ }\bibfield  {title} {\bibinfo {title} {Dynamical
  polarizability of atoms in arbitrary light fields: general theory and
  application to cesium},\ }\href {https://doi.org/10.1140/epjd/e2013-30729-x}
  {\bibfield  {journal} {\bibinfo  {journal} {The European Physical Journal D}\
  }\textbf {\bibinfo {volume} {67}},\ \bibinfo {pages} {92} (\bibinfo {year}
  {2013})}\BibitemShut {NoStop}%
\bibitem [{\citenamefont {Friedrich}\ and\ \citenamefont
  {Herschbach}(1995{\natexlab{b}})}]{1995_Friedrich_Spheroidal_Wave_Eqn_Theory_Details}%
  \BibitemOpen
  \bibfield  {author} {\bibinfo {author} {\bibfnamefont {B.}~\bibnamefont
  {Friedrich}}\ and\ \bibinfo {author} {\bibfnamefont {D.}~\bibnamefont
  {Herschbach}},\ }\bibfield  {title} {\bibinfo {title} {Polarization of
  molecules induced by intense nonresonant laser fields},\ }\href
  {https://doi.org/10.1021/j100042a051} {\bibfield  {journal} {\bibinfo
  {journal} {The Journal of Physical Chemistry}\ }\textbf {\bibinfo {volume}
  {99}},\ \bibinfo {pages} {15686} (\bibinfo {year}
  {1995}{\natexlab{b}})}\BibitemShut {NoStop}%
\bibitem [{\citenamefont
  {Friedrich}(2022)}]{2022_Friedrich_ElectroOptic_Trap_for_Molecules}%
  \BibitemOpen
  \bibfield  {author} {\bibinfo {author} {\bibfnamefont {B.}~\bibnamefont
  {Friedrich}},\ }\bibfield  {title} {\bibinfo {title} {Electro-optical trap
  for polar molecules},\ }\href {https://doi.org/10.1103/PhysRevA.105.053126}
  {\bibfield  {journal} {\bibinfo  {journal} {Physical Review A}\ }\textbf
  {\bibinfo {volume} {105}},\ \bibinfo {pages} {053126} (\bibinfo {year}
  {2022})}\BibitemShut {NoStop}%
\bibitem [{\citenamefont {Kongsted}\ and\ \citenamefont
  {Christiansen}(2007)}]{2007_Kongsted_dynamic_polarizability_methane}%
  \BibitemOpen
  \bibfield  {author} {\bibinfo {author} {\bibfnamefont {J.}~\bibnamefont
  {Kongsted}}\ and\ \bibinfo {author} {\bibfnamefont {O.}~\bibnamefont
  {Christiansen}},\ }\bibfield  {title} {\bibinfo {title} {Vibrational and
  thermal effects on the dipole polarizability of methane and carbon
  tetrachloride from vibrational structure calculations},\ }\href
  {https://doi.org/10.1063/1.2790025} {\bibfield  {journal} {\bibinfo
  {journal} {The Journal of Chemical Physics}\ }\textbf {\bibinfo {volume}
  {127}},\ \bibinfo {pages} {154315} (\bibinfo {year} {2007})}\BibitemShut
  {NoStop}%
\bibitem [{\citenamefont {Tomza}\ \emph {et~al.}(2013)\citenamefont {Tomza},
  \citenamefont {Skomorowski}, \citenamefont {Musia\l{}}, \citenamefont
  {Gonz\'{a}lez-F\'{e}rez}, \citenamefont {Koch},\ and\ \citenamefont
  {Moszynski}}]{2013_Tomza_Anisotropic_Polarizability_Trap_Depth_Heating_Rb2}%
  \BibitemOpen
  \bibfield  {author} {\bibinfo {author} {\bibfnamefont {M.}~\bibnamefont
  {Tomza}}, \bibinfo {author} {\bibfnamefont {W.}~\bibnamefont {Skomorowski}},
  \bibinfo {author} {\bibfnamefont {M.}~\bibnamefont {Musia\l{}}}, \bibinfo
  {author} {\bibfnamefont {R.}~\bibnamefont {Gonz\'{a}lez-F\'{e}rez}}, \bibinfo
  {author} {\bibfnamefont {C.~P.}\ \bibnamefont {Koch}},\ and\ \bibinfo
  {author} {\bibfnamefont {R.}~\bibnamefont {Moszynski}},\ }\bibfield  {title}
  {\bibinfo {title} {Interatomic potentials, electric properties and
  spectroscopy of the ground and excited states of the {R}b$_2$ molecule:
  \textit{ab initio} calculations and effect of a non-resonant field},\ }\href
  {https://doi.org/10.1080/00268976.2013.793835} {\bibfield  {journal}
  {\bibinfo  {journal} {Molecular Physics}\ }\textbf {\bibinfo {volume}
  {111}},\ \bibinfo {pages} {1781} (\bibinfo {year} {2013})}\BibitemShut
  {NoStop}%
\bibitem [{\citenamefont {He}\ \emph {et~al.}(2021)\citenamefont {He},
  \citenamefont {Fang}, \citenamefont {Shoshanim}, \citenamefont {Brown},\ and\
  \citenamefont
  {Rudich}}]{2020_He_Scattering_and_Absorption_Cross_Sections_Atmospheric_Gases_UV_Vis}%
  \BibitemOpen
  \bibfield  {author} {\bibinfo {author} {\bibfnamefont {Q.}~\bibnamefont
  {He}}, \bibinfo {author} {\bibfnamefont {Z.}~\bibnamefont {Fang}}, \bibinfo
  {author} {\bibfnamefont {O.}~\bibnamefont {Shoshanim}}, \bibinfo {author}
  {\bibfnamefont {S.~S.}\ \bibnamefont {Brown}},\ and\ \bibinfo {author}
  {\bibfnamefont {Y.}~\bibnamefont {Rudich}},\ }\bibfield  {title} {\bibinfo
  {title} {Scattering and absorption cross sections of atmospheric gases in the
  ultraviolet--visible wavelength range (307--725\,nm)},\ }\href
  {https://doi.org/10.5194/acp-21-14927-2021} {\bibfield  {journal} {\bibinfo
  {journal} {Atmospheric Chemistry and Physics}\ }\textbf {\bibinfo {volume}
  {21}},\ \bibinfo {pages} {14927} (\bibinfo {year} {2021})}\BibitemShut
  {NoStop}%
\bibitem [{\citenamefont {Andrews}\ and\ \citenamefont
  {Demidov}(2002)}]{2002_Andrews_an_introduction_to_laser_spectroscopy}%
  \BibitemOpen
  \bibfield  {author} {\bibinfo {author} {\bibfnamefont {D.}~\bibnamefont
  {Andrews}}\ and\ \bibinfo {author} {\bibfnamefont {A.}~\bibnamefont
  {Demidov}},\ }\href@noop {} {\emph {\bibinfo {title} {An Introduction to
  Laser Spectroscopy}}}\ (\bibinfo  {publisher} {Springer US},\ \bibinfo {year}
  {2002})\BibitemShut {NoStop}%
\bibitem [{\citenamefont {Motsch}\ \emph {et~al.}(2010)\citenamefont {Motsch},
  \citenamefont {Zeppenfeld}, \citenamefont {Pinkse},\ and\ \citenamefont
  {Rempe}}]{2010_Motsch_CavityRayleigh}%
  \BibitemOpen
  \bibfield  {author} {\bibinfo {author} {\bibfnamefont {M.}~\bibnamefont
  {Motsch}}, \bibinfo {author} {\bibfnamefont {M.}~\bibnamefont {Zeppenfeld}},
  \bibinfo {author} {\bibfnamefont {P.~W.~H.}\ \bibnamefont {Pinkse}},\ and\
  \bibinfo {author} {\bibfnamefont {G.}~\bibnamefont {Rempe}},\ }\bibfield
  {title} {\bibinfo {title} {Cavity-enhanced {R}ayleigh scattering},\ }\href
  {https://doi.org/10.1088/1367-2630/12/6/063022} {\bibfield  {journal}
  {\bibinfo  {journal} {New Journal of Physics}\ }\textbf {\bibinfo {volume}
  {12}},\ \bibinfo {pages} {063022} (\bibinfo {year} {2010})}\BibitemShut
  {NoStop}%
\bibitem [{\citenamefont {Sesko}\ \emph {et~al.}(1991)\citenamefont {Sesko},
  \citenamefont {Walker},\ and\ \citenamefont
  {Wieman}}]{1991_Sesko_Density_Limit_from_Repulsive_Scattering}%
  \BibitemOpen
  \bibfield  {author} {\bibinfo {author} {\bibfnamefont {D.~W.}\ \bibnamefont
  {Sesko}}, \bibinfo {author} {\bibfnamefont {T.~G.}\ \bibnamefont {Walker}},\
  and\ \bibinfo {author} {\bibfnamefont {C.~E.}\ \bibnamefont {Wieman}},\
  }\bibfield  {title} {\bibinfo {title} {Behavior of neutral atoms in a
  spontaneous force trap},\ }\href {https://doi.org/10.1364/JOSAB.8.000946}
  {\bibfield  {journal} {\bibinfo  {journal} {Journal of the Optical Society of
  America B}\ }\textbf {\bibinfo {volume} {8}},\ \bibinfo {pages} {946}
  (\bibinfo {year} {1991})}\BibitemShut {NoStop}%
\bibitem [{\citenamefont {Herzberg}(1950)}]{1950_Herzberg}%
  \BibitemOpen
  \bibfield  {author} {\bibinfo {author} {\bibfnamefont {G.}~\bibnamefont
  {Herzberg}},\ }\href@noop {} {\emph {\bibinfo {title} {Molecular spectra and
  molecular structure}}},\ \bibinfo {edition} {2nd}\ ed.,\ Prentice-Hall
  physics series\ (\bibinfo  {publisher} {Van Nostrand},\ \bibinfo {address}
  {New York},\ \bibinfo {year} {1950})\BibitemShut {NoStop}%
\bibitem [{\citenamefont {Penney}\ \emph {et~al.}(1974)\citenamefont {Penney},
  \citenamefont {Peters},\ and\ \citenamefont
  {Lapp}}]{1974_Penney_Rotational_Raman_Scattering}%
  \BibitemOpen
  \bibfield  {author} {\bibinfo {author} {\bibfnamefont {C.~M.}\ \bibnamefont
  {Penney}}, \bibinfo {author} {\bibfnamefont {R.~L.~S.}\ \bibnamefont
  {Peters}},\ and\ \bibinfo {author} {\bibfnamefont {M.}~\bibnamefont {Lapp}},\
  }\bibfield  {title} {\bibinfo {title} {Absolute rotational {R}aman cross
  sections for {N}$_2$, {O}$_2$, and {CO}$_2$},\ }\href
  {https://doi.org/10.1364/JOSA.64.000712} {\bibfield  {journal} {\bibinfo
  {journal} {Journal of the Optical Society of America}\ }\textbf {\bibinfo
  {volume} {64}},\ \bibinfo {pages} {712} (\bibinfo {year} {1974})}\BibitemShut
  {NoStop}%
\bibitem [{\citenamefont
  {Allemand}(1970)}]{1970_Dallemand_Depolarization_Ratio}%
  \BibitemOpen
  \bibfield  {author} {\bibinfo {author} {\bibfnamefont {C.~D.}\ \bibnamefont
  {Allemand}},\ }\bibfield  {title} {\bibinfo {title} {Depolarization ratio
  measurements in {R}aman spectrometry},\ }\href
  {https://doi.org/10.1366/000370270774371552} {\bibfield  {journal} {\bibinfo
  {journal} {Applied Spectroscopy}\ }\textbf {\bibinfo {volume} {24}},\
  \bibinfo {pages} {348} (\bibinfo {year} {1970})}\BibitemShut {NoStop}%
\bibitem [{\citenamefont {Shotton}\ and\ \citenamefont
  {Jones}(1970)}]{1970_Shotton_RRS_NO}%
  \BibitemOpen
  \bibfield  {author} {\bibinfo {author} {\bibfnamefont {K.}~\bibnamefont
  {Shotton}}\ and\ \bibinfo {author} {\bibfnamefont {W.~J.}\ \bibnamefont
  {Jones}},\ }\bibfield  {title} {\bibinfo {title} {Rotational {R}aman spectrum
  of nitric oxide},\ }\href@noop {} {\bibfield  {journal} {\bibinfo  {journal}
  {Canadian Journal of Physics}\ }\textbf {\bibinfo {volume} {48}},\ \bibinfo
  {pages} {632} (\bibinfo {year} {1970})}\BibitemShut {NoStop}%
\bibitem [{\citenamefont
  {Porto}(1966)}]{1966_Porto_Angular_Distribution_Raman_Scattering}%
  \BibitemOpen
  \bibfield  {author} {\bibinfo {author} {\bibfnamefont {S.~P.~S.}\
  \bibnamefont {Porto}},\ }\bibfield  {title} {\bibinfo {title} {Angular
  dependence and depolarization ratio of the {R}aman effect},\ }\href
  {https://doi.org/10.1364/JOSA.56.001585} {\bibfield  {journal} {\bibinfo
  {journal} {Journal of the Optical Society of America}\ }\textbf {\bibinfo
  {volume} {56}},\ \bibinfo {pages} {1585} (\bibinfo {year}
  {1966})}\BibitemShut {NoStop}%
\bibitem [{\citenamefont {Fenner}\ \emph {et~al.}(1973)\citenamefont {Fenner},
  \citenamefont {Hyatt}, \citenamefont {Kellam},\ and\ \citenamefont
  {Porto}}]{1973_Fenner_Vibrational_Raman_Cross_Sections_of_Simple_Gases}%
  \BibitemOpen
  \bibfield  {author} {\bibinfo {author} {\bibfnamefont {W.~R.}\ \bibnamefont
  {Fenner}}, \bibinfo {author} {\bibfnamefont {H.~A.}\ \bibnamefont {Hyatt}},
  \bibinfo {author} {\bibfnamefont {J.~M.}\ \bibnamefont {Kellam}},\ and\
  \bibinfo {author} {\bibfnamefont {S.~P.~S.}\ \bibnamefont {Porto}},\
  }\bibfield  {title} {\bibinfo {title} {{R}aman cross section of some simple
  gases},\ }\href {https://doi.org/10.1364/JOSA.63.000073} {\bibfield
  {journal} {\bibinfo  {journal} {Journal of the Optical Society of America}\
  }\textbf {\bibinfo {volume} {63}},\ \bibinfo {pages} {73} (\bibinfo {year}
  {1973})}\BibitemShut {NoStop}%
\bibitem [{\citenamefont {Carvalho}\ and\ \citenamefont
  {Vidal}(2022)}]{2022_Carvalho_Vibrational_Raman_Scattering_Cross_Sections}%
  \BibitemOpen
  \bibfield  {author} {\bibinfo {author} {\bibfnamefont {J.~R.}\ \bibnamefont
  {Carvalho}}\ and\ \bibinfo {author} {\bibfnamefont {L.~N.}\ \bibnamefont
  {Vidal}},\ }\bibfield  {title} {\bibinfo {title} {Calculation of absolute
  {R}aman scattering cross-sections using vibrational self-consistent
  field/vibrational configuration interaction wave functions},\ }\href
  {https://doi.org/https://doi.org/10.1002/jcc.26951} {\bibfield  {journal}
  {\bibinfo  {journal} {Journal of Computational Chemistry}\ }\textbf {\bibinfo
  {volume} {43}},\ \bibinfo {pages} {1484} (\bibinfo {year}
  {2022})}\BibitemShut {NoStop}%
\bibitem [{\citenamefont {Keldysh}\ \emph {et~al.}(1965)\citenamefont {Keldysh}
  \emph {et~al.}}]{1965_keldysh_ionization}%
  \BibitemOpen
  \bibfield  {author} {\bibinfo {author} {\bibfnamefont {L.}~\bibnamefont
  {Keldysh}} \emph {et~al.},\ }\bibfield  {title} {\bibinfo {title} {Ionization
  in the field of a strong electromagnetic wave},\ }\href@noop {} {\bibfield
  {journal} {\bibinfo  {journal} {Soviet Physics—JETP}\ }\textbf {\bibinfo
  {volume} {20}},\ \bibinfo {pages} {1307} (\bibinfo {year}
  {1965})}\BibitemShut {NoStop}%
\bibitem [{\citenamefont {Popruzhenko}\ \emph {et~al.}(2008)\citenamefont
  {Popruzhenko}, \citenamefont {Mur}, \citenamefont {Popov},\ and\
  \citenamefont
  {Bauer}}]{2008_Popruzhenko_coulomb_correction_to_Keldysh_high_freq}%
  \BibitemOpen
  \bibfield  {author} {\bibinfo {author} {\bibfnamefont {S.~V.}\ \bibnamefont
  {Popruzhenko}}, \bibinfo {author} {\bibfnamefont {V.~D.}\ \bibnamefont
  {Mur}}, \bibinfo {author} {\bibfnamefont {V.~S.}\ \bibnamefont {Popov}},\
  and\ \bibinfo {author} {\bibfnamefont {D.}~\bibnamefont {Bauer}},\ }\bibfield
   {title} {\bibinfo {title} {Strong field ionization rate for arbitrary laser
  frequencies},\ }\href {https://doi.org/10.1103/PhysRevLett.101.193003}
  {\bibfield  {journal} {\bibinfo  {journal} {Physical Review Letters}\
  }\textbf {\bibinfo {volume} {101}},\ \bibinfo {pages} {193003} (\bibinfo
  {year} {2008})}\BibitemShut {NoStop}%
\bibitem [{\citenamefont {{Perelomov}}\ \emph {et~al.}(1966)\citenamefont
  {{Perelomov}}, \citenamefont {{Popov}},\ and\ \citenamefont
  {{Terent'ev}}}]{1966_Perelomov_update_Keldysh_preexp}%
  \BibitemOpen
  \bibfield  {author} {\bibinfo {author} {\bibfnamefont {A.~M.}\ \bibnamefont
  {{Perelomov}}}, \bibinfo {author} {\bibfnamefont {V.~S.}\ \bibnamefont
  {{Popov}}},\ and\ \bibinfo {author} {\bibfnamefont {M.~V.}\ \bibnamefont
  {{Terent'ev}}},\ }\bibfield  {title} {\bibinfo {title} {Ionization of atoms
  in an alternating electric field},\ }\href@noop {} {\bibfield  {journal}
  {\bibinfo  {journal} {Soviet Journal of Experimental and Theoretical
  Physics}\ }\textbf {\bibinfo {volume} {23}},\ \bibinfo {pages} {924}
  (\bibinfo {year} {1966})}\BibitemShut {NoStop}%
\bibitem [{\citenamefont {L'Huillier}\ \emph {et~al.}(1983)\citenamefont
  {L'Huillier}, \citenamefont {Lompre}, \citenamefont {Mainfray},\ and\
  \citenamefont {Manus}}]{1983_LHuillier_MPI_noble_gases_1064}%
  \BibitemOpen
  \bibfield  {author} {\bibinfo {author} {\bibfnamefont {A.}~\bibnamefont
  {L'Huillier}}, \bibinfo {author} {\bibfnamefont {L.~A.}\ \bibnamefont
  {Lompre}}, \bibinfo {author} {\bibfnamefont {G.}~\bibnamefont {Mainfray}},\
  and\ \bibinfo {author} {\bibfnamefont {C.}~\bibnamefont {Manus}},\ }\bibfield
   {title} {\bibinfo {title} {Multiply charged ions induced by multiphoton
  absorption processes in rare-gas atoms at 1.064 \si{\micro\meter}},\ }\href
  {https://doi.org/10.1088/0022-3700/16/8/012} {\bibfield  {journal} {\bibinfo
  {journal} {Journal of Physics B: Atomic and Molecular Physics}\ }\textbf
  {\bibinfo {volume} {16}},\ \bibinfo {pages} {1363} (\bibinfo {year}
  {1983})}\BibitemShut {NoStop}%
\bibitem [{\citenamefont {Perry}\ \emph {et~al.}(1988)\citenamefont {Perry},
  \citenamefont {Landen}, \citenamefont {Sz\"oke},\ and\ \citenamefont
  {Campbell}}]{1988_Perry_multiphoton_ionisation_coefficients_noble_gases}%
  \BibitemOpen
  \bibfield  {author} {\bibinfo {author} {\bibfnamefont {M.~D.}\ \bibnamefont
  {Perry}}, \bibinfo {author} {\bibfnamefont {O.~L.}\ \bibnamefont {Landen}},
  \bibinfo {author} {\bibfnamefont {A.}~\bibnamefont {Sz\"oke}},\ and\ \bibinfo
  {author} {\bibfnamefont {E.~M.}\ \bibnamefont {Campbell}},\ }\bibfield
  {title} {\bibinfo {title} {Multiphoton ionization of the noble gases by an
  intense $10^{14}$ {W}/cm$^2$ dye laser},\ }\href
  {https://doi.org/10.1103/PhysRevA.37.747} {\bibfield  {journal} {\bibinfo
  {journal} {Physical Review A}\ }\textbf {\bibinfo {volume} {37}},\ \bibinfo
  {pages} {747} (\bibinfo {year} {1988})}\BibitemShut {NoStop}%
\bibitem [{\citenamefont {Woodbury}\ \emph {et~al.}(2020)\citenamefont
  {Woodbury}, \citenamefont {Schwartz}, \citenamefont {Rockafellow},
  \citenamefont {Wahlstrand},\ and\ \citenamefont
  {Milchberg}}]{2020_Woodbury_PRL_AbsoluteMeasuremntLaserIonization}%
  \BibitemOpen
  \bibfield  {author} {\bibinfo {author} {\bibfnamefont {D.}~\bibnamefont
  {Woodbury}}, \bibinfo {author} {\bibfnamefont {R.~M.}\ \bibnamefont
  {Schwartz}}, \bibinfo {author} {\bibfnamefont {E.}~\bibnamefont
  {Rockafellow}}, \bibinfo {author} {\bibfnamefont {J.~K.}\ \bibnamefont
  {Wahlstrand}},\ and\ \bibinfo {author} {\bibfnamefont {H.~M.}\ \bibnamefont
  {Milchberg}},\ }\bibfield  {title} {\bibinfo {title} {Absolute measurement of
  laser ionization yield in atmospheric pressure range gases over 14 decades},\
  }\href {https://doi.org/10.1103/PhysRevLett.124.013201} {\bibfield  {journal}
  {\bibinfo  {journal} {Physical Review Letters}\ }\textbf {\bibinfo {volume}
  {124}},\ \bibinfo {pages} {013201} (\bibinfo {year} {2020})}\BibitemShut
  {NoStop}%
\bibitem [{\citenamefont {Zhao}\ \emph {et~al.}(2016)\citenamefont {Zhao},
  \citenamefont {Le}, \citenamefont {Jin}, \citenamefont {Wang},\ and\
  \citenamefont {Lin}}]{2015_Zhao_molecular_ionization_rates_formula}%
  \BibitemOpen
  \bibfield  {author} {\bibinfo {author} {\bibfnamefont {S.-F.}\ \bibnamefont
  {Zhao}}, \bibinfo {author} {\bibfnamefont {A.-T.}\ \bibnamefont {Le}},
  \bibinfo {author} {\bibfnamefont {C.}~\bibnamefont {Jin}}, \bibinfo {author}
  {\bibfnamefont {X.}~\bibnamefont {Wang}},\ and\ \bibinfo {author}
  {\bibfnamefont {C.~D.}\ \bibnamefont {Lin}},\ }\bibfield  {title} {\bibinfo
  {title} {Analytical model for calibrating laser intensity in
  strong-field-ionization experiments},\ }\href
  {https://doi.org/10.1103/PhysRevA.93.023413} {\bibfield  {journal} {\bibinfo
  {journal} {Physical Review A}\ }\textbf {\bibinfo {volume} {93}},\ \bibinfo
  {pages} {023413} (\bibinfo {year} {2016})}\BibitemShut {NoStop}%
\bibitem [{\citenamefont {Lompr\'e}\ \emph {et~al.}(1985)\citenamefont
  {Lompr\'e}, \citenamefont {L'Huillier}, \citenamefont {Mainfray},\ and\
  \citenamefont {Manus}}]{1985_Lompre_multiphoton_ionisation_He_532}%
  \BibitemOpen
  \bibfield  {author} {\bibinfo {author} {\bibfnamefont {L.-A.}\ \bibnamefont
  {Lompr\'e}}, \bibinfo {author} {\bibfnamefont {A.}~\bibnamefont
  {L'Huillier}}, \bibinfo {author} {\bibfnamefont {G.}~\bibnamefont
  {Mainfray}},\ and\ \bibinfo {author} {\bibfnamefont {C.}~\bibnamefont
  {Manus}},\ }\bibfield  {title} {\bibinfo {title} {Multiphoton ionisation of
  {H}e atoms at 532 nm},\ }\href
  {https://doi.org/https://doi.org/10.1016/0375-9601(85)90350-0} {\bibfield
  {journal} {\bibinfo  {journal} {Physics Letters A}\ }\textbf {\bibinfo
  {volume} {112}},\ \bibinfo {pages} {319} (\bibinfo {year}
  {1985})}\BibitemShut {NoStop}%
\bibitem [{\citenamefont
  {Chin}(1971)}]{1971_Chin_multiphoton_ionisation_of_molecules}%
  \BibitemOpen
  \bibfield  {author} {\bibinfo {author} {\bibfnamefont {S.~L.}\ \bibnamefont
  {Chin}},\ }\bibfield  {title} {\bibinfo {title} {Multiphoton ionization of
  molecules},\ }\href {https://doi.org/10.1103/PhysRevA.4.992} {\bibfield
  {journal} {\bibinfo  {journal} {Physical Review A}\ }\textbf {\bibinfo
  {volume} {4}},\ \bibinfo {pages} {992} (\bibinfo {year} {1971})}\BibitemShut
  {NoStop}%
\bibitem [{\citenamefont {L'Huillier}\ \emph {et~al.}(1984)\citenamefont
  {L'Huillier}, \citenamefont {Mainfray},\ and\ \citenamefont
  {Johnson}}]{1983_Lhuillier_multiphoton_dissociation_ionisation_molecules}%
  \BibitemOpen
  \bibfield  {author} {\bibinfo {author} {\bibfnamefont {A.}~\bibnamefont
  {L'Huillier}}, \bibinfo {author} {\bibfnamefont {G.}~\bibnamefont
  {Mainfray}},\ and\ \bibinfo {author} {\bibfnamefont {P.}~\bibnamefont
  {Johnson}},\ }\bibfield  {title} {\bibinfo {title} {Multiphoton ionization
  versus dissociation of diatomic molecules irradiated by an intense 40 ps
  laser pulse},\ }\href
  {https://doi.org/https://doi.org/10.1016/0009-2614(84)85274-4} {\bibfield
  {journal} {\bibinfo  {journal} {Chemical Physics Letters}\ }\textbf {\bibinfo
  {volume} {103}},\ \bibinfo {pages} {447} (\bibinfo {year}
  {1984})}\BibitemShut {NoStop}%
\bibitem [{\citenamefont {Hanf}\ \emph {et~al.}(2003)\citenamefont {Hanf},
  \citenamefont {L\"{a}uter},\ and\ \citenamefont
  {Volpp}}]{2002_Hanf_UV_VUV_dissociation_CCl4}%
  \BibitemOpen
  \bibfield  {author} {\bibinfo {author} {\bibfnamefont {A.}~\bibnamefont
  {Hanf}}, \bibinfo {author} {\bibfnamefont {A.}~\bibnamefont {L\"{a}uter}},\
  and\ \bibinfo {author} {\bibfnamefont {H.-R.}\ \bibnamefont {Volpp}},\
  }\bibfield  {title} {\bibinfo {title} {Absolute chlorine atom quantum yield
  measurements in the {UV} and {VUV} gas-phase laser photolysis of {CC}l$_4$},\
  }\href {https://doi.org/https://doi.org/10.1016/S0009-2614(02)01896-1}
  {\bibfield  {journal} {\bibinfo  {journal} {Chemical Physics Letters}\
  }\textbf {\bibinfo {volume} {368}},\ \bibinfo {pages} {445} (\bibinfo {year}
  {2003})}\BibitemShut {NoStop}%
\bibitem [{\citenamefont {Heck}\ \emph {et~al.}(1996)\citenamefont {Heck},
  \citenamefont {Zare},\ and\ \citenamefont
  {Chandler}}]{1996_Heck_UV_photofragmentation_methane}%
  \BibitemOpen
  \bibfield  {author} {\bibinfo {author} {\bibfnamefont {A.~J.~R.}\
  \bibnamefont {Heck}}, \bibinfo {author} {\bibfnamefont {R.~N.}\ \bibnamefont
  {Zare}},\ and\ \bibinfo {author} {\bibfnamefont {D.~W.}\ \bibnamefont
  {Chandler}},\ }\bibfield  {title} {\bibinfo {title} {Photofragment imaging of
  methane},\ }\href {https://doi.org/10.1063/1.471214} {\bibfield  {journal}
  {\bibinfo  {journal} {The Journal of Chemical Physics}\ }\textbf {\bibinfo
  {volume} {104}},\ \bibinfo {pages} {4019} (\bibinfo {year}
  {1996})}\BibitemShut {NoStop}%
\bibitem [{\citenamefont {Xu}\ \emph {et~al.}(2006)\citenamefont {Xu},
  \citenamefont {Daigle}, \citenamefont {Luo},\ and\ \citenamefont
  {Chin}}]{2006_Xu_neutral_dissociation_methane_spectroscopy}%
  \BibitemOpen
  \bibfield  {author} {\bibinfo {author} {\bibfnamefont {H.~L.}\ \bibnamefont
  {Xu}}, \bibinfo {author} {\bibfnamefont {J.~F.}\ \bibnamefont {Daigle}},
  \bibinfo {author} {\bibfnamefont {Q.}~\bibnamefont {Luo}},\ and\ \bibinfo
  {author} {\bibfnamefont {S.~L.}\ \bibnamefont {Chin}},\ }\bibfield  {title}
  {\bibinfo {title} {Femtosecond laser-induced nonlinear spectroscopy for
  remote sensing of methane},\ }\href
  {https://doi.org/10.1007/s00340-005-2123-8} {\bibfield  {journal} {\bibinfo
  {journal} {Applied Physics B}\ }\textbf {\bibinfo {volume} {82}},\ \bibinfo
  {pages} {655} (\bibinfo {year} {2006})}\BibitemShut {NoStop}%
\bibitem [{\citenamefont {Song}\ \emph {et~al.}(2008)\citenamefont {Song},
  \citenamefont {Liu}, \citenamefont {Kong},\ and\ \citenamefont
  {Xia}}]{2008_Song_neutral_dissociation_methane}%
  \BibitemOpen
  \bibfield  {author} {\bibinfo {author} {\bibfnamefont {D.}~\bibnamefont
  {Song}}, \bibinfo {author} {\bibfnamefont {K.}~\bibnamefont {Liu}}, \bibinfo
  {author} {\bibfnamefont {F.}~\bibnamefont {Kong}},\ and\ \bibinfo {author}
  {\bibfnamefont {A.}~\bibnamefont {Xia}},\ }\bibfield  {title} {\bibinfo
  {title} {Neutral dissociation of methane in the ultra-fast laser pulse},\
  }\href {https://doi.org/10.1007/s11434-008-0232-6} {\bibfield  {journal}
  {\bibinfo  {journal} {Chinese Science Bulletin}\ }\textbf {\bibinfo {volume}
  {53}},\ \bibinfo {pages} {1946} (\bibinfo {year} {2008})}\BibitemShut
  {NoStop}%
\bibitem [{\citenamefont {Carney}\ and\ \citenamefont
  {Baer}(1981)}]{1981_Carney_mechanism_of_multiphoton_ionisation_H2S}%
  \BibitemOpen
  \bibfield  {author} {\bibinfo {author} {\bibfnamefont {T.~E.}\ \bibnamefont
  {Carney}}\ and\ \bibinfo {author} {\bibfnamefont {T.}~\bibnamefont {Baer}},\
  }\bibfield  {title} {\bibinfo {title} {The mechanism for multiphoton
  ionization of {H}$_2${S}},\ }\href {https://doi.org/10.1063/1.442607}
  {\bibfield  {journal} {\bibinfo  {journal} {The Journal of Chemical Physics}\
  }\textbf {\bibinfo {volume} {75}},\ \bibinfo {pages} {4422} (\bibinfo {year}
  {1981})}\BibitemShut {NoStop}%
\bibitem [{\citenamefont {Maitre}\ \emph {et~al.}(2020)\citenamefont {Maitre},
  \citenamefont {Scuderi}, \citenamefont {Corinti}, \citenamefont {Chiavarino},
  \citenamefont {Crestoni},\ and\ \citenamefont
  {Fornarini}}]{2020_Maitre_IRMPD_review}%
  \BibitemOpen
  \bibfield  {author} {\bibinfo {author} {\bibfnamefont {P.}~\bibnamefont
  {Maitre}}, \bibinfo {author} {\bibfnamefont {D.}~\bibnamefont {Scuderi}},
  \bibinfo {author} {\bibfnamefont {D.}~\bibnamefont {Corinti}}, \bibinfo
  {author} {\bibfnamefont {B.}~\bibnamefont {Chiavarino}}, \bibinfo {author}
  {\bibfnamefont {M.~E.}\ \bibnamefont {Crestoni}},\ and\ \bibinfo {author}
  {\bibfnamefont {S.}~\bibnamefont {Fornarini}},\ }\bibfield  {title} {\bibinfo
  {title} {Applications of infrared multiple photon dissociation ({IRMPD}) to
  the detection of posttranslational modifications},\ }\href
  {https://doi.org/10.1021/acs.chemrev.9b00395} {\bibfield  {journal} {\bibinfo
   {journal} {Chemical Reviews}\ }\textbf {\bibinfo {volume} {120}},\ \bibinfo
  {pages} {3261} (\bibinfo {year} {2020})}\BibitemShut {NoStop}%
\bibitem [{\citenamefont {Woodin}\ \emph {et~al.}(1978)\citenamefont {Woodin},
  \citenamefont {Bomse},\ and\ \citenamefont
  {Beauchamp}}]{1978_Woodin_IRMPD_original_fluence_NOT_intensity}%
  \BibitemOpen
  \bibfield  {author} {\bibinfo {author} {\bibfnamefont {R.~L.}\ \bibnamefont
  {Woodin}}, \bibinfo {author} {\bibfnamefont {D.~S.}\ \bibnamefont {Bomse}},\
  and\ \bibinfo {author} {\bibfnamefont {J.~L.}\ \bibnamefont {Beauchamp}},\
  }\bibfield  {title} {\bibinfo {title} {Multiphoton dissociation of molecules
  with low power continuous wave infrared laser radiation},\ }\href
  {https://doi.org/10.1021/ja00478a065} {\bibfield  {journal} {\bibinfo
  {journal} {Journal of the American Chemical Society}\ }\textbf {\bibinfo
  {volume} {100}},\ \bibinfo {pages} {3248} (\bibinfo {year}
  {1978})}\BibitemShut {NoStop}%
\bibitem [{\citenamefont {Black}\ \emph {et~al.}(1977)\citenamefont {Black},
  \citenamefont {Yablonovitch}, \citenamefont {Bloembergen},\ and\
  \citenamefont {Mukamel}}]{1977_Black_IRMPD_SF6_Mechanism}%
  \BibitemOpen
  \bibfield  {author} {\bibinfo {author} {\bibfnamefont {J.~G.}\ \bibnamefont
  {Black}}, \bibinfo {author} {\bibfnamefont {E.}~\bibnamefont {Yablonovitch}},
  \bibinfo {author} {\bibfnamefont {N.}~\bibnamefont {Bloembergen}},\ and\
  \bibinfo {author} {\bibfnamefont {S.}~\bibnamefont {Mukamel}},\ }\bibfield
  {title} {\bibinfo {title} {Collisionless multiphoton dissociation of
  {SF}$_6$: {A} statistical thermodynamic process},\ }\href
  {https://doi.org/10.1103/PhysRevLett.38.1131} {\bibfield  {journal} {\bibinfo
   {journal} {Physical Review Letters}\ }\textbf {\bibinfo {volume} {38}},\
  \bibinfo {pages} {1131} (\bibinfo {year} {1977})}\BibitemShut {NoStop}%
\bibitem [{\citenamefont {Harrison}\ and\ \citenamefont
  {Butcher}(1980)}]{1980_Harrison_IRMPD_Mechanism_Review_polyatomic_molecules}%
  \BibitemOpen
  \bibfield  {author} {\bibinfo {author} {\bibfnamefont {R.~G.}\ \bibnamefont
  {Harrison}}\ and\ \bibinfo {author} {\bibfnamefont {S.~R.}\ \bibnamefont
  {Butcher}},\ }\bibfield  {title} {\bibinfo {title} {Multiple photon infrared
  processes in polyatomic molecules},\ }\href
  {https://doi.org/10.1080/00107518008210939} {\bibfield  {journal} {\bibinfo
  {journal} {Contemporary Physics}\ }\textbf {\bibinfo {volume} {21}},\
  \bibinfo {pages} {19} (\bibinfo {year} {1980})}\BibitemShut {NoStop}%
\bibitem [{\citenamefont {Bloembergen}\ \emph {et~al.}(1976)\citenamefont
  {Bloembergen}, \citenamefont {Cantrell},\ and\ \citenamefont
  {Larsen}}]{1976_Book_TunableLasersAndApplications_w_chapter_on_IRMPD}%
  \BibitemOpen
  \bibfield  {author} {\bibinfo {author} {\bibfnamefont {N.}~\bibnamefont
  {Bloembergen}}, \bibinfo {author} {\bibfnamefont {C.~D.}\ \bibnamefont
  {Cantrell}},\ and\ \bibinfo {author} {\bibfnamefont {D.~M.}\ \bibnamefont
  {Larsen}},\ }\bibfield  {title} {\bibinfo {title} {Collisionless dissociation
  of polyatomic molecules by multiphoton infrared absorption},\ }in\ \href@noop
  {} {\emph {\bibinfo {booktitle} {Tunable {L}asers and {A}pplications}}},\
  \bibinfo {editor} {edited by\ \bibinfo {editor} {\bibfnamefont
  {A.}~\bibnamefont {Mooradian}}, \bibinfo {editor} {\bibfnamefont
  {T.}~\bibnamefont {Jaeger}},\ and\ \bibinfo {editor} {\bibfnamefont
  {P.}~\bibnamefont {Stokseth}}}\ (\bibinfo  {publisher} {Springer Berlin
  Heidelberg},\ \bibinfo {address} {Berlin, Heidelberg},\ \bibinfo {year}
  {1976})\ pp.\ \bibinfo {pages} {162--176}\BibitemShut {NoStop}%
\bibitem [{\citenamefont {Bloembergen}\ \emph {et~al.}(1984)\citenamefont
  {Bloembergen}, \citenamefont {Burak},\ and\ \citenamefont
  {Simpson}}]{1984_Bloembergen_IRMPE_IRMPD_small_molecules}%
  \BibitemOpen
  \bibfield  {author} {\bibinfo {author} {\bibfnamefont {N.}~\bibnamefont
  {Bloembergen}}, \bibinfo {author} {\bibfnamefont {I.}~\bibnamefont {Burak}},\
  and\ \bibinfo {author} {\bibfnamefont {T.}~\bibnamefont {Simpson}},\
  }\bibfield  {title} {\bibinfo {title} {Infrared multiphoton excitation of
  small molecules},\ }\href
  {https://doi.org/https://doi.org/10.1016/0022-2860(84)80134-9} {\bibfield
  {journal} {\bibinfo  {journal} {Journal of Molecular Structure}\ }\textbf
  {\bibinfo {volume} {113}},\ \bibinfo {pages} {69} (\bibinfo {year}
  {1984})}\BibitemShut {NoStop}%
\bibitem [{\citenamefont
  {Gupta}(2016)}]{2016_Gupta_Interaction_of_Radiation_and_Matter_and_Electronic_Spectra}%
  \BibitemOpen
  \bibfield  {author} {\bibinfo {author} {\bibfnamefont {V.~P.}\ \bibnamefont
  {Gupta}},\ }\href {https://doi.org/10.1016/C2014-0-05143-X} {\emph {\bibinfo
  {title} {Principles and Applications of Quantum Chemistry}}}\ (\bibinfo
  {publisher} {Academic Press},\ \bibinfo {address} {Boston},\ \bibinfo {year}
  {2016})\BibitemShut {NoStop}%
\bibitem [{\citenamefont
  {Giver}(1978)}]{1978_Giver_CH4_Rovib_Overtone_Spectroscopy}%
  \BibitemOpen
  \bibfield  {author} {\bibinfo {author} {\bibfnamefont {L.~P.}\ \bibnamefont
  {Giver}},\ }\bibfield  {title} {\bibinfo {title} {Intensity measurements of
  the {CH}$_4$ bands in the region 4350 {{\AA}} to 10,600 {{\AA}}},\ }\href
  {https://doi.org/https://doi.org/10.1016/0022-4073(78)90064-X} {\bibfield
  {journal} {\bibinfo  {journal} {Journal of Quantitative Spectroscopy and
  Radiative Transfer}\ }\textbf {\bibinfo {volume} {19}},\ \bibinfo {pages}
  {311} (\bibinfo {year} {1978})}\BibitemShut {NoStop}%
\bibitem [{\citenamefont {Rueda}\ \emph {et~al.}(2005)\citenamefont {Rueda},
  \citenamefont {Boyarkin}, \citenamefont {Rizzo}, \citenamefont
  {Chirokolava},\ and\ \citenamefont
  {Perry}}]{2005_Rueda_Overtone_Spectroscopy_Methanol}%
  \BibitemOpen
  \bibfield  {author} {\bibinfo {author} {\bibfnamefont {D.}~\bibnamefont
  {Rueda}}, \bibinfo {author} {\bibfnamefont {O.~V.}\ \bibnamefont {Boyarkin}},
  \bibinfo {author} {\bibfnamefont {T.~R.}\ \bibnamefont {Rizzo}}, \bibinfo
  {author} {\bibfnamefont {A.}~\bibnamefont {Chirokolava}},\ and\ \bibinfo
  {author} {\bibfnamefont {D.~S.}\ \bibnamefont {Perry}},\ }\bibfield  {title}
  {\bibinfo {title} {Vibrational overtone spectroscopy of jet-cooled methanol
  from 5000 to 14000 cm$^{-1}$},\ }\href {https://doi.org/10.1063/1.1833353}
  {\bibfield  {journal} {\bibinfo  {journal} {The Journal of Chemical Physics}\
  }\textbf {\bibinfo {volume} {122}},\ \bibinfo {pages} {044314} (\bibinfo
  {year} {2005})}\BibitemShut {NoStop}%
\bibitem [{\citenamefont {Hutzler}\ \emph {et~al.}(2012)\citenamefont
  {Hutzler}, \citenamefont {Lu},\ and\ \citenamefont
  {Doyle}}]{2012_Hutzler_Buffer_Gas_Beam_Review}%
  \BibitemOpen
  \bibfield  {author} {\bibinfo {author} {\bibfnamefont {N.~R.}\ \bibnamefont
  {Hutzler}}, \bibinfo {author} {\bibfnamefont {H.-I.}\ \bibnamefont {Lu}},\
  and\ \bibinfo {author} {\bibfnamefont {J.~M.}\ \bibnamefont {Doyle}},\
  }\bibfield  {title} {\bibinfo {title} {The buffer gas beam: {A}n intense,
  cold, and slow source for atoms and molecules},\ }\href
  {https://doi.org/10.1021/cr200362u} {\bibfield  {journal} {\bibinfo
  {journal} {Chemical Reviews}\ }\textbf {\bibinfo {volume} {112}},\ \bibinfo
  {pages} {4803} (\bibinfo {year} {2012})}\BibitemShut {NoStop}%
\bibitem [{\citenamefont {Gantner}\ \emph {et~al.}(2020)\citenamefont
  {Gantner}, \citenamefont {Koller}, \citenamefont {Wu}, \citenamefont
  {Rempe},\ and\ \citenamefont
  {Zeppenfeld}}]{2020_Gantner_low_density_buffer_gas_sim_vs_expt}%
  \BibitemOpen
  \bibfield  {author} {\bibinfo {author} {\bibfnamefont {T.}~\bibnamefont
  {Gantner}}, \bibinfo {author} {\bibfnamefont {M.}~\bibnamefont {Koller}},
  \bibinfo {author} {\bibfnamefont {X.}~\bibnamefont {Wu}}, \bibinfo {author}
  {\bibfnamefont {G.}~\bibnamefont {Rempe}},\ and\ \bibinfo {author}
  {\bibfnamefont {M.}~\bibnamefont {Zeppenfeld}},\ }\bibfield  {title}
  {\bibinfo {title} {Buffer-gas cooling of molecules in the low-density regime:
  {C}omparison between simulation and experiment},\ }\href
  {http://iopscience.iop.org/10.1088/1361-6455/ab8b42} {\bibfield  {journal}
  {\bibinfo  {journal} {Journal of Physics B: Atomic, Molecular and Optical
  Physics}\ } (\bibinfo {year} {2020})}\BibitemShut {NoStop}%
\bibitem [{\citenamefont {Beneventi}\ \emph {et~al.}(1986)\citenamefont
  {Beneventi}, \citenamefont {Casavecchia},\ and\ \citenamefont
  {Volpi}}]{1986_Beneventi_He_molecule_collision_cross_sections}%
  \BibitemOpen
  \bibfield  {author} {\bibinfo {author} {\bibfnamefont {L.}~\bibnamefont
  {Beneventi}}, \bibinfo {author} {\bibfnamefont {P.}~\bibnamefont
  {Casavecchia}},\ and\ \bibinfo {author} {\bibfnamefont {G.~G.}\ \bibnamefont
  {Volpi}},\ }\bibfield  {title} {\bibinfo {title} {High-resolution total
  differential cross sections for scattering of helium by {O}$_2$, {N}$_2$, and
  {NO}},\ }\href {https://doi.org/10.1063/1.451389} {\bibfield  {journal}
  {\bibinfo  {journal} {The Journal of Chemical Physics}\ }\textbf {\bibinfo
  {volume} {85}},\ \bibinfo {pages} {7011} (\bibinfo {year}
  {1986})}\BibitemShut {NoStop}%
\bibitem [{\citenamefont {Slankas}\ \emph {et~al.}(1979)\citenamefont
  {Slankas}, \citenamefont {Keil},\ and\ \citenamefont
  {Kuppermann}}]{1978_Slankas_300K_He_CH4_cross_section}%
  \BibitemOpen
  \bibfield  {author} {\bibinfo {author} {\bibfnamefont {J.~T.}\ \bibnamefont
  {Slankas}}, \bibinfo {author} {\bibfnamefont {M.}~\bibnamefont {Keil}},\ and\
  \bibinfo {author} {\bibfnamefont {A.}~\bibnamefont {Kuppermann}},\ }\bibfield
   {title} {\bibinfo {title} {Scattering of thermal {H}e beams by crossed
  atomic and molecular beams. {IV}. {S}pherically symmetric intermolecular
  potentials for {H}e+{CH}$_4$, {NH}$_3$, {H}$_2${O}, {SF}$_6$},\ }\href
  {https://doi.org/10.1063/1.437587} {\bibfield  {journal} {\bibinfo  {journal}
  {The Journal of Chemical Physics}\ }\textbf {\bibinfo {volume} {70}},\
  \bibinfo {pages} {1482} (\bibinfo {year} {1979})}\BibitemShut {NoStop}%
\bibitem [{\citenamefont {Au}(2013)}]{2013_Au_thesis_with_he3_cross_sections}%
  \BibitemOpen
  \bibfield  {author} {\bibinfo {author} {\bibfnamefont {Y.~S.}\ \bibnamefont
  {Au}},\ }\emph {\bibinfo {title} {Inelastic Collisions of Atomic Thorium and
  Molecular Thorium Monoxide with Cold Helium-3}},\ \href@noop {} {Ph.D.
  thesis},\ \bibinfo  {school} {Harvard University} (\bibinfo {year}
  {2013})\BibitemShut {NoStop}%
\bibitem [{\citenamefont {Landau}\ and\ \citenamefont
  {Lifshitz}(2013)}]{2013_Landau_Lifshitz_Statistical_Physics}%
  \BibitemOpen
  \bibfield  {author} {\bibinfo {author} {\bibfnamefont {L.}~\bibnamefont
  {Landau}}\ and\ \bibinfo {author} {\bibfnamefont {E.}~\bibnamefont
  {Lifshitz}},\ }\href {https://books.google.com/books?id=VzgJN-XPTRsC} {\emph
  {\bibinfo {title} {Statistical Physics}}},\ Vol.~\bibinfo {volume} {5}\
  (\bibinfo  {publisher} {Elsevier Science},\ \bibinfo {year}
  {2013})\BibitemShut {NoStop}%
\bibitem [{\citenamefont {Bause}\ \emph {et~al.}(2021)\citenamefont {Bause},
  \citenamefont {Schindewolf}, \citenamefont {Tao}, \citenamefont {Duda},
  \citenamefont {Chen}, \citenamefont {Qu\'em\'ener}, \citenamefont {Karman},
  \citenamefont {Christianen}, \citenamefont {Bloch},\ and\ \citenamefont
  {Luo}}]{2021_Bause_universal_loss_debate}%
  \BibitemOpen
  \bibfield  {author} {\bibinfo {author} {\bibfnamefont {R.}~\bibnamefont
  {Bause}}, \bibinfo {author} {\bibfnamefont {A.}~\bibnamefont {Schindewolf}},
  \bibinfo {author} {\bibfnamefont {R.}~\bibnamefont {Tao}}, \bibinfo {author}
  {\bibfnamefont {M.}~\bibnamefont {Duda}}, \bibinfo {author} {\bibfnamefont
  {X.-Y.}\ \bibnamefont {Chen}}, \bibinfo {author} {\bibfnamefont
  {G.}~\bibnamefont {Qu\'em\'ener}}, \bibinfo {author} {\bibfnamefont
  {T.}~\bibnamefont {Karman}}, \bibinfo {author} {\bibfnamefont
  {A.}~\bibnamefont {Christianen}}, \bibinfo {author} {\bibfnamefont
  {I.}~\bibnamefont {Bloch}},\ and\ \bibinfo {author} {\bibfnamefont {X.-Y.}\
  \bibnamefont {Luo}},\ }\bibfield  {title} {\bibinfo {title} {Collisions of
  ultracold molecules in bright and dark optical dipole traps},\ }\href
  {https://doi.org/10.1103/PhysRevResearch.3.033013} {\bibfield  {journal}
  {\bibinfo  {journal} {Physical Review Research}\ }\textbf {\bibinfo {volume}
  {3}},\ \bibinfo {pages} {033013} (\bibinfo {year} {2021})}\BibitemShut
  {NoStop}%
\bibitem [{\citenamefont {Hummon}\ \emph {et~al.}(2011)\citenamefont {Hummon},
  \citenamefont {Tscherbul}, \citenamefont {K\l{}os}, \citenamefont {Lu},
  \citenamefont {Tsikata}, \citenamefont {Campbell}, \citenamefont {Dalgarno},\
  and\ \citenamefont {Doyle}}]{2011_Hummon_N_NH_Collisions_Magnetic_Trap}%
  \BibitemOpen
  \bibfield  {author} {\bibinfo {author} {\bibfnamefont {M.~T.}\ \bibnamefont
  {Hummon}}, \bibinfo {author} {\bibfnamefont {T.~V.}\ \bibnamefont
  {Tscherbul}}, \bibinfo {author} {\bibfnamefont {J.}~\bibnamefont {K\l{}os}},
  \bibinfo {author} {\bibfnamefont {H.-I.}\ \bibnamefont {Lu}}, \bibinfo
  {author} {\bibfnamefont {E.}~\bibnamefont {Tsikata}}, \bibinfo {author}
  {\bibfnamefont {W.~C.}\ \bibnamefont {Campbell}}, \bibinfo {author}
  {\bibfnamefont {A.}~\bibnamefont {Dalgarno}},\ and\ \bibinfo {author}
  {\bibfnamefont {J.~M.}\ \bibnamefont {Doyle}},\ }\bibfield  {title} {\bibinfo
  {title} {Cold {N}+{NH} collisions in a magnetic trap},\ }\href
  {https://doi.org/10.1103/PhysRevLett.106.053201} {\bibfield  {journal}
  {\bibinfo  {journal} {Physical Review Letters}\ }\textbf {\bibinfo {volume}
  {106}},\ \bibinfo {pages} {053201} (\bibinfo {year} {2011})}\BibitemShut
  {NoStop}%
\bibitem [{\citenamefont {Harris}\ \emph {et~al.}(2004)\citenamefont {Harris},
  \citenamefont {Michniak}, \citenamefont {Nguyen}, \citenamefont {Brahms},
  \citenamefont {Ketterle},\ and\ \citenamefont
  {Doyle}}]{2004_Harris_cryogenic_films}%
  \BibitemOpen
  \bibfield  {author} {\bibinfo {author} {\bibfnamefont {J.~G.~E.}\
  \bibnamefont {Harris}}, \bibinfo {author} {\bibfnamefont {R.~A.}\
  \bibnamefont {Michniak}}, \bibinfo {author} {\bibfnamefont {S.~V.}\
  \bibnamefont {Nguyen}}, \bibinfo {author} {\bibfnamefont {N.}~\bibnamefont
  {Brahms}}, \bibinfo {author} {\bibfnamefont {W.}~\bibnamefont {Ketterle}},\
  and\ \bibinfo {author} {\bibfnamefont {J.~M.}\ \bibnamefont {Doyle}},\
  }\bibfield  {title} {\bibinfo {title} {Buffer gas cooling and trapping of
  atoms with small effective magnetic moments},\ }\href
  {https://doi.org/10.1209/epl/i2004-10059-y} {\bibfield  {journal} {\bibinfo
  {journal} {Europhysics Letters ({EPL})}\ }\textbf {\bibinfo {volume} {67}},\
  \bibinfo {pages} {198} (\bibinfo {year} {2004})}\BibitemShut {NoStop}%
\bibitem [{\citenamefont
  {Michniak}(2004)}]{2004_Michniak_low_eta_magnetic_trapping}%
  \BibitemOpen
  \bibfield  {author} {\bibinfo {author} {\bibfnamefont {R.}~\bibnamefont
  {Michniak}},\ }\emph {\bibinfo {title} {Enhanced buffer gas loading:
  {C}ooling and trapping of atoms with low effective magnetic moments}},\
  \href@noop {} {Ph.D. thesis},\ \bibinfo  {school} {Harvard University}
  (\bibinfo {year} {2004})\BibitemShut {NoStop}%
\bibitem [{\citenamefont {Ketterle}\ and\ \citenamefont
  {Druten}(1996)}]{1996_Ketterle_evaporative_cooling_review}%
  \BibitemOpen
  \bibfield  {author} {\bibinfo {author} {\bibfnamefont {W.}~\bibnamefont
  {Ketterle}}\ and\ \bibinfo {author} {\bibfnamefont {N.~V.}\ \bibnamefont
  {Druten}},\ }\bibfield  {title} {\bibinfo {title} {Evaporative cooling of
  trapped atoms},\ }in\ \href
  {https://doi.org/https://doi.org/10.1016/S1049-250X(08)60101-9} {\emph
  {\bibinfo {booktitle} {Advances In Atomic, Molecular, and Optical
  Physics}}},\ Vol.~\bibinfo {volume} {37},\ \bibinfo {editor} {edited by\
  \bibinfo {editor} {\bibfnamefont {B.}~\bibnamefont {Bederson}}\ and\ \bibinfo
  {editor} {\bibfnamefont {H.}~\bibnamefont {Walther}}}\ (\bibinfo  {publisher}
  {Academic Press},\ \bibinfo {year} {1996})\ pp.\ \bibinfo {pages}
  {181--236}\BibitemShut {NoStop}%
\bibitem [{\citenamefont {Bourgain}\ \emph {et~al.}(2013)\citenamefont
  {Bourgain}, \citenamefont {Pellegrino}, \citenamefont {Fuhrmanek},
  \citenamefont {Sortais},\ and\ \citenamefont
  {Browaeys}}]{2013_Bourgain_Evaporative_Cooling_Small_Number_of_Molecules_Dipole_Trap}%
  \BibitemOpen
  \bibfield  {author} {\bibinfo {author} {\bibfnamefont {R.}~\bibnamefont
  {Bourgain}}, \bibinfo {author} {\bibfnamefont {J.}~\bibnamefont
  {Pellegrino}}, \bibinfo {author} {\bibfnamefont {A.}~\bibnamefont
  {Fuhrmanek}}, \bibinfo {author} {\bibfnamefont {Y.~R.~P.}\ \bibnamefont
  {Sortais}},\ and\ \bibinfo {author} {\bibfnamefont {A.}~\bibnamefont
  {Browaeys}},\ }\bibfield  {title} {\bibinfo {title} {Evaporative cooling of a
  small number of atoms in a single-beam microscopic dipole trap},\ }\href
  {https://doi.org/10.1103/PhysRevA.88.023428} {\bibfield  {journal} {\bibinfo
  {journal} {Physical Review A}\ }\textbf {\bibinfo {volume} {88}},\ \bibinfo
  {pages} {023428} (\bibinfo {year} {2013})}\BibitemShut {NoStop}%
\bibitem [{\citenamefont {O'Hara}\ \emph {et~al.}(2001)\citenamefont {O'Hara},
  \citenamefont {Gehm}, \citenamefont {Granade},\ and\ \citenamefont
  {Thomas}}]{2001_Ohara_Evaporative_Cooling_Time_Scaling_Theory}%
  \BibitemOpen
  \bibfield  {author} {\bibinfo {author} {\bibfnamefont {K.~M.}\ \bibnamefont
  {O'Hara}}, \bibinfo {author} {\bibfnamefont {M.~E.}\ \bibnamefont {Gehm}},
  \bibinfo {author} {\bibfnamefont {S.~R.}\ \bibnamefont {Granade}},\ and\
  \bibinfo {author} {\bibfnamefont {J.~E.}\ \bibnamefont {Thomas}},\ }\bibfield
   {title} {\bibinfo {title} {Scaling laws for evaporative cooling in
  time-dependent optical traps},\ }\href
  {https://doi.org/10.1103/PhysRevA.64.051403} {\bibfield  {journal} {\bibinfo
  {journal} {Physical Review A}\ }\textbf {\bibinfo {volume} {64}},\ \bibinfo
  {pages} {051403(R)} (\bibinfo {year} {2001})}\BibitemShut {NoStop}%
\bibitem [{\citenamefont {Lee}\ \emph {et~al.}(2006)\citenamefont {Lee},
  \citenamefont {Kim},\ and\ \citenamefont
  {Cho}}]{2006_Lee_bichromatic_cavity_remove_standing_wave_pattern}%
  \BibitemOpen
  \bibfield  {author} {\bibinfo {author} {\bibfnamefont {S.~K.}\ \bibnamefont
  {Lee}}, \bibinfo {author} {\bibfnamefont {J.~J.}\ \bibnamefont {Kim}},\ and\
  \bibinfo {author} {\bibfnamefont {D.}~\bibnamefont {Cho}},\ }\bibfield
  {title} {\bibinfo {title} {Transformable optical dipole trap using a
  phase-modulated standing wave},\ }\href
  {https://doi.org/10.1103/PhysRevA.74.063401} {\bibfield  {journal} {\bibinfo
  {journal} {Physical Review A}\ }\textbf {\bibinfo {volume} {74}},\ \bibinfo
  {pages} {063401} (\bibinfo {year} {2006})}\BibitemShut {NoStop}%
\bibitem [{\citenamefont {Edmunds}\ and\ \citenamefont
  {Barker}(2014)}]{2014_Edmunds_Selective_Heating_Med_Finesse_Cavity_Argon_Trap}%
  \BibitemOpen
  \bibfield  {author} {\bibinfo {author} {\bibfnamefont {P.~D.}\ \bibnamefont
  {Edmunds}}\ and\ \bibinfo {author} {\bibfnamefont {P.~F.}\ \bibnamefont
  {Barker}},\ }\bibfield  {title} {\bibinfo {title} {Trapping cold ground state
  argon atoms},\ }\href {https://doi.org/10.1103/PhysRevLett.113.183001}
  {\bibfield  {journal} {\bibinfo  {journal} {Physical Review Letters}\
  }\textbf {\bibinfo {volume} {113}},\ \bibinfo {pages} {183001} (\bibinfo
  {year} {2014})}\BibitemShut {NoStop}%
\bibitem [{\citenamefont {Boesl}\ and\ \citenamefont
  {Zimmermann}(2021)}]{2021_Boesl_REMPIReview}%
  \BibitemOpen
  \bibfield  {author} {\bibinfo {author} {\bibfnamefont {U.}~\bibnamefont
  {Boesl}}\ and\ \bibinfo {author} {\bibfnamefont {R.}~\bibnamefont
  {Zimmermann}},\ }\bibinfo {title} {Fundamentals and mechanisms of
  resonance-enhanced multiphoton ionization ({REMPI}) in vacuum and its
  application in molecular spectroscopy},\ in\ \href
  {https://doi.org/https://doi.org/10.1002/9783527682201.ch2} {\emph {\bibinfo
  {booktitle} {Photoionization and Photo‐Induced Processes in Mass
  Spectrometry}}}\ (\bibinfo  {publisher} {John Wiley \& Sons, Ltd},\ \bibinfo
  {year} {2021})\ Chap.~\bibinfo {chapter} {2}, pp.\ \bibinfo {pages}
  {23--88}\BibitemShut {NoStop}%
\bibitem [{\citenamefont {Yamaguchi}\ \emph {et~al.}(2012)\citenamefont
  {Yamaguchi}, \citenamefont {Moriyama}, \citenamefont {Ide}, \citenamefont
  {Ito}, \citenamefont {Matsuda},\ and\ \citenamefont
  {Niimi}}]{2012_Yamaguchi_REMPI_N2_rotational_temperature}%
  \BibitemOpen
  \bibfield  {author} {\bibinfo {author} {\bibfnamefont {H.}~\bibnamefont
  {Yamaguchi}}, \bibinfo {author} {\bibfnamefont {T.}~\bibnamefont {Moriyama}},
  \bibinfo {author} {\bibfnamefont {K.}~\bibnamefont {Ide}}, \bibinfo {author}
  {\bibfnamefont {J.}~\bibnamefont {Ito}}, \bibinfo {author} {\bibfnamefont
  {Y.}~\bibnamefont {Matsuda}},\ and\ \bibinfo {author} {\bibfnamefont
  {T.}~\bibnamefont {Niimi}},\ }\bibfield  {title} {\bibinfo {title}
  {Measurement of the rotational temperature in a nitrogen molecular beam by
  {REMPI}},\ }\href {https://doi.org/10.1063/1.4769697} {\bibfield  {journal}
  {\bibinfo  {journal} {AIP Conference Proceedings}\ }\textbf {\bibinfo
  {volume} {1501}},\ \bibinfo {pages} {1350} (\bibinfo {year}
  {2012})}\BibitemShut {NoStop}%
\bibitem [{\citenamefont {Ashfold}\ and\ \citenamefont
  {Western}(2017)}]{2017_Ashford_Multiphoton_Spectroscopy_Applications}%
  \BibitemOpen
  \bibfield  {author} {\bibinfo {author} {\bibfnamefont {M.}~\bibnamefont
  {Ashfold}}\ and\ \bibinfo {author} {\bibfnamefont {C.}~\bibnamefont
  {Western}},\ }\bibfield  {title} {\bibinfo {title} {Multiphoton spectroscopy,
  applications},\ }in\ \href
  {https://doi.org/https://doi.org/10.1016/B978-0-12-409547-2.05032-0} {\emph
  {\bibinfo {booktitle} {Encyclopedia of Spectroscopy and Spectrometry}}},\
  \bibinfo {editor} {edited by\ \bibinfo {editor} {\bibfnamefont {J.~C.}\
  \bibnamefont {Lindon}}, \bibinfo {editor} {\bibfnamefont {G.~E.}\
  \bibnamefont {Tranter}},\ and\ \bibinfo {editor} {\bibfnamefont {D.~W.}\
  \bibnamefont {Koppenaal}}}\ (\bibinfo  {publisher} {Academic Press},\
  \bibinfo {address} {Oxford},\ \bibinfo {year} {2017})\ \bibinfo {edition}
  {3rd}\ ed.,\ pp.\ \bibinfo {pages} {954--961}\BibitemShut {NoStop}%
\bibitem [{\citenamefont {Nolde}\ \emph {et~al.}(2005)\citenamefont {Nolde},
  \citenamefont {Weitzel},\ and\ \citenamefont
  {Western}}]{2005_Nolde_REMPI_NH3}%
  \BibitemOpen
  \bibfield  {author} {\bibinfo {author} {\bibfnamefont {M.}~\bibnamefont
  {Nolde}}, \bibinfo {author} {\bibfnamefont {K.-M.}\ \bibnamefont {Weitzel}},\
  and\ \bibinfo {author} {\bibfnamefont {C.~M.}\ \bibnamefont {Western}},\
  }\bibfield  {title} {\bibinfo {title} {The resonance enhanced multiphoton
  ionisation spectroscopy of ammonia isotopomers {{NH}}{\textsubscript{3}},
  {{NH}}{\textsubscript{2}}{{D}}, {{NHD}}{\textsubscript{2}} and
  {{ND}}{\textsubscript{3}}},\ }\href {https://doi.org/10.1039/B417835C}
  {\bibfield  {journal} {\bibinfo  {journal} {Physical Chemistry Chemical
  Physics}\ }\textbf {\bibinfo {volume} {7}},\ \bibinfo {pages} {1527}
  (\bibinfo {year} {2005})}\BibitemShut {NoStop}%
\bibitem [{\citenamefont {Johnson}(1976)}]{1976_Johnson_REMPI_Benzene}%
  \BibitemOpen
  \bibfield  {author} {\bibinfo {author} {\bibfnamefont {P.~M.}\ \bibnamefont
  {Johnson}},\ }\bibfield  {title} {\bibinfo {title} {The multiphoton
  ionization spectrum of benzene},\ }\href {https://doi.org/10.1063/1.431983}
  {\bibfield  {journal} {\bibinfo  {journal} {The Journal of Chemical Physics}\
  }\textbf {\bibinfo {volume} {64}},\ \bibinfo {pages} {4143} (\bibinfo {year}
  {1976})}\BibitemShut {NoStop}%
\bibitem [{\citenamefont {Xue}\ \emph {et~al.}(2000)\citenamefont {Xue},
  \citenamefont {Chen},\ and\ \citenamefont {Dai}}]{2000_Xue_REMPI_SO2}%
  \BibitemOpen
  \bibfield  {author} {\bibinfo {author} {\bibfnamefont {B.}~\bibnamefont
  {Xue}}, \bibinfo {author} {\bibfnamefont {Y.}~\bibnamefont {Chen}},\ and\
  \bibinfo {author} {\bibfnamefont {H.-L.}\ \bibnamefont {Dai}},\ }\bibfield
  {title} {\bibinfo {title} {Observation of the singlet-triplet pair of the 4p
  {R}ydberg state and assignment of the {R}ydberg series of {SO}$_2$},\ }\href
  {https://doi.org/10.1063/1.480787} {\bibfield  {journal} {\bibinfo  {journal}
  {The Journal of Chemical Physics}\ }\textbf {\bibinfo {volume} {112}},\
  \bibinfo {pages} {2210} (\bibinfo {year} {2000})}\BibitemShut {NoStop}%
\bibitem [{\citenamefont {Meijer}\ \emph {et~al.}(1986)\citenamefont {Meijer},
  \citenamefont {ter Meulen}, \citenamefont {Andresen},\ and\ \citenamefont
  {Bath}}]{1986_Meijer_REMPI_H2O}%
  \BibitemOpen
  \bibfield  {author} {\bibinfo {author} {\bibfnamefont {G.}~\bibnamefont
  {Meijer}}, \bibinfo {author} {\bibfnamefont {J.~J.}\ \bibnamefont {ter
  Meulen}}, \bibinfo {author} {\bibfnamefont {P.}~\bibnamefont {Andresen}},\
  and\ \bibinfo {author} {\bibfnamefont {A.}~\bibnamefont {Bath}},\ }\bibfield
  {title} {\bibinfo {title} {Sensitive quantum state selective detection of
  {H}$_2${O} and {D}$_2${O} by (2+1)-resonance enhanced multiphoton
  ionization},\ }\href {https://doi.org/10.1063/1.451845} {\bibfield  {journal}
  {\bibinfo  {journal} {The Journal of Chemical Physics}\ }\textbf {\bibinfo
  {volume} {85}},\ \bibinfo {pages} {6914} (\bibinfo {year}
  {1986})}\BibitemShut {NoStop}%
\bibitem [{\citenamefont {Philis}(2007)}]{2007_Philis_REMPIMethanolEthanol}%
  \BibitemOpen
  \bibfield  {author} {\bibinfo {author} {\bibfnamefont {J.~G.}\ \bibnamefont
  {Philis}},\ }\bibfield  {title} {\bibinfo {title} {Resonance-enhanced
  multiphoton ionization spectra of jet-cooled methanol and ethanol},\ }\href
  {https://doi.org/https://doi.org/10.1016/j.cplett.2007.10.089} {\bibfield
  {journal} {\bibinfo  {journal} {Chemical Physics Letters}\ }\textbf {\bibinfo
  {volume} {449}},\ \bibinfo {pages} {291} (\bibinfo {year}
  {2007})}\BibitemShut {NoStop}%
\bibitem [{\citenamefont {Dogariu}\ \emph {et~al.}(2011)\citenamefont
  {Dogariu}, \citenamefont {Stein}, \citenamefont {Glaser},\ and\ \citenamefont
  {Miles}}]{2011_Dogariu_REMPI_NO_SF2}%
  \BibitemOpen
  \bibfield  {author} {\bibinfo {author} {\bibfnamefont {A.}~\bibnamefont
  {Dogariu}}, \bibinfo {author} {\bibfnamefont {C.}~\bibnamefont {Stein}},
  \bibinfo {author} {\bibfnamefont {A.}~\bibnamefont {Glaser}},\ and\ \bibinfo
  {author} {\bibfnamefont {R.~B.}\ \bibnamefont {Miles}},\ }\bibfield  {title}
  {\bibinfo {title} {{Long range trace detection by radar REMPI}},\ }in\ \href
  {https://doi.org/10.1117/12.883982} {\emph {\bibinfo {booktitle} {Advanced
  {E}nvironmental, {C}hemical, and {B}iological {S}ensing {T}echnologies
  {VIII}}}},\ Vol.\ \bibinfo {volume} {8024},\ \bibinfo {editor} {edited by\
  \bibinfo {editor} {\bibfnamefont {T.}~\bibnamefont {Vo-Dinh}}, \bibinfo
  {editor} {\bibfnamefont {R.~A.}\ \bibnamefont {Lieberman}},\ and\ \bibinfo
  {editor} {\bibfnamefont {G.}~\bibnamefont {Gauglitz}}},\ \bibinfo
  {organization} {International Society for Optics and Photonics}\ (\bibinfo
  {publisher} {SPIE},\ \bibinfo {year} {2011})\ pp.\ \bibinfo {pages} {88 --
  96}\BibitemShut {NoStop}%
\bibitem [{\citenamefont {de~Beer}\ \emph {et~al.}(1991)\citenamefont
  {de~Beer}, \citenamefont {Koopmans}, \citenamefont {de~Lange}, \citenamefont
  {Wang},\ and\ \citenamefont {Chupka}}]{1991_deBeer_REMPI_OH}%
  \BibitemOpen
  \bibfield  {author} {\bibinfo {author} {\bibfnamefont {E.}~\bibnamefont
  {de~Beer}}, \bibinfo {author} {\bibfnamefont {M.~P.}\ \bibnamefont
  {Koopmans}}, \bibinfo {author} {\bibfnamefont {C.~A.}\ \bibnamefont
  {de~Lange}}, \bibinfo {author} {\bibfnamefont {Y.}~\bibnamefont {Wang}},\
  and\ \bibinfo {author} {\bibfnamefont {W.~A.}\ \bibnamefont {Chupka}},\
  }\bibfield  {title} {\bibinfo {title} {(2+1) resonance-enhanced multiphoton
  ionization-photoelectron spectroscopy of the {OH} radical},\ }\href
  {https://doi.org/10.1063/1.460150} {\bibfield  {journal} {\bibinfo  {journal}
  {The Journal of Chemical Physics}\ }\textbf {\bibinfo {volume} {94}},\
  \bibinfo {pages} {7634} (\bibinfo {year} {1991})}\BibitemShut {NoStop}%
\bibitem [{\citenamefont {Streibel}\ and\ \citenamefont
  {Zimmermann}(2014)}]{2014_Streibel_REMPI_TOFMS_review}%
  \BibitemOpen
  \bibfield  {author} {\bibinfo {author} {\bibfnamefont {T.}~\bibnamefont
  {Streibel}}\ and\ \bibinfo {author} {\bibfnamefont {R.}~\bibnamefont
  {Zimmermann}},\ }\bibfield  {title} {\bibinfo {title} {Resonance-enhanced
  multiphoton ionization mass spectrometry ({REMPI-MS}): {A}pplications for
  process analysis},\ }\href
  {https://doi.org/10.1146/annurev-anchem-062012-092648} {\bibfield  {journal}
  {\bibinfo  {journal} {Annual Review of Analytical Chemistry}\ }\textbf
  {\bibinfo {volume} {7}},\ \bibinfo {pages} {361} (\bibinfo {year} {2014})},\
  \bibinfo {note} {pMID: 25014345}\BibitemShut {NoStop}%
\bibitem [{\citenamefont {Zhang}\ \emph {et~al.}(2021)\citenamefont {Zhang},
  \citenamefont {Shneider},\ and\ \citenamefont
  {Miles}}]{2021_Zhang_REMPI_review_microwave_scattering}%
  \BibitemOpen
  \bibfield  {author} {\bibinfo {author} {\bibfnamefont {Z.}~\bibnamefont
  {Zhang}}, \bibinfo {author} {\bibfnamefont {M.~N.}\ \bibnamefont
  {Shneider}},\ and\ \bibinfo {author} {\bibfnamefont {R.~B.}\ \bibnamefont
  {Miles}},\ }\bibfield  {title} {\bibinfo {title} {Coherent microwave
  scattering from resonance enhanced multi-photon ionization (radar {REMPI}): a
  review},\ }\href {https://doi.org/10.1088/1361-6595/ac2350} {\bibfield
  {journal} {\bibinfo  {journal} {Plasma Sources Science and Technology}\
  }\textbf {\bibinfo {volume} {30}},\ \bibinfo {pages} {103001} (\bibinfo
  {year} {2021})}\BibitemShut {NoStop}%
\bibitem [{\citenamefont {Levshakov}\ \emph {et~al.}(2011)\citenamefont
  {Levshakov}, \citenamefont {Kozlov},\ and\ \citenamefont
  {Reimers}}]{2011_Levshakov_CH3_fundamental_constants}%
  \BibitemOpen
  \bibfield  {author} {\bibinfo {author} {\bibfnamefont {S.~A.}\ \bibnamefont
  {Levshakov}}, \bibinfo {author} {\bibfnamefont {M.~G.}\ \bibnamefont
  {Kozlov}},\ and\ \bibinfo {author} {\bibfnamefont {D.}~\bibnamefont
  {Reimers}},\ }\bibfield  {title} {\bibinfo {title} {Methanol as a tracer of
  fundamental constants},\ }\href {https://doi.org/10.1088/0004-637x/738/1/26}
  {\bibfield  {journal} {\bibinfo  {journal} {The Astrophysical Journal}\
  }\textbf {\bibinfo {volume} {738}},\ \bibinfo {pages} {26} (\bibinfo {year}
  {2011})}\BibitemShut {NoStop}%
\bibitem [{\citenamefont {Yang}\ \emph {et~al.}(2019)\citenamefont {Yang},
  \citenamefont {Huang}, \citenamefont {Xiao}, \citenamefont {Chen},
  \citenamefont {Wang}, \citenamefont {Dai}, \citenamefont {Lique},
  \citenamefont {Alexander}, \citenamefont {Sun}, \citenamefont {Zhang},
  \citenamefont {Yang},\ and\ \citenamefont
  {Neumark}}]{2019_Yang_Cold_Interstallar_Clouds}%
  \BibitemOpen
  \bibfield  {author} {\bibinfo {author} {\bibfnamefont {T.}~\bibnamefont
  {Yang}}, \bibinfo {author} {\bibfnamefont {L.}~\bibnamefont {Huang}},
  \bibinfo {author} {\bibfnamefont {C.}~\bibnamefont {Xiao}}, \bibinfo {author}
  {\bibfnamefont {J.}~\bibnamefont {Chen}}, \bibinfo {author} {\bibfnamefont
  {T.}~\bibnamefont {Wang}}, \bibinfo {author} {\bibfnamefont {D.}~\bibnamefont
  {Dai}}, \bibinfo {author} {\bibfnamefont {F.}~\bibnamefont {Lique}}, \bibinfo
  {author} {\bibfnamefont {M.~H.}\ \bibnamefont {Alexander}}, \bibinfo {author}
  {\bibfnamefont {Z.}~\bibnamefont {Sun}}, \bibinfo {author} {\bibfnamefont
  {D.~H.}\ \bibnamefont {Zhang}}, \bibinfo {author} {\bibfnamefont
  {X.}~\bibnamefont {Yang}},\ and\ \bibinfo {author} {\bibfnamefont {D.~M.}\
  \bibnamefont {Neumark}},\ }\bibfield  {title} {\bibinfo {title} {Enhanced
  reactivity of fluorine with \textit{para}-hydrogen in cold interstellar
  clouds by resonance-induced quantum tunnelling},\ }\href
  {https://doi.org/10.1038/s41557-019-0280-3} {\bibfield  {journal} {\bibinfo
  {journal} {Nature Chemistry}\ }\textbf {\bibinfo {volume} {11}},\ \bibinfo
  {pages} {744} (\bibinfo {year} {2019})}\BibitemShut {NoStop}%
\bibitem [{\citenamefont {Shannon}\ \emph {et~al.}(2013)\citenamefont
  {Shannon}, \citenamefont {Blitz}, \citenamefont {Goddard},\ and\
  \citenamefont {Heard}}]{2013_Shannon_OH_CH3OH_Cold_Chemistry}%
  \BibitemOpen
  \bibfield  {author} {\bibinfo {author} {\bibfnamefont {R.~J.}\ \bibnamefont
  {Shannon}}, \bibinfo {author} {\bibfnamefont {M.~A.}\ \bibnamefont {Blitz}},
  \bibinfo {author} {\bibfnamefont {A.}~\bibnamefont {Goddard}},\ and\ \bibinfo
  {author} {\bibfnamefont {D.~E.}\ \bibnamefont {Heard}},\ }\bibfield  {title}
  {\bibinfo {title} {Accelerated chemistry in the reaction between the hydroxyl
  radical and methanol at interstellar temperatures facilitated by
  tunnelling},\ }\href {https://doi.org/10.1038/nchem.1692} {\bibfield
  {journal} {\bibinfo  {journal} {Nature Chemistry}\ }\textbf {\bibinfo
  {volume} {5}},\ \bibinfo {pages} {745} (\bibinfo {year} {2013})}\BibitemShut
  {NoStop}%
\bibitem [{\citenamefont {Christianen}\ \emph
  {et~al.}(2019{\natexlab{a}})\citenamefont {Christianen}, \citenamefont
  {Karman},\ and\ \citenamefont
  {Groenenboom}}]{2019_Christianen_Karman_RRKM_DOS_Calculation}%
  \BibitemOpen
  \bibfield  {author} {\bibinfo {author} {\bibfnamefont {A.}~\bibnamefont
  {Christianen}}, \bibinfo {author} {\bibfnamefont {T.}~\bibnamefont
  {Karman}},\ and\ \bibinfo {author} {\bibfnamefont {G.~C.}\ \bibnamefont
  {Groenenboom}},\ }\bibfield  {title} {\bibinfo {title} {Quasiclassical method
  for calculating the density of states of ultracold collision complexes},\
  }\href {https://doi.org/10.1103/PhysRevA.100.032708} {\bibfield  {journal}
  {\bibinfo  {journal} {Physical Review A}\ }\textbf {\bibinfo {volume}
  {100}},\ \bibinfo {pages} {032708} (\bibinfo {year}
  {2019}{\natexlab{a}})}\BibitemShut {NoStop}%
\bibitem [{\citenamefont {Snyder}\ \emph {et~al.}(2005)\citenamefont {Snyder},
  \citenamefont {Lovas}, \citenamefont {Hollis}, \citenamefont {Friedel},
  \citenamefont {Jewell}, \citenamefont {Remijan}, \citenamefont {Ilyushin},
  \citenamefont {Alekseev},\ and\ \citenamefont
  {Dyubko}}]{2005_Snyder_Interstellar_Glycine_Controversy}%
  \BibitemOpen
  \bibfield  {author} {\bibinfo {author} {\bibfnamefont {L.~E.}\ \bibnamefont
  {Snyder}}, \bibinfo {author} {\bibfnamefont {F.~J.}\ \bibnamefont {Lovas}},
  \bibinfo {author} {\bibfnamefont {J.~M.}\ \bibnamefont {Hollis}}, \bibinfo
  {author} {\bibfnamefont {D.~N.}\ \bibnamefont {Friedel}}, \bibinfo {author}
  {\bibfnamefont {P.~R.}\ \bibnamefont {Jewell}}, \bibinfo {author}
  {\bibfnamefont {A.}~\bibnamefont {Remijan}}, \bibinfo {author} {\bibfnamefont
  {V.~V.}\ \bibnamefont {Ilyushin}}, \bibinfo {author} {\bibfnamefont {E.~A.}\
  \bibnamefont {Alekseev}},\ and\ \bibinfo {author} {\bibfnamefont {S.~F.}\
  \bibnamefont {Dyubko}},\ }\bibfield  {title} {\bibinfo {title} {A rigorous
  attempt to verify interstellar glycine},\ }\href
  {https://doi.org/10.1086/426677} {\bibfield  {journal} {\bibinfo  {journal}
  {The Astrophysical Journal}\ }\textbf {\bibinfo {volume} {619}},\ \bibinfo
  {pages} {914} (\bibinfo {year} {2005})}\BibitemShut {NoStop}%
\bibitem [{\citenamefont
  {Tennyson}(2016)}]{2016_Tennyson_Comp_Chem_Rovibrational_Calculations_Better_Than_Measurement}%
  \BibitemOpen
  \bibfield  {author} {\bibinfo {author} {\bibfnamefont {J.}~\bibnamefont
  {Tennyson}},\ }\bibfield  {title} {\bibinfo {title} {Perspective: {A}ccurate
  ro-vibrational calculations on small molecules},\ }\href
  {https://doi.org/10.1063/1.4962907} {\bibfield  {journal} {\bibinfo
  {journal} {The Journal of Chemical Physics}\ }\textbf {\bibinfo {volume}
  {145}},\ \bibinfo {pages} {120901} (\bibinfo {year} {2016})}\BibitemShut
  {NoStop}%
\bibitem [{\citenamefont {Beyer}\ \emph {et~al.}(2017)\citenamefont {Beyer},
  \citenamefont {Maisenbacher}, \citenamefont {Matveev}, \citenamefont {Pohl},
  \citenamefont {Khabarova}, \citenamefont {Grinin}, \citenamefont {Lamour},
  \citenamefont {Yost}, \citenamefont {Hänsch}, \citenamefont {Kolachevsky},\
  and\ \citenamefont {Udem}}]{Beyer2017}%
  \BibitemOpen
  \bibfield  {author} {\bibinfo {author} {\bibfnamefont {A.}~\bibnamefont
  {Beyer}}, \bibinfo {author} {\bibfnamefont {L.}~\bibnamefont {Maisenbacher}},
  \bibinfo {author} {\bibfnamefont {A.}~\bibnamefont {Matveev}}, \bibinfo
  {author} {\bibfnamefont {R.}~\bibnamefont {Pohl}}, \bibinfo {author}
  {\bibfnamefont {K.}~\bibnamefont {Khabarova}}, \bibinfo {author}
  {\bibfnamefont {A.}~\bibnamefont {Grinin}}, \bibinfo {author} {\bibfnamefont
  {T.}~\bibnamefont {Lamour}}, \bibinfo {author} {\bibfnamefont {D.~C.}\
  \bibnamefont {Yost}}, \bibinfo {author} {\bibfnamefont {T.~W.}\ \bibnamefont
  {Hänsch}}, \bibinfo {author} {\bibfnamefont {N.}~\bibnamefont
  {Kolachevsky}},\ and\ \bibinfo {author} {\bibfnamefont {T.}~\bibnamefont
  {Udem}},\ }\bibfield  {title} {\bibinfo {title} {The {{Rydberg}} constant and
  proton size from atomic hydrogen},\ }\href
  {https://doi.org/10.1126/science.aah6677} {\bibfield  {journal} {\bibinfo
  {journal} {Science}\ }\textbf {\bibinfo {volume} {358}},\ \bibinfo {pages}
  {79} (\bibinfo {year} {2017})}\BibitemShut {NoStop}%
\bibitem [{\citenamefont {Grinin}\ \emph {et~al.}(2020)\citenamefont {Grinin},
  \citenamefont {Matveev}, \citenamefont {Yost}, \citenamefont {Maisenbacher},
  \citenamefont {Wirthl}, \citenamefont {Pohl}, \citenamefont {Hänsch},\ and\
  \citenamefont {Udem}}]{Grinin2020}%
  \BibitemOpen
  \bibfield  {author} {\bibinfo {author} {\bibfnamefont {A.}~\bibnamefont
  {Grinin}}, \bibinfo {author} {\bibfnamefont {A.}~\bibnamefont {Matveev}},
  \bibinfo {author} {\bibfnamefont {D.~C.}\ \bibnamefont {Yost}}, \bibinfo
  {author} {\bibfnamefont {L.}~\bibnamefont {Maisenbacher}}, \bibinfo {author}
  {\bibfnamefont {V.}~\bibnamefont {Wirthl}}, \bibinfo {author} {\bibfnamefont
  {R.}~\bibnamefont {Pohl}}, \bibinfo {author} {\bibfnamefont {T.~W.}\
  \bibnamefont {Hänsch}},\ and\ \bibinfo {author} {\bibfnamefont
  {T.}~\bibnamefont {Udem}},\ }\bibfield  {title} {\bibinfo {title} {Two-photon
  frequency comb spectroscopy of atomic hydrogen},\ }\href
  {https://doi.org/10.1126/science.abc7776} {\bibfield  {journal} {\bibinfo
  {journal} {Science}\ }\textbf {\bibinfo {volume} {370}},\ \bibinfo {pages}
  {1061} (\bibinfo {year} {2020})}\BibitemShut {NoStop}%
\bibitem [{\citenamefont {Ye}\ \emph {et~al.}(2008)\citenamefont {Ye},
  \citenamefont {Kimble},\ and\ \citenamefont
  {Katori}}]{2008_Ye_Magic_Wavelengths}%
  \BibitemOpen
  \bibfield  {author} {\bibinfo {author} {\bibfnamefont {J.}~\bibnamefont
  {Ye}}, \bibinfo {author} {\bibfnamefont {H.~J.}\ \bibnamefont {Kimble}},\
  and\ \bibinfo {author} {\bibfnamefont {H.}~\bibnamefont {Katori}},\
  }\bibfield  {title} {\bibinfo {title} {Quantum state engineering and
  precision metrology using state-insensitive light traps},\ }\href
  {https://doi.org/10.1126/science.1148259} {\bibfield  {journal} {\bibinfo
  {journal} {Science}\ }\textbf {\bibinfo {volume} {320}},\ \bibinfo {pages}
  {1734} (\bibinfo {year} {2008})}\BibitemShut {NoStop}%
\bibitem [{\citenamefont {Kondov}\ \emph {et~al.}(2019)\citenamefont {Kondov},
  \citenamefont {Lee}, \citenamefont {Leung}, \citenamefont {Liedl},
  \citenamefont {Majewska}, \citenamefont {Moszynski},\ and\ \citenamefont
  {Zelevinsky}}]{2019_Kondov_Magic_Wavelength}%
  \BibitemOpen
  \bibfield  {author} {\bibinfo {author} {\bibfnamefont {S.~S.}\ \bibnamefont
  {Kondov}}, \bibinfo {author} {\bibfnamefont {C.-H.}\ \bibnamefont {Lee}},
  \bibinfo {author} {\bibfnamefont {K.~H.}\ \bibnamefont {Leung}}, \bibinfo
  {author} {\bibfnamefont {C.}~\bibnamefont {Liedl}}, \bibinfo {author}
  {\bibfnamefont {I.}~\bibnamefont {Majewska}}, \bibinfo {author}
  {\bibfnamefont {R.}~\bibnamefont {Moszynski}},\ and\ \bibinfo {author}
  {\bibfnamefont {T.}~\bibnamefont {Zelevinsky}},\ }\bibfield  {title}
  {\bibinfo {title} {Molecular lattice clock with long vibrational coherence},\
  }\href {https://doi.org/10.1038/s41567-019-0632-3} {\bibfield  {journal}
  {\bibinfo  {journal} {Nature Physics}\ }\textbf {\bibinfo {volume} {15}},\
  \bibinfo {pages} {1118} (\bibinfo {year} {2019})}\BibitemShut {NoStop}%
\bibitem [{\citenamefont {Leung}\ \emph {et~al.}(2023)\citenamefont {Leung},
  \citenamefont {Iritani}, \citenamefont {Tiberi}, \citenamefont {Majewska},
  \citenamefont {Borkowski}, \citenamefont {Moszynski},\ and\ \citenamefont
  {Zelevinsky}}]{2023_Leung_Zelevinski_Sr2_Clock_Magic_wavelength}%
  \BibitemOpen
  \bibfield  {author} {\bibinfo {author} {\bibfnamefont {K.~H.}\ \bibnamefont
  {Leung}}, \bibinfo {author} {\bibfnamefont {B.}~\bibnamefont {Iritani}},
  \bibinfo {author} {\bibfnamefont {E.}~\bibnamefont {Tiberi}}, \bibinfo
  {author} {\bibfnamefont {I.}~\bibnamefont {Majewska}}, \bibinfo {author}
  {\bibfnamefont {M.}~\bibnamefont {Borkowski}}, \bibinfo {author}
  {\bibfnamefont {R.}~\bibnamefont {Moszynski}},\ and\ \bibinfo {author}
  {\bibfnamefont {T.}~\bibnamefont {Zelevinsky}},\ }\bibfield  {title}
  {\bibinfo {title} {Terahertz vibrational molecular clock with systematic
  uncertainty at the ${10}^{\ensuremath{-}14}$ level},\ }\href
  {https://doi.org/10.1103/PhysRevX.13.011047} {\bibfield  {journal} {\bibinfo
  {journal} {Phys. Rev. X}\ }\textbf {\bibinfo {volume} {13}},\ \bibinfo
  {pages} {011047} (\bibinfo {year} {2023})}\BibitemShut {NoStop}%
\bibitem [{\citenamefont {J\'{o}\'{z}wiak}\ and\ \citenamefont
  {Wcis\l{}o}(2022)}]{2022_Jozwiak_Magic_Wavelength_Rovibrational_H2}%
  \BibitemOpen
  \bibfield  {author} {\bibinfo {author} {\bibfnamefont {H.}~\bibnamefont
  {J\'{o}\'{z}wiak}}\ and\ \bibinfo {author} {\bibfnamefont {P.}~\bibnamefont
  {Wcis\l{}o}},\ }\bibfield  {title} {\bibinfo {title} {Magic wavelength for a
  rovibrational transition in molecular hydrogen},\ }\href
  {https://doi.org/10.1038/s41598-022-18159-y} {\bibfield  {journal} {\bibinfo
  {journal} {Scientific Reports}\ }\textbf {\bibinfo {volume} {12}},\ \bibinfo
  {pages} {14529} (\bibinfo {year} {2022})}\BibitemShut {NoStop}%
\bibitem [{\citenamefont {Boyd}(2020)}]{2008_Boyd_Nonlinear_Optics}%
  \BibitemOpen
  \bibfield  {author} {\bibinfo {author} {\bibfnamefont {R.}~\bibnamefont
  {Boyd}},\ }\href@noop {} {\emph {\bibinfo {title} {Nonlinear Optics}}},\
  \bibinfo {edition} {4th}\ ed.\ (\bibinfo  {publisher} {Academic Press},\
  \bibinfo {address} {London},\ \bibinfo {year} {2020})\BibitemShut {NoStop}%
\bibitem [{\citenamefont {Abramowitz}\ and\ \citenamefont
  {Stegun}(1964)}]{1965_Abramowitz_Stegun}%
  \BibitemOpen
  \bibfield  {author} {\bibinfo {author} {\bibfnamefont {M.}~\bibnamefont
  {Abramowitz}}\ and\ \bibinfo {author} {\bibfnamefont {I.~A.}\ \bibnamefont
  {Stegun}},\ }\href@noop {} {\emph {\bibinfo {title} {Handbook of mathematical
  functions, with formulas, graphs, and mathematical tables}}},\ Dover {B}ooks
  on {I}ntermediate and {A}dvanced {M}athematics\ (\bibinfo  {publisher} {Dover
  Publications},\ \bibinfo {address} {New York},\ \bibinfo {year}
  {1964})\BibitemShut {NoStop}%
\bibitem [{\citenamefont {Virtanen}\ \emph {et~al.}(2020)\citenamefont
  {Virtanen}, \citenamefont {Gommers}, \citenamefont {Oliphant}, \citenamefont
  {Haberland}, \citenamefont {Reddy}, \citenamefont {Cournapeau}, \citenamefont
  {Burovski}, \citenamefont {Peterson}, \citenamefont {Weckesser},
  \citenamefont {Bright}, \citenamefont {{van der Walt}}, \citenamefont
  {Brett}, \citenamefont {Wilson}, \citenamefont {Millman}, \citenamefont
  {Mayorov}, \citenamefont {Nelson}, \citenamefont {Jones}, \citenamefont
  {Kern}, \citenamefont {Larson}, \citenamefont {Carey}, \citenamefont {Polat},
  \citenamefont {Feng}, \citenamefont {Moore}, \citenamefont {{VanderPlas}},
  \citenamefont {Laxalde}, \citenamefont {Perktold}, \citenamefont {Cimrman},
  \citenamefont {Henriksen}, \citenamefont {Quintero}, \citenamefont {Harris},
  \citenamefont {Archibald}, \citenamefont {Ribeiro}, \citenamefont
  {Pedregosa}, \citenamefont {{van Mulbregt}},\ and\ \citenamefont {{SciPy 1.0
  Contributors}}}]{Scipy}%
  \BibitemOpen
  \bibfield  {author} {\bibinfo {author} {\bibfnamefont {P.}~\bibnamefont
  {Virtanen}}, \bibinfo {author} {\bibfnamefont {R.}~\bibnamefont {Gommers}},
  \bibinfo {author} {\bibfnamefont {T.~E.}\ \bibnamefont {Oliphant}}, \bibinfo
  {author} {\bibfnamefont {M.}~\bibnamefont {Haberland}}, \bibinfo {author}
  {\bibfnamefont {T.}~\bibnamefont {Reddy}}, \bibinfo {author} {\bibfnamefont
  {D.}~\bibnamefont {Cournapeau}}, \bibinfo {author} {\bibfnamefont
  {E.}~\bibnamefont {Burovski}}, \bibinfo {author} {\bibfnamefont
  {P.}~\bibnamefont {Peterson}}, \bibinfo {author} {\bibfnamefont
  {W.}~\bibnamefont {Weckesser}}, \bibinfo {author} {\bibfnamefont
  {J.}~\bibnamefont {Bright}}, \bibinfo {author} {\bibfnamefont {S.~J.}\
  \bibnamefont {{van der Walt}}}, \bibinfo {author} {\bibfnamefont
  {M.}~\bibnamefont {Brett}}, \bibinfo {author} {\bibfnamefont
  {J.}~\bibnamefont {Wilson}}, \bibinfo {author} {\bibfnamefont {K.~J.}\
  \bibnamefont {Millman}}, \bibinfo {author} {\bibfnamefont {N.}~\bibnamefont
  {Mayorov}}, \bibinfo {author} {\bibfnamefont {A.~R.~J.}\ \bibnamefont
  {Nelson}}, \bibinfo {author} {\bibfnamefont {E.}~\bibnamefont {Jones}},
  \bibinfo {author} {\bibfnamefont {R.}~\bibnamefont {Kern}}, \bibinfo {author}
  {\bibfnamefont {E.}~\bibnamefont {Larson}}, \bibinfo {author} {\bibfnamefont
  {C.~J.}\ \bibnamefont {Carey}}, \bibinfo {author} {\bibfnamefont
  {{\.I}.}~\bibnamefont {Polat}}, \bibinfo {author} {\bibfnamefont
  {Y.}~\bibnamefont {Feng}}, \bibinfo {author} {\bibfnamefont {E.~W.}\
  \bibnamefont {Moore}}, \bibinfo {author} {\bibfnamefont {J.}~\bibnamefont
  {{VanderPlas}}}, \bibinfo {author} {\bibfnamefont {D.}~\bibnamefont
  {Laxalde}}, \bibinfo {author} {\bibfnamefont {J.}~\bibnamefont {Perktold}},
  \bibinfo {author} {\bibfnamefont {R.}~\bibnamefont {Cimrman}}, \bibinfo
  {author} {\bibfnamefont {I.}~\bibnamefont {Henriksen}}, \bibinfo {author}
  {\bibfnamefont {E.~A.}\ \bibnamefont {Quintero}}, \bibinfo {author}
  {\bibfnamefont {C.~R.}\ \bibnamefont {Harris}}, \bibinfo {author}
  {\bibfnamefont {A.~M.}\ \bibnamefont {Archibald}}, \bibinfo {author}
  {\bibfnamefont {A.~H.}\ \bibnamefont {Ribeiro}}, \bibinfo {author}
  {\bibfnamefont {F.}~\bibnamefont {Pedregosa}}, \bibinfo {author}
  {\bibfnamefont {P.}~\bibnamefont {{van Mulbregt}}},\ and\ \bibinfo {author}
  {\bibnamefont {{SciPy 1.0 Contributors}}},\ }\bibfield  {title} {\bibinfo
  {title} {{{SciPy} 1.0: {F}undamental Algorithms for Scientific Computing in
  Python}},\ }\href {https://doi.org/10.1038/s41592-019-0686-2} {\bibfield
  {journal} {\bibinfo  {journal} {Nature Methods}\ }\textbf {\bibinfo {volume}
  {17}},\ \bibinfo {pages} {261} (\bibinfo {year} {2020})}\BibitemShut
  {NoStop}%
\bibitem [{\citenamefont {Archibong}\ and\ \citenamefont
  {Thakkar}(1994)}]{1994_Archibong_hyperpolarizability_N2}%
  \BibitemOpen
  \bibfield  {author} {\bibinfo {author} {\bibfnamefont {E.~F.}\ \bibnamefont
  {Archibong}}\ and\ \bibinfo {author} {\bibfnamefont {A.~J.}\ \bibnamefont
  {Thakkar}},\ }\bibfield  {title} {\bibinfo {title} {{Static
  hyperpolarizability of N$_2$}},\ }\href {https://doi.org/10.1063/1.466890}
  {\bibfield  {journal} {\bibinfo  {journal} {The Journal of Chemical Physics}\
  }\textbf {\bibinfo {volume} {100}},\ \bibinfo {pages} {7471} (\bibinfo {year}
  {1994})}\BibitemShut {NoStop}%
\bibitem [{\citenamefont {Luo}\ \emph {et~al.}(1995)\citenamefont {Luo},
  \citenamefont {\r{A}gren}, \citenamefont {Minaev},\ and\ \citenamefont
  {J{\o}rgensen}}]{1995_Luo_hyperpolarizability_O2}%
  \BibitemOpen
  \bibfield  {author} {\bibinfo {author} {\bibfnamefont {Y.}~\bibnamefont
  {Luo}}, \bibinfo {author} {\bibfnamefont {H.}~\bibnamefont {\r{A}gren}},
  \bibinfo {author} {\bibfnamefont {B.}~\bibnamefont {Minaev}},\ and\ \bibinfo
  {author} {\bibfnamefont {P.}~\bibnamefont {J{\o}rgensen}},\ }\bibfield
  {title} {\bibinfo {title} {The hyperpolarizability of molecular oxygen},\
  }\href {https://doi.org/https://doi.org/10.1016/0166-1280(94)04095-A}
  {\bibfield  {journal} {\bibinfo  {journal} {Journal of Molecular Structure:
  THEOCHEM}\ }\textbf {\bibinfo {volume} {336}},\ \bibinfo {pages} {61}
  (\bibinfo {year} {1995})}\BibitemShut {NoStop}%
\bibitem [{\citenamefont
  {Maroulis}(1996)}]{1996_Maroulis_hyperpolarizability_CO}%
  \BibitemOpen
  \bibfield  {author} {\bibinfo {author} {\bibfnamefont {G.}~\bibnamefont
  {Maroulis}},\ }\bibfield  {title} {\bibinfo {title} {Electric polarizability
  and hyperpolarizability of carbon monoxide},\ }\href
  {https://doi.org/10.1021/jp960412n} {\bibfield  {journal} {\bibinfo
  {journal} {The Journal of Physical Chemistry}\ }\textbf {\bibinfo {volume}
  {100}},\ \bibinfo {pages} {13466} (\bibinfo {year} {1996})}\BibitemShut
  {NoStop}%
\bibitem [{\citenamefont {Fern\'{a}ndez}\ \emph {et~al.}(1998)\citenamefont
  {Fern\'{a}ndez}, \citenamefont {Coriani},\ and\ \citenamefont
  {Rizzo}}]{1998_Fernandez_hyperpolarizability_HCl_HBr}%
  \BibitemOpen
  \bibfield  {author} {\bibinfo {author} {\bibfnamefont {B.}~\bibnamefont
  {Fern\'{a}ndez}}, \bibinfo {author} {\bibfnamefont {S.}~\bibnamefont
  {Coriani}},\ and\ \bibinfo {author} {\bibfnamefont {A.}~\bibnamefont
  {Rizzo}},\ }\bibfield  {title} {\bibinfo {title} {{MCSCF} polarizability and
  hyperpolarizabilities of {HCl} and {HBr}},\ }\href
  {https://doi.org/https://doi.org/10.1016/S0009-2614(98)00355-8} {\bibfield
  {journal} {\bibinfo  {journal} {Chemical Physics Letters}\ }\textbf {\bibinfo
  {volume} {288}},\ \bibinfo {pages} {677} (\bibinfo {year}
  {1998})}\BibitemShut {NoStop}%
\bibitem [{\citenamefont {Fernandez}\ and\ \citenamefont
  {Shelton}(2020)}]{2020_Fernandez_CS2_hyperpolarizability}%
  \BibitemOpen
  \bibfield  {author} {\bibinfo {author} {\bibfnamefont {R.~N.}\ \bibnamefont
  {Fernandez}}\ and\ \bibinfo {author} {\bibfnamefont {D.~P.}\ \bibnamefont
  {Shelton}},\ }\bibfield  {title} {\bibinfo {title} {Hyperpolarizability
  dispersion measured for cs2 vapor},\ }\href
  {https://doi.org/10.1364/JOSAB.394315} {\bibfield  {journal} {\bibinfo
  {journal} {J. Opt. Soc. Am. B}\ }\textbf {\bibinfo {volume} {37}},\ \bibinfo
  {pages} {1769} (\bibinfo {year} {2020})}\BibitemShut {NoStop}%
\bibitem [{\citenamefont {Maroulis}\ and\ \citenamefont
  {Thakkar}(1988)}]{1988_Maroulis_higher_order_multipole_polarizabilities_N2}%
  \BibitemOpen
  \bibfield  {author} {\bibinfo {author} {\bibfnamefont {G.}~\bibnamefont
  {Maroulis}}\ and\ \bibinfo {author} {\bibfnamefont {A.~J.}\ \bibnamefont
  {Thakkar}},\ }\bibfield  {title} {\bibinfo {title} {{Multipole moments,
  polarizabilities, and hyperpolarizabilities for N$_2$ from fourth‐order
  many‐body perturbation theory calculations}},\ }\href
  {https://doi.org/10.1063/1.454327} {\bibfield  {journal} {\bibinfo  {journal}
  {The Journal of Chemical Physics}\ }\textbf {\bibinfo {volume} {88}},\
  \bibinfo {pages} {7623} (\bibinfo {year} {1988})}\BibitemShut {NoStop}%
\bibitem [{\citenamefont {Davies}(1954)}]{1953_Davies_Surface_Roughness}%
  \BibitemOpen
  \bibfield  {author} {\bibinfo {author} {\bibfnamefont {H.}~\bibnamefont
  {Davies}},\ }\bibfield  {title} {\bibinfo {title} {The reflection of
  electromagnetic waves from a rough surface},\ }\href
  {https://digital-library.theiet.org/content/journals/10.1049/pi-4.1954.0025}
  {\bibfield  {journal} {\bibinfo  {journal} {Proceedings of the IEE - Part IV:
  Institution Monographs}\ }\textbf {\bibinfo {volume} {101}},\ \bibinfo
  {pages} {209} (\bibinfo {year} {1954})}\BibitemShut {NoStop}%
\bibitem [{\citenamefont {Bennett}\ and\ \citenamefont
  {Porteus}(1961)}]{1961_Bennet_Surface_Roughness}%
  \BibitemOpen
  \bibfield  {author} {\bibinfo {author} {\bibfnamefont {H.~E.}\ \bibnamefont
  {Bennett}}\ and\ \bibinfo {author} {\bibfnamefont {J.~O.}\ \bibnamefont
  {Porteus}},\ }\bibfield  {title} {\bibinfo {title} {Relation between surface
  roughness and specular reflectance at normal incidence},\ }\href
  {https://doi.org/10.1364/JOSA.51.000123} {\bibfield  {journal} {\bibinfo
  {journal} {Journal of the Optical Society of America}\ }\textbf {\bibinfo
  {volume} {51}},\ \bibinfo {pages} {123} (\bibinfo {year} {1961})}\BibitemShut
  {NoStop}%
\bibitem [{\citenamefont {Steyerl}\ \emph {et~al.}(1991)\citenamefont
  {Steyerl}, \citenamefont {Malik},\ and\ \citenamefont
  {Iyengar}}]{1991_Steyerl_surface_roughness}%
  \BibitemOpen
  \bibfield  {author} {\bibinfo {author} {\bibfnamefont {A.}~\bibnamefont
  {Steyerl}}, \bibinfo {author} {\bibfnamefont {S.}~\bibnamefont {Malik}},\
  and\ \bibinfo {author} {\bibfnamefont {L.}~\bibnamefont {Iyengar}},\
  }\bibfield  {title} {\bibinfo {title} {Specular and diffuse reflection and
  refraction at surfaces},\ }\href
  {https://doi.org/https://doi.org/10.1016/0921-4526(91)90034-C} {\bibfield
  {journal} {\bibinfo  {journal} {Physica B: Condensed Matter}\ }\textbf
  {\bibinfo {volume} {173}},\ \bibinfo {pages} {47} (\bibinfo {year}
  {1991})}\BibitemShut {NoStop}%
\bibitem [{\citenamefont {Savard}\ \emph {et~al.}(1997)\citenamefont {Savard},
  \citenamefont {O'Hara},\ and\ \citenamefont
  {Thomas}}]{1997_Savard_heating_from_laser_noise}%
  \BibitemOpen
  \bibfield  {author} {\bibinfo {author} {\bibfnamefont {T.~A.}\ \bibnamefont
  {Savard}}, \bibinfo {author} {\bibfnamefont {K.~M.}\ \bibnamefont {O'Hara}},\
  and\ \bibinfo {author} {\bibfnamefont {J.~E.}\ \bibnamefont {Thomas}},\
  }\bibfield  {title} {\bibinfo {title} {Laser-noise-induced heating in far-off
  resonance optical traps},\ }\href {https://doi.org/10.1103/PhysRevA.56.R1095}
  {\bibfield  {journal} {\bibinfo  {journal} {Physical Review A}\ }\textbf
  {\bibinfo {volume} {56}},\ \bibinfo {pages} {R1095} (\bibinfo {year}
  {1997})}\BibitemShut {NoStop}%
\bibitem [{\citenamefont {Gardiner}\ \emph {et~al.}(2000)\citenamefont
  {Gardiner}, \citenamefont {Ye}, \citenamefont {Nagerl},\ and\ \citenamefont
  {Kimble}}]{2000_Gardiner_Savard_analysis_worse}%
  \BibitemOpen
  \bibfield  {author} {\bibinfo {author} {\bibfnamefont {C.~W.}\ \bibnamefont
  {Gardiner}}, \bibinfo {author} {\bibfnamefont {J.}~\bibnamefont {Ye}},
  \bibinfo {author} {\bibfnamefont {H.~C.}\ \bibnamefont {Nagerl}},\ and\
  \bibinfo {author} {\bibfnamefont {H.~J.}\ \bibnamefont {Kimble}},\ }\bibfield
   {title} {\bibinfo {title} {Evaluation of heating effects on atoms trapped in
  an optical trap},\ }\href {https://doi.org/10.1103/PhysRevA.61.045801}
  {\bibfield  {journal} {\bibinfo  {journal} {Physical Review A}\ }\textbf
  {\bibinfo {volume} {61}},\ \bibinfo {pages} {045801} (\bibinfo {year}
  {2000})}\BibitemShut {NoStop}%
\bibitem [{\citenamefont {Christianen}\ \emph
  {et~al.}(2019{\natexlab{b}})\citenamefont {Christianen}, \citenamefont
  {Zwierlein}, \citenamefont {Groenenboom},\ and\ \citenamefont
  {Karman}}]{2019_Christianen_Karman_Two-body_Loss_ab_initio_molpro}%
  \BibitemOpen
  \bibfield  {author} {\bibinfo {author} {\bibfnamefont {A.}~\bibnamefont
  {Christianen}}, \bibinfo {author} {\bibfnamefont {M.~W.}\ \bibnamefont
  {Zwierlein}}, \bibinfo {author} {\bibfnamefont {G.~C.}\ \bibnamefont
  {Groenenboom}},\ and\ \bibinfo {author} {\bibfnamefont {T.}~\bibnamefont
  {Karman}},\ }\bibfield  {title} {\bibinfo {title} {Photoinduced two-body loss
  of ultracold molecules},\ }\href
  {https://doi.org/10.1103/PhysRevLett.123.123402} {\bibfield  {journal}
  {\bibinfo  {journal} {Physical Review Letters}\ }\textbf {\bibinfo {volume}
  {123}},\ \bibinfo {pages} {123402} (\bibinfo {year}
  {2019}{\natexlab{b}})}\BibitemShut {NoStop}%
\bibitem [{\citenamefont {Hellmann}\ \emph {et~al.}(2008)\citenamefont
  {Hellmann}, \citenamefont {Bich},\ and\ \citenamefont
  {Vogel}}]{2008_Hellmann_CH4_CH4_interaction_potential}%
  \BibitemOpen
  \bibfield  {author} {\bibinfo {author} {\bibfnamefont {R.}~\bibnamefont
  {Hellmann}}, \bibinfo {author} {\bibfnamefont {E.}~\bibnamefont {Bich}},\
  and\ \bibinfo {author} {\bibfnamefont {E.}~\bibnamefont {Vogel}},\ }\bibfield
   {title} {\bibinfo {title} {\textit{Ab initio} intermolecular potential
  energy surface and second pressure virial coefficients of methane},\ }\href
  {https://doi.org/10.1063/1.2932103} {\bibfield  {journal} {\bibinfo
  {journal} {The Journal of Chemical Physics}\ }\textbf {\bibinfo {volume}
  {128}},\ \bibinfo {pages} {214303} (\bibinfo {year} {2008})}\BibitemShut
  {NoStop}%
\bibitem [{\citenamefont
  {Hellmann}(2013)}]{2013_Hellmann_N2_N2_interaction_potential}%
  \BibitemOpen
  \bibfield  {author} {\bibinfo {author} {\bibfnamefont {R.}~\bibnamefont
  {Hellmann}},\ }\bibfield  {title} {\bibinfo {title} {\textit{Ab initio}
  potential energy surface for the nitrogen molecule pair and thermophysical
  properties of nitrogen gas},\ }\href
  {https://doi.org/10.1080/00268976.2012.726379} {\bibfield  {journal}
  {\bibinfo  {journal} {Molecular Physics}\ }\textbf {\bibinfo {volume}
  {111}},\ \bibinfo {pages} {387} (\bibinfo {year} {2013})}\BibitemShut
  {NoStop}%
\bibitem [{\citenamefont {Ospelkaus}\ \emph {et~al.}(2010)\citenamefont
  {Ospelkaus}, \citenamefont {Ni}, \citenamefont {Wang}, \citenamefont
  {de~Miranda}, \citenamefont {Neyenhuis}, \citenamefont {Qu\'{e}m\'{e}ner},
  \citenamefont {Julienne}, \citenamefont {Bohn}, \citenamefont {Jin},\ and\
  \citenamefont {Ye}}]{2009_Ospelkaus_universal_loss}%
  \BibitemOpen
  \bibfield  {author} {\bibinfo {author} {\bibfnamefont {S.}~\bibnamefont
  {Ospelkaus}}, \bibinfo {author} {\bibfnamefont {K.}~\bibnamefont {Ni}},
  \bibinfo {author} {\bibfnamefont {D.}~\bibnamefont {Wang}}, \bibinfo {author}
  {\bibfnamefont {M.~H.~G.}\ \bibnamefont {de~Miranda}}, \bibinfo {author}
  {\bibfnamefont {B.}~\bibnamefont {Neyenhuis}}, \bibinfo {author}
  {\bibfnamefont {G.}~\bibnamefont {Qu\'{e}m\'{e}ner}}, \bibinfo {author}
  {\bibfnamefont {P.~S.}\ \bibnamefont {Julienne}}, \bibinfo {author}
  {\bibfnamefont {J.~L.}\ \bibnamefont {Bohn}}, \bibinfo {author}
  {\bibfnamefont {D.~S.}\ \bibnamefont {Jin}},\ and\ \bibinfo {author}
  {\bibfnamefont {J.}~\bibnamefont {Ye}},\ }\bibfield  {title} {\bibinfo
  {title} {Quantum-state controlled chemical reactions of ultracold
  potassium-rubidium molecules},\ }\href
  {https://doi.org/10.1126/science.1184121} {\bibfield  {journal} {\bibinfo
  {journal} {Science}\ }\textbf {\bibinfo {volume} {327}},\ \bibinfo {pages}
  {853} (\bibinfo {year} {2010})}\BibitemShut {NoStop}%
\bibitem [{\citenamefont
  {Julienne}(2009)}]{2009_Julienne_C6_def_and_inelastic_losses}%
  \BibitemOpen
  \bibfield  {author} {\bibinfo {author} {\bibfnamefont {P.~S.}\ \bibnamefont
  {Julienne}},\ }\bibfield  {title} {\bibinfo {title} {Ultracold molecules from
  ultracold atoms: a case study with the {KR}b molecule},\ }\href
  {https://doi.org/10.1039/B820917K} {\bibfield  {journal} {\bibinfo  {journal}
  {Faraday Discussions}\ }\textbf {\bibinfo {volume} {142}},\ \bibinfo {pages}
  {361} (\bibinfo {year} {2009})}\BibitemShut {NoStop}%
\bibitem [{\citenamefont {Tao}\ and\ \citenamefont
  {Rappe}(2016)}]{2016_Tao_C_6_coefficients_Simple_SFA_Model}%
  \BibitemOpen
  \bibfield  {author} {\bibinfo {author} {\bibfnamefont {J.}~\bibnamefont
  {Tao}}\ and\ \bibinfo {author} {\bibfnamefont {A.~M.}\ \bibnamefont
  {Rappe}},\ }\bibfield  {title} {\bibinfo {title} {Communication: {A}ccurate
  higher-order van der {W}aals coefficients between molecules from a model
  dynamic multipole polarizability},\ }\href
  {https://doi.org/10.1063/1.4940397} {\bibfield  {journal} {\bibinfo
  {journal} {The Journal of Chemical Physics}\ }\textbf {\bibinfo {volume}
  {144}},\ \bibinfo {pages} {031102} (\bibinfo {year} {2016})}\BibitemShut
  {NoStop}%
\bibitem [{\citenamefont
  {Frommhold}(1994)}]{1994_Frommhold_Collision_Induced_Absorption_in_Gases}%
  \BibitemOpen
  \bibfield  {author} {\bibinfo {author} {\bibfnamefont {L.}~\bibnamefont
  {Frommhold}},\ }\href {https://doi.org/10.1017/CBO9780511524523} {\emph
  {\bibinfo {title} {Collision-induced Absorption in Gases}}},\ Cambridge
  Monographs on Atomic, Molecular and Chemical Physics\ (\bibinfo  {publisher}
  {Cambridge University Press},\ \bibinfo {year} {1994})\BibitemShut {NoStop}%
\bibitem [{\citenamefont {Karman}\ \emph {et~al.}(2018)\citenamefont {Karman},
  \citenamefont {Koenis}, \citenamefont {Banerjee}, \citenamefont {Parker},
  \citenamefont {Gordon}, \citenamefont {van~der Avoird}, \citenamefont
  {van~der Zande},\ and\ \citenamefont
  {Groenenboom}}]{2018_Karman_O2_N2_collisional_absorption}%
  \BibitemOpen
  \bibfield  {author} {\bibinfo {author} {\bibfnamefont {T.}~\bibnamefont
  {Karman}}, \bibinfo {author} {\bibfnamefont {M.~A.~J.}\ \bibnamefont
  {Koenis}}, \bibinfo {author} {\bibfnamefont {A.}~\bibnamefont {Banerjee}},
  \bibinfo {author} {\bibfnamefont {D.~H.}\ \bibnamefont {Parker}}, \bibinfo
  {author} {\bibfnamefont {I.~E.}\ \bibnamefont {Gordon}}, \bibinfo {author}
  {\bibfnamefont {A.}~\bibnamefont {van~der Avoird}}, \bibinfo {author}
  {\bibfnamefont {W.~J.}\ \bibnamefont {van~der Zande}},\ and\ \bibinfo
  {author} {\bibfnamefont {G.~C.}\ \bibnamefont {Groenenboom}},\ }\bibfield
  {title} {\bibinfo {title} {{O}$_2$-{O}$_2$ and {O}$_2$-{N}$_2$
  collision-induced absorption mechanisms unravelled},\ }\href
  {https://doi.org/10.1038/s41557-018-0015-x} {\bibfield  {journal} {\bibinfo
  {journal} {Nature Chemistry}\ }\textbf {\bibinfo {volume} {10}},\ \bibinfo
  {pages} {549} (\bibinfo {year} {2018})}\BibitemShut {NoStop}%
\bibitem [{\citenamefont {Belikov}\ and\ \citenamefont
  {Sharafutdinov}(1995)}]{1995_Belikov_N2_He_Inelastic_Rate}%
  \BibitemOpen
  \bibfield  {author} {\bibinfo {author} {\bibfnamefont {A.~E.}\ \bibnamefont
  {Belikov}}\ and\ \bibinfo {author} {\bibfnamefont {R.~G.}\ \bibnamefont
  {Sharafutdinov}},\ }\bibfield  {title} {\bibinfo {title} {Rotational
  relaxation time in free jets of {H}e + {N}$_2$ mixtures},\ }\href
  {https://doi.org/https://doi.org/10.1016/0009-2614(95)00617-D} {\bibfield
  {journal} {\bibinfo  {journal} {Chemical Physics Letters}\ }\textbf {\bibinfo
  {volume} {241}},\ \bibinfo {pages} {209} (\bibinfo {year}
  {1995})}\BibitemShut {NoStop}%
\bibitem [{\citenamefont {Miller}\ and\ \citenamefont
  {Andres}(1967)}]{1967_Miller_N2_N2_inelastic_cross_section}%
  \BibitemOpen
  \bibfield  {author} {\bibinfo {author} {\bibfnamefont {D.~R.}\ \bibnamefont
  {Miller}}\ and\ \bibinfo {author} {\bibfnamefont {R.~P.}\ \bibnamefont
  {Andres}},\ }\bibfield  {title} {\bibinfo {title} {Rotational relaxation of
  molecular nitrogen},\ }\href {https://doi.org/10.1063/1.1841233} {\bibfield
  {journal} {\bibinfo  {journal} {The Journal of Chemical Physics}\ }\textbf
  {\bibinfo {volume} {46}},\ \bibinfo {pages} {3418} (\bibinfo {year}
  {1967})}\BibitemShut {NoStop}%
\bibitem [{\citenamefont {Belikov}\ \emph {et~al.}(1988)\citenamefont
  {Belikov}, \citenamefont {Solov'ev}, \citenamefont {Sukhinin},\ and\
  \citenamefont {Sharafutdinov}}]{1988_Belikov_N2_N2_inelastic_cross_section}%
  \BibitemOpen
  \bibfield  {author} {\bibinfo {author} {\bibfnamefont {A.~E.}\ \bibnamefont
  {Belikov}}, \bibinfo {author} {\bibfnamefont {I.~Y.}\ \bibnamefont
  {Solov'ev}}, \bibinfo {author} {\bibfnamefont {G.~I.}\ \bibnamefont
  {Sukhinin}},\ and\ \bibinfo {author} {\bibfnamefont {R.~G.}\ \bibnamefont
  {Sharafutdinov}},\ }\bibfield  {title} {\bibinfo {title} {Rotational
  relaxation time of nitrogen},\ }\href {https://doi.org/10.1007/BF00857905}
  {\bibfield  {journal} {\bibinfo  {journal} {Journal of Applied Mechanics and
  Technical Physics}\ }\textbf {\bibinfo {volume} {29}},\ \bibinfo {pages}
  {630} (\bibinfo {year} {1988})}\BibitemShut {NoStop}%
\bibitem [{\citenamefont
  {Morgan}(1975)}]{1975_Morgan_Breakdown_ionisation_review}%
  \BibitemOpen
  \bibfield  {author} {\bibinfo {author} {\bibfnamefont {C.~G.}\ \bibnamefont
  {Morgan}},\ }\bibfield  {title} {\bibinfo {title} {Laser-induced breakdown of
  gases},\ }\href {https://doi.org/10.1088/0034-4885/38/5/002} {\bibfield
  {journal} {\bibinfo  {journal} {Reports on Progress in Physics}\ }\textbf
  {\bibinfo {volume} {38}},\ \bibinfo {pages} {621} (\bibinfo {year}
  {1975})}\BibitemShut {NoStop}%
\bibitem [{\citenamefont {Ali}(1983)}]{1983_Ali_breakdown_threshold}%
  \BibitemOpen
  \bibfield  {author} {\bibinfo {author} {\bibfnamefont {A.}~\bibnamefont
  {Ali}},\ }\href@noop {} {\emph {\bibinfo {title} {On laser air breakdown,
  threshold power and laser generated channel length}}},\ \bibinfo {type}
  {Tech. Rep.}\ (\bibinfo  {institution} {Naval {R}esearch {L}ab, {W}ashington,
  {DC}},\ \bibinfo {year} {1983})\BibitemShut {NoStop}%
\bibitem [{\citenamefont {Rosen}\ and\ \citenamefont
  {Weyl}(1987)}]{1987_Rosen_breakdown_N2_noble_gases}%
  \BibitemOpen
  \bibfield  {author} {\bibinfo {author} {\bibfnamefont {D.~I.}\ \bibnamefont
  {Rosen}}\ and\ \bibinfo {author} {\bibfnamefont {G.}~\bibnamefont {Weyl}},\
  }\bibfield  {title} {\bibinfo {title} {Laser-induced breakdown in nitrogen
  and the rare gases at 0.53 and 0.357 \si{\micro\meter}},\ }\href
  {https://doi.org/10.1088/0022-3727/20/10/009} {\bibfield  {journal} {\bibinfo
   {journal} {Journal of Physics D: Applied Physics}\ }\textbf {\bibinfo
  {volume} {20}},\ \bibinfo {pages} {1264} (\bibinfo {year}
  {1987})}\BibitemShut {NoStop}%
\bibitem [{\citenamefont {Radziemski}\ and\ \citenamefont
  {Cremers}(1989)}]{1989_Radziemski_laser_induced_plasmas_and_applications}%
  \BibitemOpen
  \bibfield  {author} {\bibinfo {author} {\bibfnamefont {L.~J.}\ \bibnamefont
  {Radziemski}}\ and\ \bibinfo {author} {\bibfnamefont {D.~A.}\ \bibnamefont
  {Cremers}},\ }\href {https://doi.org/10.1201/9781003066200} {\emph {\bibinfo
  {title} {Laser-induced plasmas and applications.}}},\ Optical engineering:
  21\ (\bibinfo  {publisher} {M. Dekker},\ \bibinfo {year} {1989})\BibitemShut
  {NoStop}%
\bibitem [{\citenamefont {Isaacs}\ \emph {et~al.}(2016)\citenamefont {Isaacs},
  \citenamefont {Miao},\ and\ \citenamefont
  {Sprangle}}]{2016_Isaacs_breakdown_thresholds}%
  \BibitemOpen
  \bibfield  {author} {\bibinfo {author} {\bibfnamefont {J.}~\bibnamefont
  {Isaacs}}, \bibinfo {author} {\bibfnamefont {C.}~\bibnamefont {Miao}},\ and\
  \bibinfo {author} {\bibfnamefont {P.}~\bibnamefont {Sprangle}},\ }\bibfield
  {title} {\bibinfo {title} {Remote monostatic detection of radioactive
  material by laser-induced breakdown},\ }\href
  {https://doi.org/10.1063/1.4943404} {\bibfield  {journal} {\bibinfo
  {journal} {Physics of Plasmas}\ }\textbf {\bibinfo {volume} {23}},\ \bibinfo
  {pages} {033507} (\bibinfo {year} {2016})}\BibitemShut {NoStop}%
\bibitem [{\citenamefont {Brehme}(1971)}]{1971_Brehme_multiphoton_inv_bms}%
  \BibitemOpen
  \bibfield  {author} {\bibinfo {author} {\bibfnamefont {H.}~\bibnamefont
  {Brehme}},\ }\bibfield  {title} {\bibinfo {title} {Laser-induced multiphoton
  processes in $e^-$-$p$ scattering},\ }\href
  {https://doi.org/10.1103/PhysRevC.3.837} {\bibfield  {journal} {\bibinfo
  {journal} {Physical Review C}\ }\textbf {\bibinfo {volume} {3}},\ \bibinfo
  {pages} {837} (\bibinfo {year} {1971})}\BibitemShut {NoStop}%
\bibitem [{\citenamefont {Seely}\ and\ \citenamefont
  {Harris}(1973)}]{1972_Seely_Inverse_Bremsstrahlung_Full}%
  \BibitemOpen
  \bibfield  {author} {\bibinfo {author} {\bibfnamefont {J.~F.}\ \bibnamefont
  {Seely}}\ and\ \bibinfo {author} {\bibfnamefont {E.~G.}\ \bibnamefont
  {Harris}},\ }\bibfield  {title} {\bibinfo {title} {Heating of a plasma by
  multiphoton inverse bremsstrahlung},\ }\href
  {https://doi.org/10.1103/PhysRevA.7.1064} {\bibfield  {journal} {\bibinfo
  {journal} {Physical Review A}\ }\textbf {\bibinfo {volume} {7}},\ \bibinfo
  {pages} {1064} (\bibinfo {year} {1973})}\BibitemShut {NoStop}%
\bibitem [{\citenamefont {Cavaliere}\ \emph {et~al.}(1980)\citenamefont
  {Cavaliere}, \citenamefont {Ferrante},\ and\ \citenamefont
  {Leone}}]{1980_Cavaliere_light_assisted_e_impact_ionisation}%
  \BibitemOpen
  \bibfield  {author} {\bibinfo {author} {\bibfnamefont {P.}~\bibnamefont
  {Cavaliere}}, \bibinfo {author} {\bibfnamefont {G.}~\bibnamefont
  {Ferrante}},\ and\ \bibinfo {author} {\bibfnamefont {C.}~\bibnamefont
  {Leone}},\ }\bibfield  {title} {\bibinfo {title} {Particle-atom ionising
  collisions in the presence of a laser radiation field},\ }\href
  {https://doi.org/10.1088/0022-3700/13/22/021} {\bibfield  {journal} {\bibinfo
   {journal} {Journal of Physics B: Atomic and Molecular Physics}\ }\textbf
  {\bibinfo {volume} {13}},\ \bibinfo {pages} {4495} (\bibinfo {year}
  {1980})}\BibitemShut {NoStop}%
\bibitem [{\citenamefont {Zarcone}\ \emph {et~al.}(1983)\citenamefont
  {Zarcone}, \citenamefont {Moores},\ and\ \citenamefont
  {McDowell}}]{Zarcone_1983_light_assisted_impact_ionisation_He}%
  \BibitemOpen
  \bibfield  {author} {\bibinfo {author} {\bibfnamefont {M.}~\bibnamefont
  {Zarcone}}, \bibinfo {author} {\bibfnamefont {D.~L.}\ \bibnamefont
  {Moores}},\ and\ \bibinfo {author} {\bibfnamefont {M.~R.~C.}\ \bibnamefont
  {McDowell}},\ }\bibfield  {title} {\bibinfo {title} {Laser-assisted electron
  impact ionisation of helium at 256.5 {eV}},\ }\href
  {https://doi.org/10.1088/0022-3700/16/2/001} {\bibfield  {journal} {\bibinfo
  {journal} {Journal of Physics B: Atomic and Molecular Physics}\ }\textbf
  {\bibinfo {volume} {16}},\ \bibinfo {pages} {L11} (\bibinfo {year}
  {1983})}\BibitemShut {NoStop}%
\bibitem [{\citenamefont {Brunger}\ \emph {et~al.}(1992)\citenamefont
  {Brunger}, \citenamefont {Buckman}, \citenamefont {Allen}, \citenamefont
  {McCarthy},\ and\ \citenamefont
  {Ratnavelu}}]{1992_Brunger_He_e_elastic_cross_section}%
  \BibitemOpen
  \bibfield  {author} {\bibinfo {author} {\bibfnamefont {M.~J.}\ \bibnamefont
  {Brunger}}, \bibinfo {author} {\bibfnamefont {S.~J.}\ \bibnamefont
  {Buckman}}, \bibinfo {author} {\bibfnamefont {L.~J.}\ \bibnamefont {Allen}},
  \bibinfo {author} {\bibfnamefont {I.~E.}\ \bibnamefont {McCarthy}},\ and\
  \bibinfo {author} {\bibfnamefont {K.}~\bibnamefont {Ratnavelu}},\ }\bibfield
  {title} {\bibinfo {title} {Elastic electron scattering from helium: Absolute
  experimental cross sections, theory and derived interaction potentials},\
  }\href {https://doi.org/10.1088/0953-4075/25/8/016} {\bibfield  {journal}
  {\bibinfo  {journal} {Journal of Physics B: Atomic and Molecular Physics}\
  }\textbf {\bibinfo {volume} {25}},\ \bibinfo {pages} {1823} (\bibinfo {year}
  {1992})}\BibitemShut {NoStop}%
\bibitem [{\citenamefont {Ralchenko}\ \emph {et~al.}(2000)\citenamefont
  {Ralchenko}, \citenamefont {Janev}, \citenamefont {Kato}, \citenamefont
  {Fursa}, \citenamefont {Bray},\ and\ \citenamefont {{de
  Heer}}}]{2000_Ralchenko_electron_helium_cross_sections}%
  \BibitemOpen
  \bibfield  {author} {\bibinfo {author} {\bibfnamefont {Y.}~\bibnamefont
  {Ralchenko}}, \bibinfo {author} {\bibfnamefont {R.~K.}\ \bibnamefont
  {Janev}}, \bibinfo {author} {\bibfnamefont {T.}~\bibnamefont {Kato}},
  \bibinfo {author} {\bibfnamefont {D.~V.}\ \bibnamefont {Fursa}}, \bibinfo
  {author} {\bibfnamefont {I.}~\bibnamefont {Bray}},\ and\ \bibinfo {author}
  {\bibfnamefont {F.}~\bibnamefont {{de Heer}}},\ }\href
  {http://inis.iaea.org/search/search.aspx?orig_q=RN:32019416} {\emph {\bibinfo
  {title} {Cross section database for collision processes of helium atom with
  charged particles. {I}. {E}lectron impact processes}}},\ \bibinfo {type}
  {Tech. Rep.}\ \bibinfo {number} {{NIFS-DATA--59}}\ (\bibinfo  {institution}
  {NIFS},\ \bibinfo {address} {Japan},\ \bibinfo {year} {2000})\BibitemShut
  {NoStop}%
\bibitem [{\citenamefont {Montague}\ \emph {et~al.}(1984)\citenamefont
  {Montague}, \citenamefont {Harrison},\ and\ \citenamefont
  {Smith}}]{1984_Montague_He_e_ionisation}%
  \BibitemOpen
  \bibfield  {author} {\bibinfo {author} {\bibfnamefont {R.~G.}\ \bibnamefont
  {Montague}}, \bibinfo {author} {\bibfnamefont {M.~F.~A.}\ \bibnamefont
  {Harrison}},\ and\ \bibinfo {author} {\bibfnamefont {A.~C.~H.}\ \bibnamefont
  {Smith}},\ }\bibfield  {title} {\bibinfo {title} {A measurement of the cross
  section for ionisation of helium by electron impact using a fast crossed beam
  technique},\ }\href {https://doi.org/10.1088/0022-3700/17/16/012} {\bibfield
  {journal} {\bibinfo  {journal} {Journal of Physics B: Atomic and Molecular
  Physics}\ }\textbf {\bibinfo {volume} {17}},\ \bibinfo {pages} {3295}
  (\bibinfo {year} {1984})}\BibitemShut {NoStop}%
\bibitem [{\citenamefont {Song}\ \emph {et~al.}(2015)\citenamefont {Song},
  \citenamefont {Yoon}, \citenamefont {Cho}, \citenamefont {Itikawa},
  \citenamefont {Karwasz}, \citenamefont {Kokoouline}, \citenamefont
  {Nakamura},\ and\ \citenamefont
  {Tennyson}}]{2014_Song_Cross_Sections_for_electron_CH4_collisions}%
  \BibitemOpen
  \bibfield  {author} {\bibinfo {author} {\bibfnamefont {M.-Y.}\ \bibnamefont
  {Song}}, \bibinfo {author} {\bibfnamefont {J.-S.}\ \bibnamefont {Yoon}},
  \bibinfo {author} {\bibfnamefont {H.}~\bibnamefont {Cho}}, \bibinfo {author}
  {\bibfnamefont {Y.}~\bibnamefont {Itikawa}}, \bibinfo {author} {\bibfnamefont
  {G.~P.}\ \bibnamefont {Karwasz}}, \bibinfo {author} {\bibfnamefont
  {V.}~\bibnamefont {Kokoouline}}, \bibinfo {author} {\bibfnamefont
  {Y.}~\bibnamefont {Nakamura}},\ and\ \bibinfo {author} {\bibfnamefont
  {J.}~\bibnamefont {Tennyson}},\ }\bibfield  {title} {\bibinfo {title} {Cross
  sections for electron collisions with methane},\ }\href
  {https://doi.org/10.1063/1.4918630} {\bibfield  {journal} {\bibinfo
  {journal} {Journal of Physical and Chemical Reference Data}\ }\textbf
  {\bibinfo {volume} {44}},\ \bibinfo {pages} {023101} (\bibinfo {year}
  {2015})}\BibitemShut {NoStop}%
\bibitem [{\citenamefont {Fuss}\ \emph {et~al.}(2010)\citenamefont {Fuss},
  \citenamefont {{n}oz}, \citenamefont {Oller}, \citenamefont {Blanco},
  \citenamefont {Hubin-Franskin}, \citenamefont {Almeida}, \citenamefont {{a}o
  Vieira},\ and\ \citenamefont
  {Garc{\'i}a}}]{2010_Fuss_e_impact_CH4_cross_sections}%
  \BibitemOpen
  \bibfield  {author} {\bibinfo {author} {\bibfnamefont {M.}~\bibnamefont
  {Fuss}}, \bibinfo {author} {\bibfnamefont {A.~M.}\ \bibnamefont {{n}oz}},
  \bibinfo {author} {\bibfnamefont {J.}~\bibnamefont {Oller}}, \bibinfo
  {author} {\bibfnamefont {F.}~\bibnamefont {Blanco}}, \bibinfo {author}
  {\bibfnamefont {M.-J.}\ \bibnamefont {Hubin-Franskin}}, \bibinfo {author}
  {\bibfnamefont {D.}~\bibnamefont {Almeida}}, \bibinfo {author} {\bibfnamefont
  {P.~L.}\ \bibnamefont {{a}o Vieira}},\ and\ \bibinfo {author} {\bibfnamefont
  {G.}~\bibnamefont {Garc{\'i}a}},\ }\bibfield  {title} {\bibinfo {title}
  {Electron–methane interaction model for the energy range 0.1-10000 e{V}},\
  }\href {https://doi.org/https://doi.org/10.1016/j.cplett.2009.12.097}
  {\bibfield  {journal} {\bibinfo  {journal} {Chemical Physics Letters}\
  }\textbf {\bibinfo {volume} {486}},\ \bibinfo {pages} {110} (\bibinfo {year}
  {2010})}\BibitemShut {NoStop}%
\bibitem [{\citenamefont {Kim}\ \emph {et~al.}(2000)\citenamefont {Kim},
  \citenamefont {Johnson},\ and\ \citenamefont
  {Rudd}}]{2000_Kim_BED_ionisation_cross_section}%
  \BibitemOpen
  \bibfield  {author} {\bibinfo {author} {\bibfnamefont {Y.-K.}\ \bibnamefont
  {Kim}}, \bibinfo {author} {\bibfnamefont {W.~R.}\ \bibnamefont {Johnson}},\
  and\ \bibinfo {author} {\bibfnamefont {M.~E.}\ \bibnamefont {Rudd}},\
  }\bibfield  {title} {\bibinfo {title} {Cross sections for singly differential
  and total ionization of helium by electron impact},\ }\href
  {https://doi.org/10.1103/PhysRevA.61.034702} {\bibfield  {journal} {\bibinfo
  {journal} {Physical Review A}\ }\textbf {\bibinfo {volume} {61}},\ \bibinfo
  {pages} {034702} (\bibinfo {year} {2000})}\BibitemShut {NoStop}%
\bibitem [{\citenamefont {Hwang}\ \emph {et~al.}(1996)\citenamefont {Hwang},
  \citenamefont {Kim},\ and\ \citenamefont
  {Rudd}}]{1996_Hwang_BED_ionisation_cross_section}%
  \BibitemOpen
  \bibfield  {author} {\bibinfo {author} {\bibfnamefont {W.}~\bibnamefont
  {Hwang}}, \bibinfo {author} {\bibfnamefont {Y.}~\bibnamefont {Kim}},\ and\
  \bibinfo {author} {\bibfnamefont {M.~E.}\ \bibnamefont {Rudd}},\ }\bibfield
  {title} {\bibinfo {title} {New model for electron-impact ionization cross
  sections of molecules},\ }\href {https://doi.org/10.1063/1.471116} {\bibfield
   {journal} {\bibinfo  {journal} {The Journal of Chemical Physics}\ }\textbf
  {\bibinfo {volume} {104}},\ \bibinfo {pages} {2956} (\bibinfo {year}
  {1996})}\BibitemShut {NoStop}%
\bibitem [{\citenamefont {Itikawa}(2006)}]{2006_Itikawa_e_N2_cross_sections}%
  \BibitemOpen
  \bibfield  {author} {\bibinfo {author} {\bibfnamefont {Y.}~\bibnamefont
  {Itikawa}},\ }\bibfield  {title} {\bibinfo {title} {Cross sections for
  electron collisions with nitrogen molecules},\ }\href
  {https://doi.org/10.1063/1.1937426} {\bibfield  {journal} {\bibinfo
  {journal} {Journal of Physical and Chemical Reference Data}\ }\textbf
  {\bibinfo {volume} {35}},\ \bibinfo {pages} {31} (\bibinfo {year}
  {2006})}\BibitemShut {NoStop}%
\bibitem [{\citenamefont {Itikawa}(2009)}]{2008_Itikawa_e_O2_cross_sections}%
  \BibitemOpen
  \bibfield  {author} {\bibinfo {author} {\bibfnamefont {Y.}~\bibnamefont
  {Itikawa}},\ }\bibfield  {title} {\bibinfo {title} {Cross sections for
  electron collisions with oxygen molecules},\ }\href
  {https://doi.org/10.1063/1.3025886} {\bibfield  {journal} {\bibinfo
  {journal} {Journal of Physical and Chemical Reference Data}\ }\textbf
  {\bibinfo {volume} {38}},\ \bibinfo {pages} {1} (\bibinfo {year}
  {2009})}\BibitemShut {NoStop}%
\end{thebibliography}%

\end{document}